\documentclass[onecolumn,notitlepage,nofootinbib]{revtex4-1}
\pdfoutput=1

\usepackage{graphicx}
\usepackage{latexsym}
\usepackage{amssymb}
\usepackage{multirow}
\usepackage{amsmath}
\usepackage{mathrsfs}  
\usepackage{xcolor}
\usepackage{footmisc}
\usepackage[T1]{fontenc}
\usepackage[utf8]{inputenc}
\usepackage{hyperref}
\usepackage{textalpha}
\usepackage{subcaption}

\newcommand{\GRAPPA}{Gravitation Astroparticle Physics Amsterdam (GRAPPA), Institute for Theoretical Physics Amsterdam and Delta Institute for Theoretical Physics, University of Amsterdam, Science Park 904, 1098 XH Amsterdam, The Netherlands}

\newcommand{\INFN}{INFN - Sezione di Napoli, Complesso Univ. Monte S. Angelo, I-80126 Napoli, Italy}

\newcommand{\UNI}{Dipartimento di Fisica "Ettore Pancini", Universit\'a degli studi di Napoli "Federico II", Complesso Univ. Monte S. Angelo, I-80126 Napoli, Italy}

\begin{document}

\title{Decaying dark matter at IceCube and its signature on High Energy gamma experiments}

\author{Marco Chianese}
\email{m.chianese@uva.nl}
\affiliation{\GRAPPA}

\author{Damiano F. G. Fiorillo}
\email{dfgfiorillo@na.infn.it}
\affiliation{\INFN}
\affiliation{\UNI}

\author{Gennaro Miele}
\email{miele@na.infn.it}
\affiliation{\INFN}
\affiliation{\UNI}

\author{Stefano Morisi}
\email{smorisi@na.infn.t}
\affiliation{\INFN}
\affiliation{\UNI}

\author{Ofelia Pisanti}
\email{pisanti@na.infn.it}
\affiliation{\INFN}
\affiliation{\UNI}

\begin{abstract}
The origin of neutrino flux observed in IceCube is still mainly unknown. Typically two flux components are assumed, namely:  atmospheric neutrinos and an unknown astrophysical term. In principle the latter could also contain a top-down contribution coming for example from decaying dark matter. In this case one should also expect prompt and secondary gamma's as well. This leads to the possibility of a multimessenger analysis based on the simultaneous comparison of the Dark Matter hypothesis both with neutrino and high energy gamma rays data. In this paper, we analyze, for different decaying Dark Matter channels, the $7.5$ years IceCube HESE data, and compare the results with previous exclusion limits coming from Fermi data. Finally,  we test whether the Dark Matter hypothesis could be further scrutinised by using forthcoming high energy gamma rays experiments.
\end{abstract}

\maketitle

\section{Introduction}

After the IceCube recent measurements of ultra-high energy (UHE) neutrinos at TeV-PeV  \cite{Aartsen:2013jdh,Aartsen:2014gkd,Aartsen:2017mau} novel possibilities have opened up for astroparticle physics. We are specially interested for the present work in the High-Energy Starting Events (HESE) data set. While it is clear that these neutrinos cannot be explained in terms of the atmospheric neutrino background, their astrophysical origin is still not fully understood. The astrophysical processes which are expected to be the source of these highly energetic neutrinos occur through the decay of charged pions, in turn originated in hadronic~\cite{Loeb:2006tw,Murase:2013rfa,Tamborra:2014xia,Bechtol:2015uqb} and photo-hadronic~\cite{Winter:2013cla,Murase:2015xka} interactions. Together with neutrinos, gamma rays are produced as well. This simultaneous production has recently been confirmed by the multi-messenger observation of the coincident gamma-ray and neutrino emission from the flaring blazar TXS 0506+056~\cite{IceCube:2018cha,IceCube:2018dnn,Ahnen:2018mvi}. It has been nevertheless pointed out that blazar flares can only contribute up to about 10\% of the total observed neutrino flux~\cite{Murase:2018iyl}. Moreover, other analyses of spatial and temporal correlations with gamma-rays~\cite{Adrian-Martinez:2015ver, Aartsen:2016oji, Aartsen:2018fpd,Aartsen:2018ywr} and for angular clustering~\cite{Murase:2016gly, Aartsen:2014ivk,Ando:2017xcb,Mertsch:2016hcd,Dekker:2018cqu} have placed strong constraints on the contribution of several extragalactic astrophysical sources. Allowed astrophysical candidates are therefore hidden cosmic-ray accelerators for which the gamma-ray flux is highly suppressed~\cite{Murase:2015xka,Senno:2015tsn}. The neutrino flux, which is therefore expected to be the superposition of many astrophysical sources, can be parameterized as a diffuse power-law spectrum $E_\nu^{-\gamma}$.

At the same time there are however tensions among different IceCube data samples. Under the assumption of a single power-law flux, the latest 10-year through-going (TG) muon events lead to a spectral index of $2.28^{+0.08}_{-0.09}$ as best-fit~\cite{ICRC_TG}. This is mildly compatible with the theoretical expectations from a neutrino flux originating by photohadronic production from charged particles accelerated through the Fermi mechanism \cite{Waxman:1998yy} and with the measurements of the blazar TXS 0506+056, which suggest a hard power-law flux with a spectral index in the range $2.0\div2.3$~\cite{IceCube:2018dnn}. However, the latest $7.5$ years High Energy Starting Events (HESE) have the best-fit $\gamma=2.89^{+0.20}_{-0.19}$~\cite{ICRC_HESE}.

A spectral index of about $2.3$ is in tension with this result. This might suggest the presence of two components that dominate the diffuse neutrino flux at different energies with different angular properties~\cite{Aartsen:2017mau, Aartsen:2015knd}. This last conclusion is favored by the observation that the TG data sample collects muon neutrinos from the Northern hemisphere only, with energies larger that 200~TeV. The HESE events cover instead the whole sky, so they are sensitive to the galactic centre of the Milky Way; further, they are collected in an energy range starting already at 20 TeV. Investigations in the framework of two different power law sources have led to the conclusion that the data are compatible with the sum of a hard isotropic extragalactic neutrino flux and an additional softer one with a potential galactic origin~\cite{Chen:2014gxa,Palladino:2016zoe,Vincent:2016nut,Palladino:2016xsy,Anchordoqui:2016ewn,Palladino:2018evm,Sui:2018bbh}. This two-component hypothesis is at the same time supported by the first combined analysis of IceCube and ANTARES data~\cite{Chianese:2017jfa}. It turns out that both IceCube and ANTARES telescopes have measured in the same energy range (about 40--200 TeV) a slight excess with respect to an astrophysical power-law flux deduced by TG data ($\gamma \leq 2.2$), after the background subtraction~\cite{Chianese:2017jfa,Chianese:2016opp,Chianese:2016kpu,Chianese:2017nwe}.

An alternative source for this diffuse UHE neutrino flux is decaying Dark Matter (DM)~\cite{Chianese:2016opp,Chianese:2016kpu,Chianese:2017nwe,Anisimov:2008gg,Feldstein:2013kka,Esmaili:2013gha,Bai:2013nga,Ema:2013nda,Esmaili:2014rma,Bhattacharya:2014vwa, Higaki:2014dwa,Rott:2014kfa,Ema:2014ufa, Murase:2015gea,Dudas:2014bca,Fong:2014bsa,Aisati:2015vma,Ko:2015nma,Dev:2016qbd,Fiorentin:2016avj,DiBari:2016guw,Anchordoqui:2015lqa,Dev:2016uxj,Chianese:2016smc,Borah:2017xgm,Boucenna:2015tra,Hiroshima:2017hmy,Bhattacharya:2017jaw,Bhattacharya:2019ucd,Kopp:2015bfa}. While another available source might be identified with annihilating DM~\cite{Chianese:2016opp,Feldstein:2013kka,Zavala:2014dla,ElAisati:2017ppn,Sui:2018bbh,Kachelriess:2018rty,Chianese:2018ijk,Bhattacharya:2019ucd}, the unitarity limit leads in general to small neutrinos fluxes which are therefore not detectable. Bounds are available in previous studies both for decaying \cite{Aartsen:2018mxl,Abeysekara:2017jxs} and annihilating DM \cite{Albert:2016emp,Gozzini:2019esk}. Analyses mainly devoted to the neutrino energy spectrum are able to distinguish among different DM decay channels. Moreover, different DM models are further constrained through the informations coming from gamma-ray observations. The current data (neutrinos and gamma-rays) show that hadronic final states are strongly disfavored or excluded, while leptonic ones are slightly disfavored or allowed~\cite{Cohen:2016uyg,Blanco:2018esa}. The most favorable case is a heavy DM candidate coupled only to neutrinos.

The foregoing observations have shown that gamma-rays constraints, mainly coming from observations of the Fermi-LAT experiment, have been critical in excluding or disfavoring various DM decay channels. It is therefore all the more interesting to find out whether gamma-rays experiments in different energy ranges would be able to put further constraints on the existence of decaying DM fluxes. While the lower energy range (up to few hundreds GeV) is mainly explored by Fermi LAT \cite{Atwood:2009ez}, the higher energy range (from about $100$~TeV) is explored by air shower experiments \cite{Kalashev:2016cre} and has already been used in exploring constraints on annihilating or decaying Dark Matter through data by Pierre Auger \cite{Aab:2015bza}, CASA-MIA \cite{Chantell:1997gs} and KASCADE \cite{Kang:2015gpa}; a future experiment along these lines will be LHAASO \cite{Cao:2010zz}. We are much interested in the intermediate range. Here the primary sources of information come from Cherenkov telescopes experiments such as the bounds coming from MAGIC \cite{Acciari:2018sjn}, HESS \cite{Rinchiuso:2019rrh}, Veritas \cite{Zitzer:2017xlo}, HAWC \cite{Abeysekara:2017jxs} and CARPET-3 \cite{Dzhappuev:2018bnl}. HAWC in particular has been able to put bounds on decaying Dark Matter, which are not, however, competitive with the Fermi-LAT ones. Data coming from these experiments have mainly been used in point-source investigations of Dark Matter decays and annihilation. A crucial development in this class will be the future CTA \cite{Ong:2019zyq}, which will overcome the previously mentioned Cherenkov experiments by an order of magnitude in its sensitivity. A number of studies on Dark Matter detection at CTA are already available, both for annihilation \cite{Ibarra:2015tya,Silverwood:2014yza,Hryczuk:2019nql} and for decay \cite{Pierre:2014tra}. 
It is further interesting to notice that an excess in the TeV region has also been observed by Fermi LAT \cite{Neronov:2018ibl,Neronov:2019ncc,Ackermann:2014usa}. This excess has been interpreted in terms of a decaying Dark Matter signal as well.

The primary aim of this paper is to provide an updated analysis of the IceCube data at $7.5$ years in the two-component hypothesis of an astrophysical spectrum, parameterized as a power law with a spectral index assumed as a fit parameter in the range between $1.5$ and $5$, and a diffuse spectrum produced from the decaying Dark Matter. We will further explore the possibilities of constraining the decaying Dark Matter hypothesis both through diffuse and through point-like investigations in the region of interest to IceCube.

The analysis we have performed is described in the following sections. In Section II we discuss the computation of the expected number of neutrino events at IceCube coming from decaying Dark Matter. In Section III the same ideas are generalized to the expected gamma fluxes on Earth: this case is more complicated, due to the presence of photon absorption and secondary production. In Section IV we describe the statistical procedure which has been adopted to compare the theoretical expectations with the latest HESE IceCube data, and the results of such analysis. Finally, in Section V we perform an investigation of the detection possibilities for the gamma rays fluxes both from diffuse and point-like searches. In Section~VI, we draw our conclusions.

\section{Neutrinos from decaying Dark Matter}\label{sec2}

The neutrino flux from decaying Dark Matter unavoidably depends upon the decay mechanism, and, specifically, upon the final products of the decay. A complete analysis of the possible models which can be constructed is beyond the scope of this paper \cite{Bertone:2004pz}. A practical way to proceed is to consider the decays of Dark Matter into a pair of Standard Model particle and antiparticle. In the following we will therefore consider separately the following decaying channels:
\begin{eqnarray}
\chi&\to& \ell \overline{\ell}  \qquad {\rm with} \qquad \ell=e,\mu,\tau,\nu\nonumber\\
 &\to & q\overline{q} \qquad {\rm with} \qquad q =u,c,t,b\nonumber\\
 &\to & ZZ,\,W^+W^-,hh,\,gg\nonumber
\end{eqnarray}
where we have denoted by $\chi$ the Dark Matter candidate. We do not consider here the $\gamma\gamma$ decay channel, which is severely constrained by gamma-ray experiments. The $d$ and $s$ quark channels give essentially the same results as the $u$ channel. Neutrino fluxes from Dark Matter decay are generated both inside the Galaxy and from extragalactic Dark Matter sources:
\begin{equation}
\phi_\chi = \phi_{\rm G}+\phi_{\rm EG}
\end{equation}
While for the former we can assume a Navarro-Frank-White (NFW) profile density, for the latter we have to assume an isotropic distribution in accordance with the Cosmological Principle.  The relevance of the choice in the Galactic Dark Matter profile density is not as strong as in the analysis of annihilating Dark Matter, due to the dependence in the latter on the square of the Dark Matter density. More detailed analyses of the effects of the profile choice can be found in \cite{Cirelli:2010xx,Pierre:2014tra}. The Galactic component of the neutrino flux can then be represented as:
\begin{equation}
\frac{d^2\phi_G}{dEd\Omega}=\frac{1}{4\pi m_{\rm DM} \tau} \frac{dN_{\nu+\bar{\nu}}}{dE} \int ds \, \rho(s,l,b) 
\end{equation}
where $m_{\rm DM}$ is the Dark Matter mass, $\tau$ is the Dark Matter lifetime, $dN_{\nu+\bar{\nu}}/dE$ is the differential spectrum of neutrinos produced per Dark Matter decay and the integral is taken over the line-of-sight. The angular dependence is parameterized through the density dependence on the Galactic coordinates, dictated by the NFW profile:
\begin{equation}
\rho(r(s,l,b))=\rho_0 \left[ \frac{r}{r_c}\left(1+\frac{r}{r_c}\right)^2 \right]^{-1}
\end{equation}
where:
\begin{equation}
r(s,l,b)=\sqrt{s^2+R_s^2-2sR_s \cos b \cos l}
\end{equation}
The numerical values for the parameters appearing in the equations are $\rho_0=0.33$~$\rm{GeV cm^{-3}}$, $r_c=20$~$\rm{kpc}$, $R_s=8.5$~$\rm{kpc}$. The extragalactic component is in the same way obtained from summing the neutrinos produced from different cosmological distances parameterized by the redshift. Here we have to take into account the physical difference between the energy at the moment of production and detection of the neutrino, as well as the difference between the decay rate at production and detection of neutrinos. The result for the flux is: 
\begin{equation}
\frac{d\phi_{\rm EG}}{dEd\Omega}=\frac{\Omega_{\chi} \rho_{cr}}{4\pi m_{\rm DM} \tau} \int \frac{dz}{H(z)} \frac{dN_{\nu+\bar{\nu}}(E(1+z))}{dE} 
\end{equation}
where $H(z)=H_0 \sqrt{\Omega_{\Lambda}+\Omega_m (1+z)^3}$ is the Hubble parameter, $\Omega_i=\rho_i/\rho_{cr}$ and $\rho_{cr}$ is the critical density of the Universe.
The fluxes have to be computed for all three flavors separately, since, as we will see, they have to be treated in different ways in the analysis of the neutrino detection. We also take into account the neutrino oscillations. We are now going to describe in more detail the computation of the neutrino spectra per decay.

The spectrum of particles produced in a decay or a collision of Dark Matter particles is typically obtained using simulation programs such as HERWIG \cite{Bahr:2008pv} or PYTHIA \cite{Sjostrand:2014zea}. For Dark Matter masses lower than about $100$ $\rm{TeV}$ a package named PPPC is available \cite{Cirelli:2010xx} which already contains the spectra interpolated for most of the decay channels. A critical factor for such high masses of the Dark Matter is the showering which immediately follows the initial decay. While the electromagnetic part of the shower is amenable to analytical treatment and generally included in the softwares for simulation, and the hadronic part  is already included both in HERWIG and PYTHIA, the weak showering is harder to describe. The reason is that, while in the first two cases the Sudakov double logarithms appearing in the probabilities for gauge boson emission cancel with the virtual corrections for the process with no gauge boson emission, in the weak case the $W$ boson emission changes the final state, so that the two processes lead to physically distinguishable final states. The appearance of double logarithms renders the perturbation results invalid already at energies of about $100$ $\rm{TeV}$, and in fact the results of \cite{Cirelli:2010xx}, obtained through perturbation theory, are only valid up to this energy. The higher energy range requires a resummation to all orders of the process. This has been partly implemented \footnote{With some limitations which are described in detail in \cite{Christiansen:2014kba}.} in PYTHIA. Different approach has been given for instance in \cite{Kachelriess:2018rty, Berezinsky:2002hq}. 

\begin{figure}[h!]
\begin{center}
\includegraphics[width=0.32\textwidth]{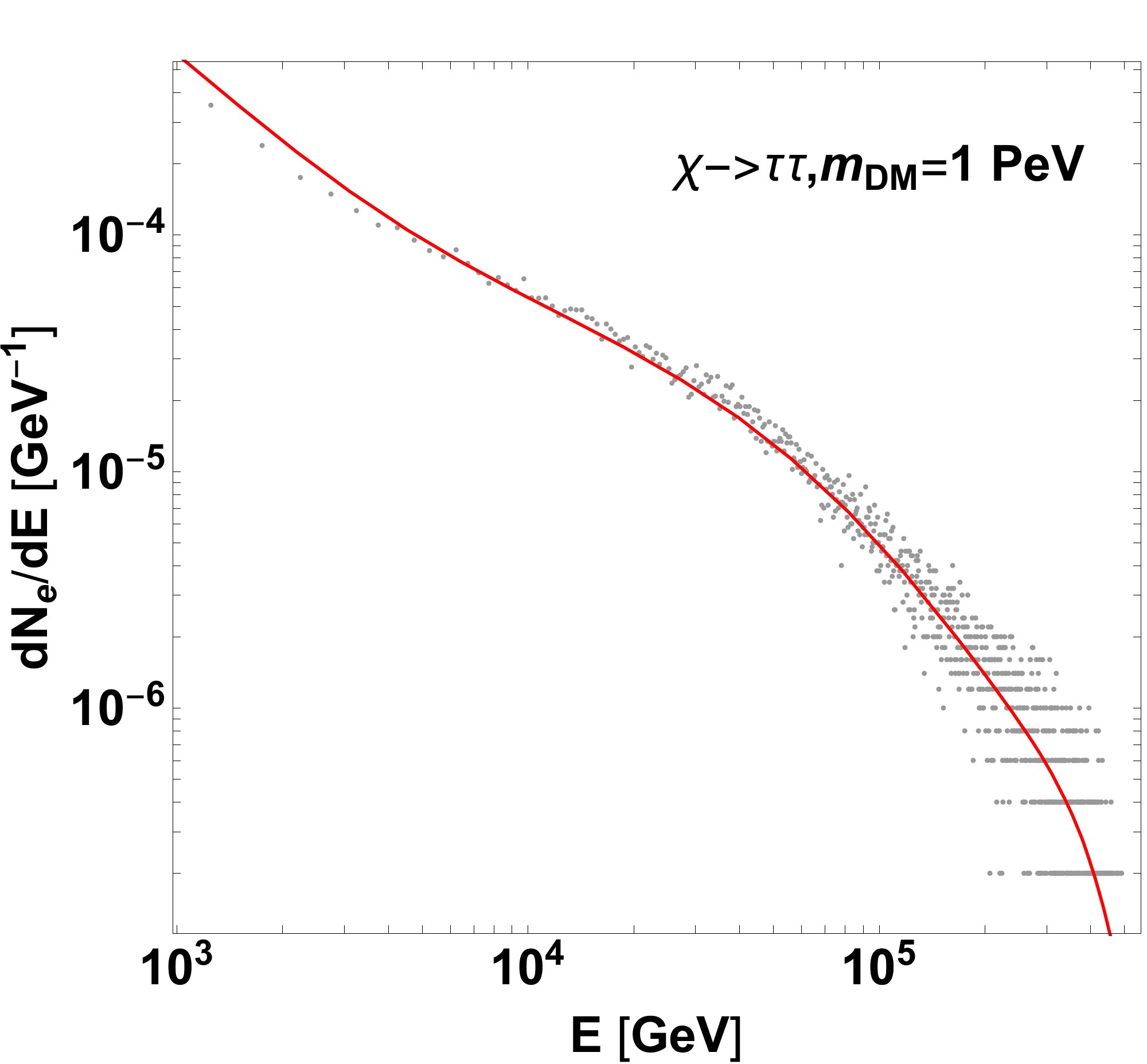}
\includegraphics[width=0.32\textwidth]{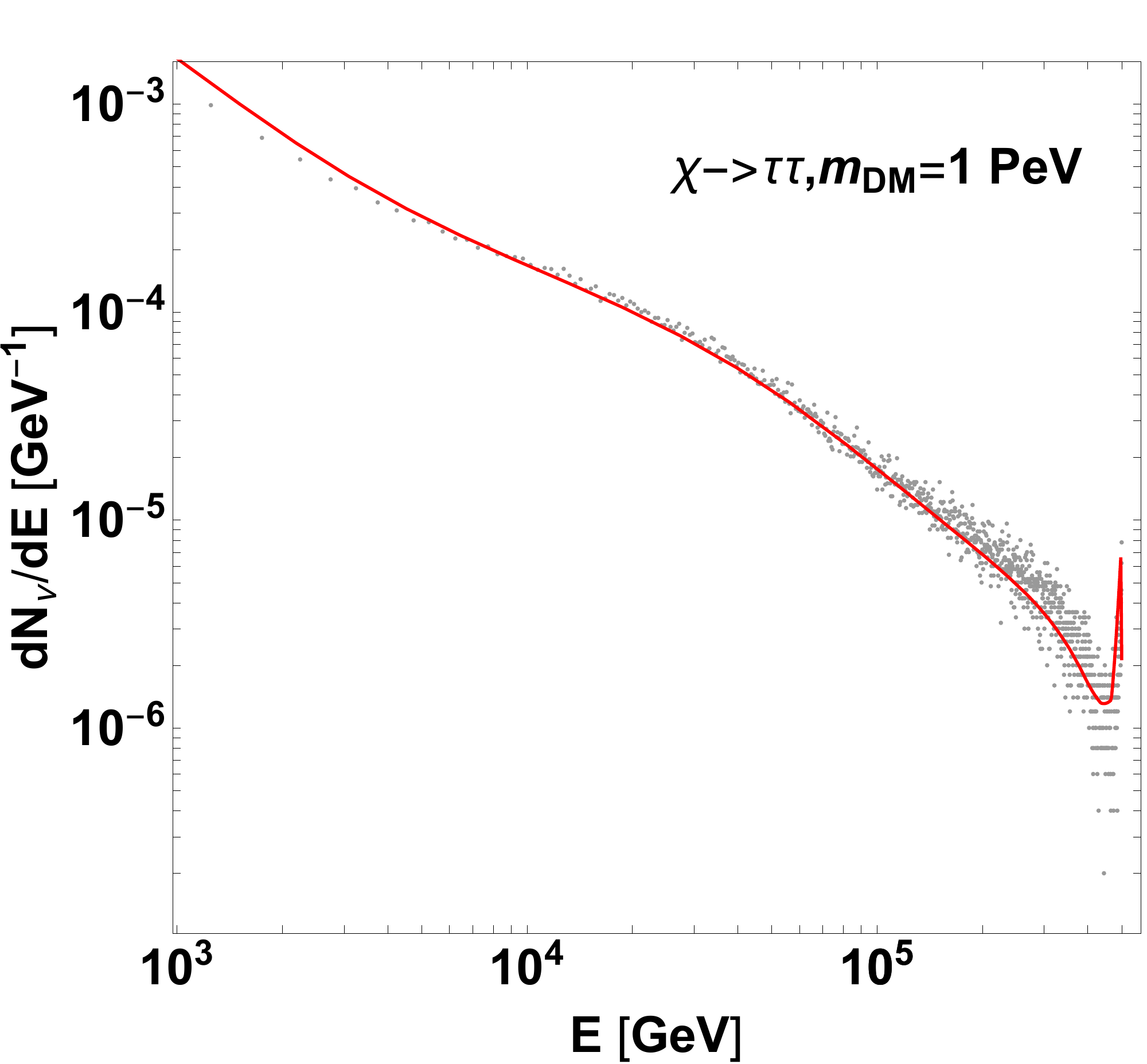}
\includegraphics[width=0.32\textwidth]{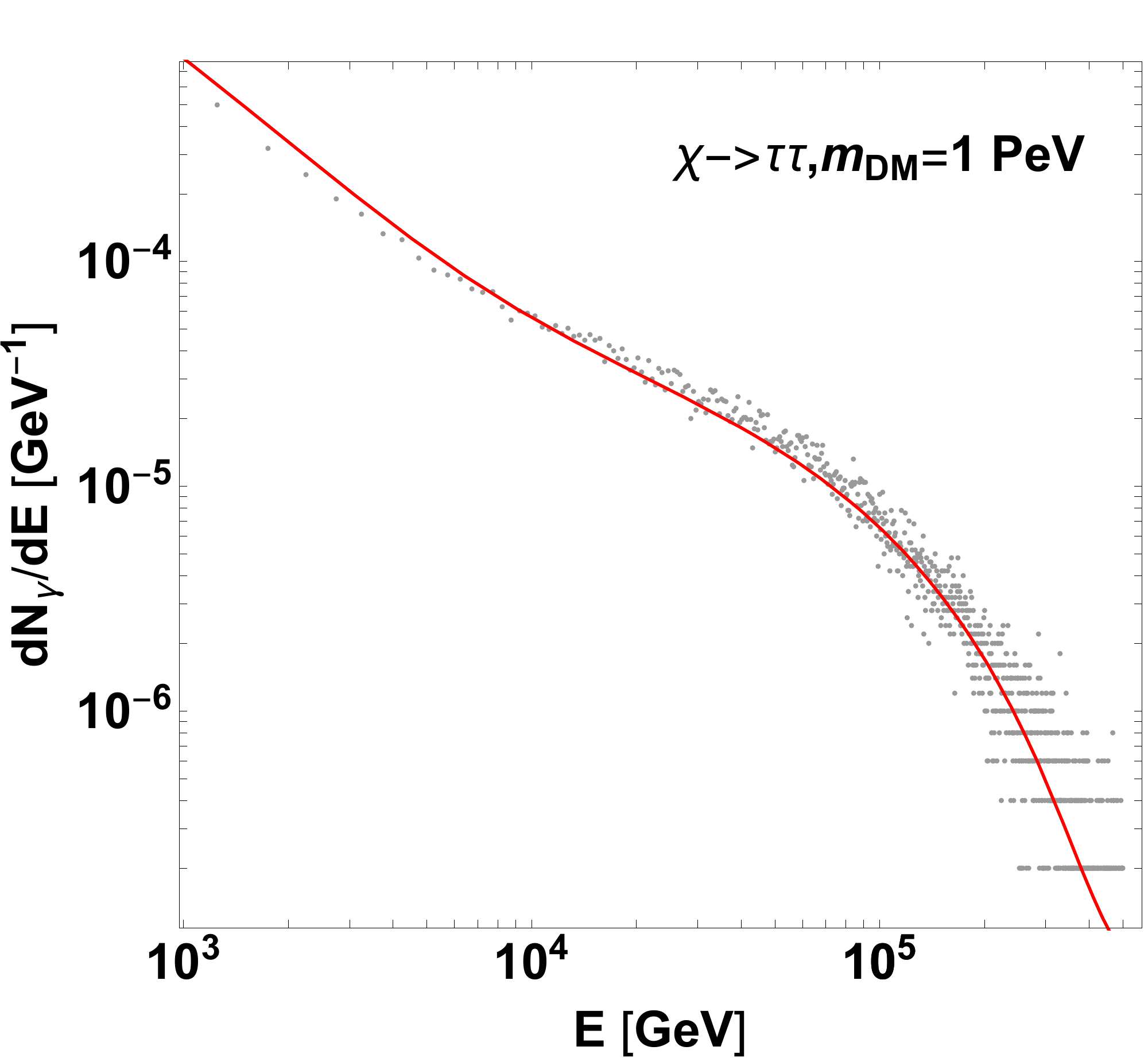}
\end{center}
\caption{Comparison between Pythia (gray points) and PPPC \cite{Cirelli:2010xx} (red line) spectra obtained using rescaling for a Dark Matter mass of $1$ $\rm{PeV}$ and decay channel into $\tau$: electron, neutrino and gamma spectra.}\label{fig1}
\end{figure}

Although the resummation of logarithms is not a trivial procedure, a guess which can be made based on pure dimensional analysis is that these differential spectra only depends on the ratio $E/m_{\rm DM}$. In fact, the only other quantity coming from weak physics and entering the result could be the mass $M_W$ of $W$ boson, but at energies much higher than this mass we would expect the result to be independent of this parameter. Of course the guess is too naive, since logarithms depend singularly on the ratio $m_{\rm DM}/M_W$ and might give rise to anomalous behaviors. We have therefore tested this hypothesis by comparing the spectra obtained using PYTHIA at masses of orders up to the $\rm{PeV}$ and the same spectra obtained by applying the scaling hypothesis to the spectra in \cite{Cirelli:2010xx} (see figure \ref{fig1}). The comparison shows that the scaling hypothesis is justified and we have therefore used it in our calculations. This amounts to using the form
\begin{equation}
\frac{dN_i}{dE}(m_{\rm DM},E)=\frac{dN_i}{dE}\left(M_\Lambda,\frac{E}{m_{\rm DM}}M_\Lambda\right)\frac{M_\Lambda}{m_{\rm DM}}
\end{equation}
as energy spectrum for Dark Matter masses $m_{\rm DM}$ larger than a cutoff mass $M_\Lambda = 200$~TeV, which is the upper limit of the interpolated spectra provided in \cite{Cirelli:2010xx}.

\section{Gamma rays from decaying dark matter}

A further source of information from decaying Dark Matter might come from the photons produced during the decay and in the subsequent propagation of particles from their decay place. We here discuss this aspect of our analysis.
The production of gamma rays from Dark Matter decays has already been studied in various works (see, e.g., \cite{Cohen:2016uyg}, \cite{Esmaili:2015xpa} and, at lower energies, \cite{Buch:2015iya}). 

This production can be divided, as in the case of neutrinos, in a Galactic and an Extragalatic component. There are, however, two main differences between photons and neutrinos. On the one hand, while the latter have roughly no interaction whatsoever during their propagation to Earth, the former are able, at high energies, to scatter off photons and electrons, accounting for an effective attenuation which, while for the Galactic component starts giving sensible consequences at electron energies higher than about $100$ $\rm{TeV}$ \cite{Esmaili:2015xpa}, is already important at lower energies for the Extragalactic component. On the other hand, while neutrinos are only produced during Dark Matter decays or in the electroweak and hadronic cascade which immediately follows, photons can also be produced by secondary electrons and positrons by the mechanisms of bremsstrahlung, synchrotron radiation (in the Galactic magnetic field) and Inverse Compton from various sources. A numerical estimate shows that at high energies the relevant process is Inverse Compton. Furthermore, the analysis in \cite{Esmaili:2015xpa} shows that at such high energies the most important target for Inverse Compton is the Cosmic Microwave Background (CMB). A summary of all the relevant components is given in Table\,\ref{tab}.
\begin{table}[h!]
\begin{center}
\begin{tabular}{|c|c|c|}
\hline
 & Primary&Secondary  \\
\hline
\hline
 Galactic & Prompt* & B, S,  IC by SL+IR, IC  by CMB* \\
\hline
Extragalactic & Prompt with attenuation* & Extra-Galactic IC by CMB with attenuation* \\ 
\hline
\end{tabular}\caption{Contributions to the gamma spectrum: Bremsstrahlung (B), Synchrotron (S), Inverse Compton (IC) on Cosmic Microwave Background (CMB), Inverse Compton on Star Light (SL) and Infrared (IR) photons. The components which we take into account are denoted by a star.}
\label{tab}
\end{center}
\end{table}
We can therefore make the simplifying assumption that the four main sources of gamma rays are the so called prompt gamma rays, produced directly during the Dark Matter decay, both in Galaxy and outside Galaxy; and the secondary gamma rays, produced by Inverse Compton of the electrons and positrons produced during the Dark Matter decay.

\subsection{Prompt}

The prompt Galactic spectrum is obtained in the same way as the prompt neutrino spectrum described previously. Thus, we have:
\begin{equation}
\frac{d^2\phi_{\rm G}}{dEd\Omega}=\frac{1}{4\pi m_{\rm DM} \tau} \frac{dN_{\gamma}}{dE} \int ds \, \rho(s,l,b) 
\end{equation}
For the prompt Extragalactic spectrum the neutrino form needs to be corrected by an attenuation factor, coming from collisions $\gamma-\gamma$ and $\gamma-e$. This attenuation factor ${\rm Att}(E,z)$ as a function of the photon detected energy and the redshift has been obtained in \cite{Cirelli:2010xx}. The Extragalactic prompt spectrum can therefore be expressed as:

\begin{equation}
\frac{d^2\phi_{\rm EG}}{dEd\Omega}=\frac{\Omega_{\chi} \rho_{cr}}{4\pi m_{\rm DM} \tau} \int \frac{dz}{H(z)} {\rm Att}(E, z) \frac{dN_{\gamma}(E(1+z))}{dE} 
\end{equation}

\subsection{Secondary}

Both the Galactic and the Extragalactic secondary gamma ray spectra are produced by electrons and positrons, and therefore depend critically upon the distributions of the latter. Such distribution should in principle be obtained by solving the kinetic equations for the positrons in the Galaxy and outside the Galaxy. For these equations we can require the equilibrium solution, since the decaying Dark Matter can be regarded on cosmic time scales as a steady source of particles. As is known from the theory of cosmic rays, these equations would have the schematic form:
\begin{equation} \label{diffusion}
\nabla \cdot \left(D\nabla n(\mathbf{x},E)\right)-\frac{\partial}{\partial E} \left( b(E) n(\mathbf{x},E)\right)=q 
\end{equation}
Here $D$ is a diffusion constant originated from the scattering of positrons on the Alfvèn waves and magnetic field fluctuations, $b(E)$ is the energy loss of the electrons, which at high energies is mainly connected with synchrotron and Inverse Compton losses, $q$ is the source term, in our case described by the injected spectrum of electrons and positrons from decaying Dark Matter and $n(E)$ is the number density of electrons per unit energy interval. While the diffusion constant is known to weakly increase with energy through a power law with exponent smaller than $1$ (see, e.g., \cite{Buch:2015iya}), the energy loss components increase much more rapidly. For example, the synchrotron energy loss grows as the square of energy, and the Inverse Compton losses increase in a slightly lower fashion. We can draw the conclusions that the diffusion in the energy space is much more important than the spatial diffusion, and therefore we can neglect the first term in \eqref{diffusion}. The solution of the resulting equation is then:
\begin{equation} \label{equi}
n(\mathbf{x},E)=\frac{1}{b(E)}\int_{E}^{m_{\rm DM}/2} q(E') dE'
\end{equation}
The energy losses due to Inverse Compton on CMB are:
\begin{eqnarray}
b_{\rm IC}(E)=3\sigma_T \int_0^{+\infty} \epsilon d\epsilon \int_{\frac{1}{4\gamma^2}}^1 dq \quad n(\epsilon,\mathbf{x}) \frac{(4\gamma^2 - \Gamma_\epsilon) q -1}{(q+\Gamma_\epsilon q)^3} \nonumber \\ \left[ 2q\ln q + q +1-2q^2+\frac{(\Gamma_\epsilon q)^2 (1-q)}{2(1+\Gamma_\epsilon q)}\right]
\end{eqnarray}
where, using the notation of \cite{Esmaili:2015xpa}, $\Gamma_\epsilon = 4\epsilon\gamma/m_e$ and $\gamma$ is the Lorentz factor of the electron. The quantity $\sigma_T$ is the Thomson cross section. The synchrotron energy losses are \cite{Esmaili:2015xpa}:
\begin{equation}
b_{\rm SYN}(E)=\frac{4\sigma_T E^2}{3m_e^2} \frac{\mathbf{B}^2(\mathbf{x})}{2}
\end{equation}
where for the Galactic magnetic field we have adopted the parameterization
\begin{equation}
B(r,z)=B_0 \exp\left\{-\frac{|r-R_{sun}|}{r_B}-\frac{|z|}{z_B}\right\}
\end{equation}
with $B_0=4.78$ $\rm{\mu G}$, $R_{sun}=8.3$ $\rm{kpc}$, $r_B=10$ $\rm{kpc}$, $z_B=2$ $\rm{kpc}$, and $r$ and $z$ being the radial and vertical distances from the Galactic center, respectively. We have here neglected the turbulent halo magnetic field. This of course has its most important effects at the high energies, where the synchrotron energy losses become dominant over the Inverse Compton losses. However, as shown in \cite{Esmaili:2015xpa}, these effects become relevant at photon energies of order $100$ $\rm{TeV}$.

The Inverse Compton spectrum $d^2\phi_{\rm GIC}/dEd\Omega$ will now be the convolution of this electron distribution $n(E)$ given in \eqref{equi} with the Inverse Compton emissivity on CMB, namely:
\begin{eqnarray}
P_{\rm IC}(E_\gamma,E_e)=\frac{3\sigma_T E_\gamma}{4\gamma^2} \int_{1/4\gamma^2}^1 dq \left[1-\frac{1}{4q\gamma^2 (1-\epsilon)}\right] \frac{n(E^0_\gamma)}{q} \nonumber \\ \left[2q\ln q +q +1-2q^2+\frac{\epsilon^2 (1-q)}{2(1-\epsilon)}\right]
\end{eqnarray}
where $\epsilon=E_\gamma/E_e$, $E^0_\gamma=m_e^2 E_\gamma [4qE_e(E_e-E_\gamma)]^{-1}$ and $n(E)$ is the number density of CMB photons per unit energy, which is homogeneous. For the Galactic secondary spectrum, the result is:
\begin{equation}
\frac{d^2\phi_{\rm IC}^{\rm G}}{dEd\Omega}=\frac{1}{2\pi E_\gamma m_{\rm DM} \tau} \int_0^{+\infty} ds \, \rho(r(s,l,b)) \int_{E_\gamma}^{m_{\rm DM}/2} dE_e \frac{P_{IC} (E_\gamma,E_e)}{b(E_e)} \int_{E_e}^{m_{\rm DM}/2} \frac{dN_e}{dE'} dE'
\end{equation}
For the Extragalactic secondary component the only modification to take into account is that the integral over the line-of-sight is converted into an integral over the redshift, and the CMB temperature is redshifted, as well as the photon energies. The final result is therefore:
\begin{eqnarray}
\frac{d^2\phi_{\rm IC}^{\rm E}}{dEd\Omega}=\frac{\rho_{cr} \Omega_\chi}{2\pi E_\gamma m_{\rm DM} \tau} \int_0^{+\infty} \frac{dz}{H(z)(1+z)} \int_{E_\gamma (1+z)}^{m_{\rm DM}/2} dE_e \frac{P_{IC} (E_\gamma (1+z),E_e, z)}{b(E_e)} \nonumber \\ \int_{E_e}^{m_{\rm DM}/2} \frac{dN_e}{dE'} dE'
\end{eqnarray}
where we have introduced the explicit dependence of $P_{\rm IC}(E_\gamma,E_e,z)$ through the CMB temperature dependence.

\section{Fitting $7.5$ years HESE IceCube data} \label{secIC}

It is here assumed that neutrinos observed at IceCube arise from a double component (together with the atmospheric neutrino background):
\begin{equation}
\phi=\phi_{\rm Astro} +\phi_\chi
\end{equation}
where $\phi_\chi$ has been discussed above while the astrophysical spectrum is parameterized as a power law spectrum isotropic and independent of flavors 
\begin{equation}
\frac{d^2\phi_{\rm Astro}}{dEd\Omega}=\phi_0 \left(\frac{E}{100 \, {\rm TeV}}\right)^{-\gamma}
\end{equation} 
where $\gamma$ is the spectral index and $\phi_0$ is the normalization of the power law.

Having delineated the neutrino flux arriving at Earth, we now discuss the conversion of this flux into the number of events detected by IceCube. The complete procedure would require the implementation of the stochastic relation between the true energy of each neutrino and the electromagnetic energy deposited in the detector, as described in detail in \cite{Palomares-Ruiz:2015mka}. We have instead followed the procedure of connecting these two energies through a deterministic function which gives the expected true energy given the deposited one. This function has been obtained in \cite{DAmico:2017dwq} for the three different cases of charged current shower events, track events and neutral current shower events. We report in Table\,\ref{tab1} the numerical values of the expected value for the true neutrino energy corresponding to some benchmark values of the deposited neutrino energy, as given in \cite{DAmico:2017dwq}:
\begin{table}[h!]
\begin{center}
\begin{tabular}{|l|c|c|c|}
\hline
$E_{dep} (\rm{TeV})$ & $E^{T}_\nu (\rm{TeV})$ & $E^{S}_\nu (\rm{TeV})$ &  $E^{NC}_\nu (\rm{TeV})$\\
\hline
100 & 278 & 139 & 294 \\
200 & 570 & 271& 531\\
300 & 830 & 389 & 736   \\
400 & 1040 & 515 & 912  \\
\hline
\end{tabular}\caption{Relation between deposited energies and expected true energies of the incident neutrino for some benchmark energy values in the three topologies of Track (T), Shower (S) and Neutral Current (NC) events.}\label{tab1}
\end{center}
\end{table}

The procedure of prediction of the number of events per unit time and energy interval is therefore as follows. For a certain detected energy interval we have obtained the corresponding true energy interval for all three topologies of events mentioned above. For each such topology we have integrated the expected neutrino flux for the three flavors by the effective area $A_{eff}(E,\Omega)$ (function of the true energy and the arrival direction of neutrino) provided by the IceCube Collaboration \cite{Aartsen:2013jdh}  over the mapped energy interval. In the case of the Galactic flux, which is anisotropic due to the inhomogeneous distribution in the Galaxy plane, we have integrated over the solid angle using the angular dependent effective areas provided by the Collaboration. The weight assigned to each topology for different flavors is obtained by the procedure delineated in \cite{Palladino:2015zua}. The probabilities for each flavor $\alpha=e,\mu,\tau$ to produce one of the three topologies, namely shower (S), track (T) and neutral current (NC) showers are given by 
\begin{eqnarray}
p^{e,\tau}_S &=& \frac{\sigma_{CC}^{e,\tau} M^{e,\tau}_{CC}}{\sigma_{NC}^{e,\tau} M_{NC}+\sigma_{CC}^{e,\tau} M^{e,\tau}_{CC}} ,\qquad  
p^{\mu}_T = \frac{\sigma_{CC}^{\mu} M^{\mu}_{CC}}{\sigma_{NC}^{\mu} M_{NC}+\sigma_{CC}^{\mu} M^{\mu}_{CC}}, \nonumber \\ \qquad 
p^{e,\tau}_{NC} &=& 1-p^{e,\tau}_{S},\qquad 
p^{\mu}_{NC} = 1-p^{\mu}_T
\end{eqnarray}
where $\sigma_{CC}$ and $\sigma_{NC}$ are the charged and neutral current cross section \cite{Gandhi:1998ri} and $M_{NC}$, $M_{CC}^{e,\mu,\tau}$ are the effective detector masses \cite{Aartsen:2013jdh}. So given a generic set of three neutrino fluxes $\phi_\alpha$ the number of track and shower events in a certain energy interval is given by
\begin{eqnarray}
N_T&=& \int d\Omega \int dE \, \frac{d^2\phi_{\mu}}{dE d\Omega} \, p^{\mu}_{T} \, A_{eff} (E,\Omega),  \nonumber \\
N_S&=& \int d\Omega \int dE \, \sum_{\alpha=e,\tau} \, \frac{d^2\phi_{\alpha}}{dE d\Omega} \, p^{\alpha}_{S} \, A_{eff} (E,\Omega),  \\
N_{NC}&=& \int d\Omega \int dE \, \sum_{\alpha=e,\mu,\tau} \, \frac{d^2\phi_{\alpha}}{dE d\Omega} \, p^{\alpha}_{NC} \, A_{eff} (E,\Omega).\nonumber 
\end{eqnarray} 
For the decaying dark matter case, the three neutrino fluxes $\phi_\alpha$ are function the specific decaying channel considered.

The fitting procedure of a generic neutrino input spectrum is then to use the methods discussed above to predict the expected number of events in each bin. We can then assume a multi Poisson distribution for each of the energy bin at energies above $60$~TeV, in agreement with the methods used by the IceCube Collaboration \cite{Ishihara:2015vqt}, to adapt the method of maximum likelihood. Let us denote the total number of expected events from astrophysical and Dark Matter sources in each energy bin by $\mu_i$, obtained according to the procedure described above, and by $b_i$ the expected number of background events according to the IceCube Collaboration. Furthermore, let $n_i$ be the number of measured events in each energy bin. Then the likelihood function adopted is:
\begin{equation} \label{multipoisson}
\mathcal{L}=\prod_i (\mu_i+b_i)^{n_i} \frac{e^{-(\mu_i+b_i)}}{n_i!}
\end{equation}
In this way we have a complete procedure for obtaining the number of total events expected from Dark Matter decay in an energy interval. In order to compare with the experimental data, of course, it is necessary to have an estimate of the background. This has been done by using the same binning as the one given by the IceCube Collaboration, which provides in graphical form for each bin the expected number of background events~\cite{ICRC_HESE}.

The computation of the likelihood for a given channel and given values of the parameters is most easily accomplished by observing that $\tau$ and $\phi_0$ just serve as normalization respectively of the Dark Matter and astrophysical fluxes, whose non trivial dependence is therefore upon $m_{\rm DM}$ and $\gamma$. We have therefore obtained, for a reference normalization flux $\tilde{\phi}_0$, the expected number of astrophysical events for each energy bin as a function of the spectral index taken in a range between $1.5$ and $5$, and, in the same way, we have obtained, for a reference Dark Matter decay lifetime $\tilde{\tau}$, the expected number of Dark Matter events for each energy bin. The fluxes for a generic normalization flux $\phi_0$ and a generic lifetime $\tau$ are therefore obtained from these by multiplying the factors $\phi_0/\tilde{\phi}_0$ and $\tilde{\tau}/\tau$ respectively.

\subsection{One-component fit: pure astrophysical power law}

A first analysis which can be done with the tools developed in the previous sections is the analysis of the IceCube data in the hypothesis of the pure astrophysical power law spectrum, which is the cumulative spectrum from the astrophysical sources. For this case with no Dark Matter, we find as best fit values $\gamma=3.02\pm 0.13$ and $\phi_0=\left(1.9\pm 0.3\right)\times 10^{-15}$ $\rm{TeV^{-1}}$ $\rm{cm^{-2}}$ $\rm{s^{-1}}$ $\rm{sr^{-1}}$. The uncertainties provided here are the $68\%$ confidence levels.

\subsection{Two-component fit: astrophysical power law and decaying dark matter}

We now describe the results obtained for the cases of two components. The statistical analysis of this case can be performed with the aid of the maximum likelihood method, with the likelihood specified above. The determination of the confidence regions can be performed once the distribution of the likelihood as a test statistic is known. A theorem which is often adopted in this context is the Wilks theorem \cite{Wilks:1938dza}, which states that the distribution of the log-likelihood ratio 
\begin{equation}
\Lambda=2\ln{\frac{\mathcal{L}(\hat{\theta})}{\mathcal{L}(\theta_0)}}
\end{equation}
between the likelihood evaluated at its maximum ($\hat{\theta}$) and the maximum of the likelihood in a region which is being tested ($\theta_0$) is a chi-squared distribution with a number of degrees of freedom equal to the difference in the number of the free parameters between the null hypothesis and the whole parameter space. The hypothesis under which this theorem is proved is that the likelihood possesses a single maximum and behaves like a Gaussian distribution around this point. As will be seen below, this hypothesis is not realized in the distributions we find. Nevertheless, it can be proven that this hypothesis can be considerably relaxed. In fact, it turns out that the likelihood distribution is a chi-squared one for each point of the parameter space which sufficiently well reproduce, in its mean value, the experimental data. This implies that each set of parameters which produces an expected flux comparable to the data within the statistical uncertainties possesses a chi-squared distribution for the likelihood ratio, and this result does not depend on the geometry of the parameter space. Therefore, we can safely adopt the results of the theorem even if we are beyond the range of application of the theorem as originally stated. Furthermore, the Neyman-Pearson lemma grants us that, at a fixed level of significativity, this is the most powerful test statistic which we can adopt.
\begin{figure}[t!]
\begin{center}
\begin{subfigure}[b]{0.3\textwidth}
\includegraphics[width=\textwidth]{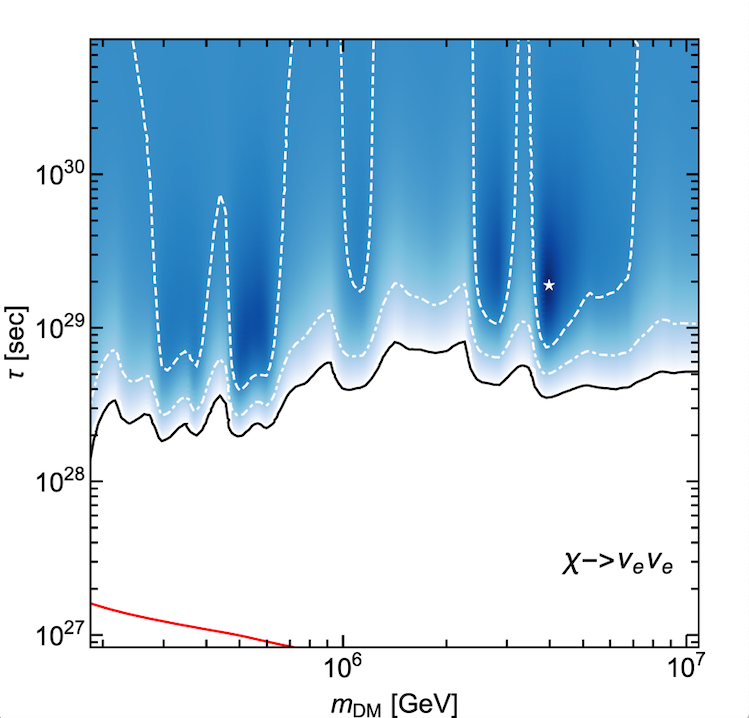}
\caption{Channel $\nu_e\nu_e$}
\end{subfigure}
\begin{subfigure}[b]{0.3\textwidth}
\includegraphics[width=\textwidth]{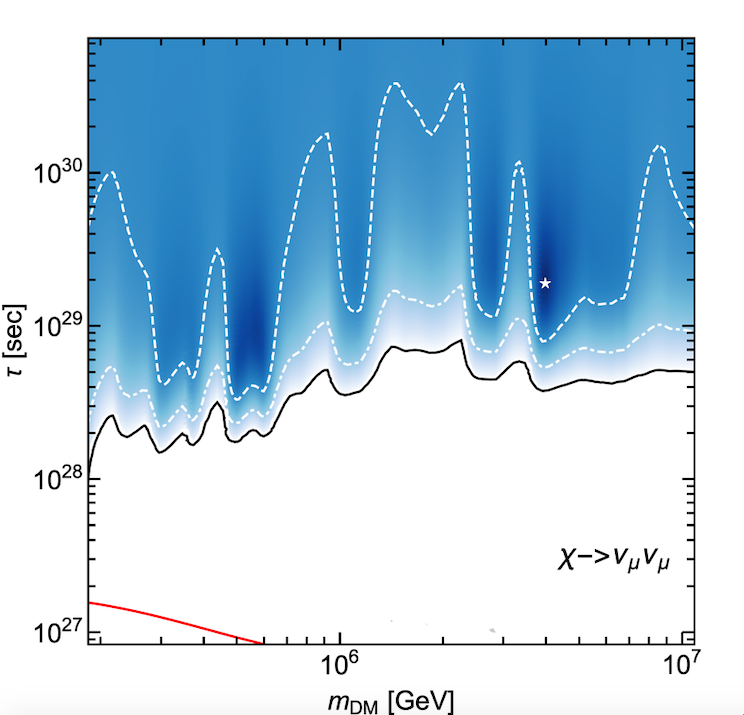}
\caption{Channel $\nu_\mu\nu_\mu$ }
\end{subfigure}
\begin{subfigure}[b]{0.335\textwidth}
\includegraphics[width=\textwidth]{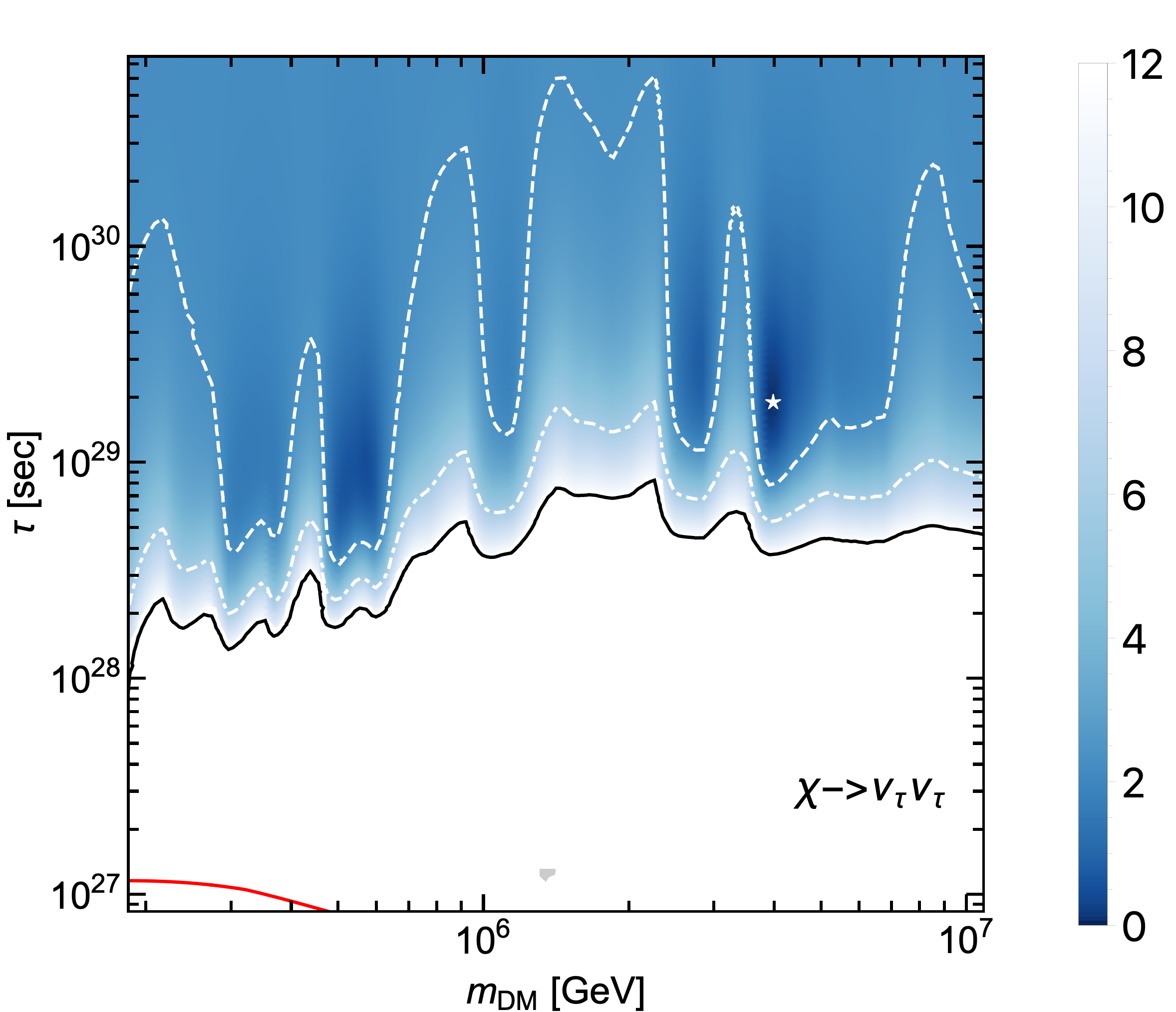}
\caption{Channel $\nu_\tau\nu_\tau$ }
\end{subfigure}
\vskip2.mm
\begin{subfigure}[b]{0.3\textwidth}
\includegraphics[width=\textwidth]{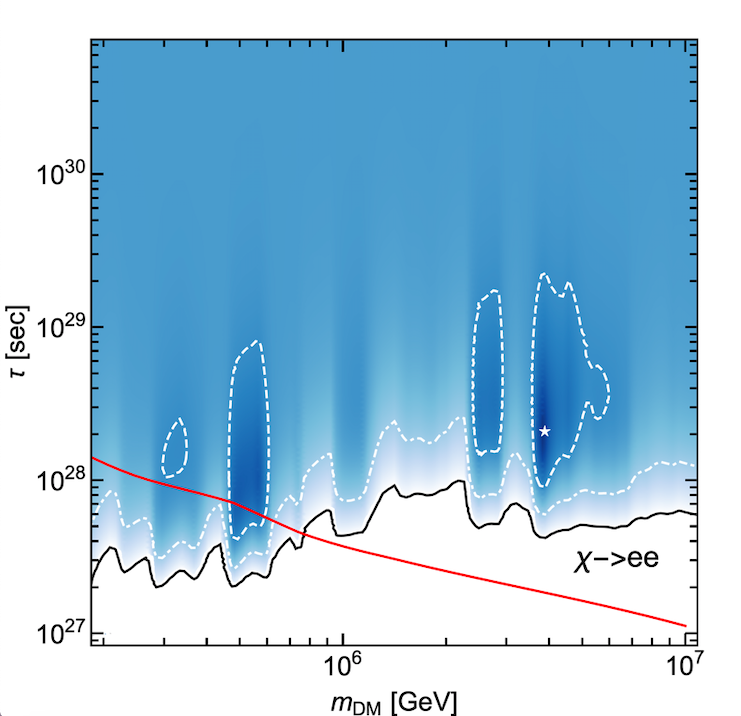}
\caption{Channel $e^+ e^-$}
\end{subfigure}
\begin{subfigure}[b]{0.3\textwidth}
\includegraphics[width=\textwidth]{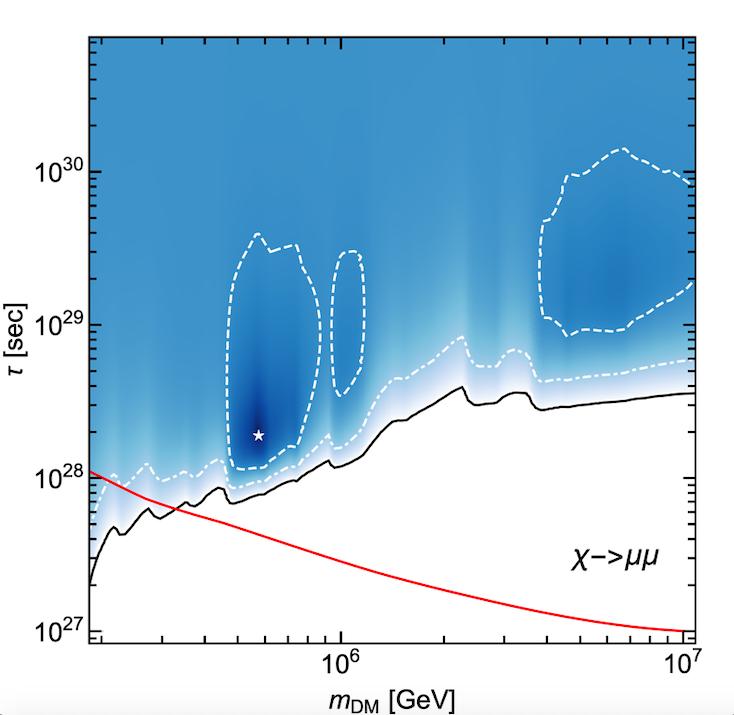}
\caption{Channel $\mu^+ \mu^-$}
\end{subfigure}
\begin{subfigure}[b]{0.3\textwidth}
\includegraphics[width=\textwidth]{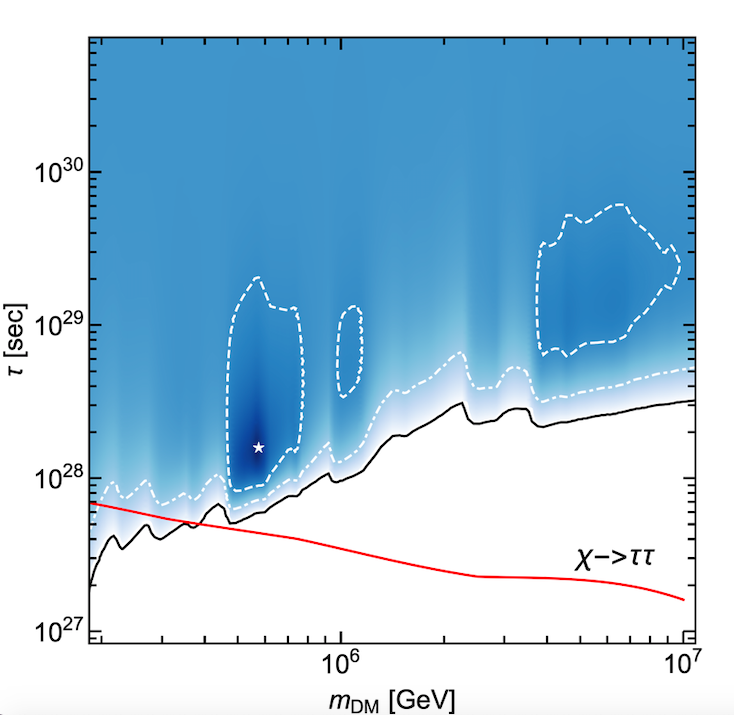}
\caption{Channel $\tau^+ \tau^-$}
\end{subfigure}
\end{center}
\caption{Likelihood contours for the leptonic channels in the $m_{\rm DM}$-$\tau$ plane. The red line represents the exclusion region coming from the Fermi-LAT data (the excluded region is below the line). The dashed lines are the $68\%$ confidence levels, the dot-dashed lines are the $95\%$ confidence levels and the continuous lines are the $99.7\%$ confidence levels.}\label{figtaum}
\end{figure}
\begin{figure}[t!]
\begin{center}
\begin{subfigure}[b]{0.3\textwidth}
\includegraphics[width=\textwidth]{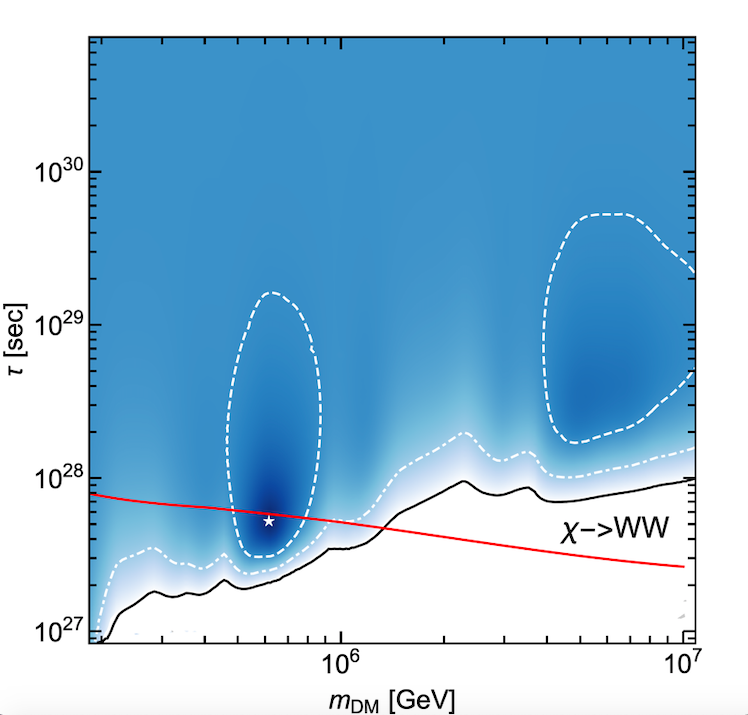}
\caption{Channel $W^+W^-$}
\end{subfigure}
\begin{subfigure}[b]{0.3\textwidth}
\includegraphics[width=\textwidth]{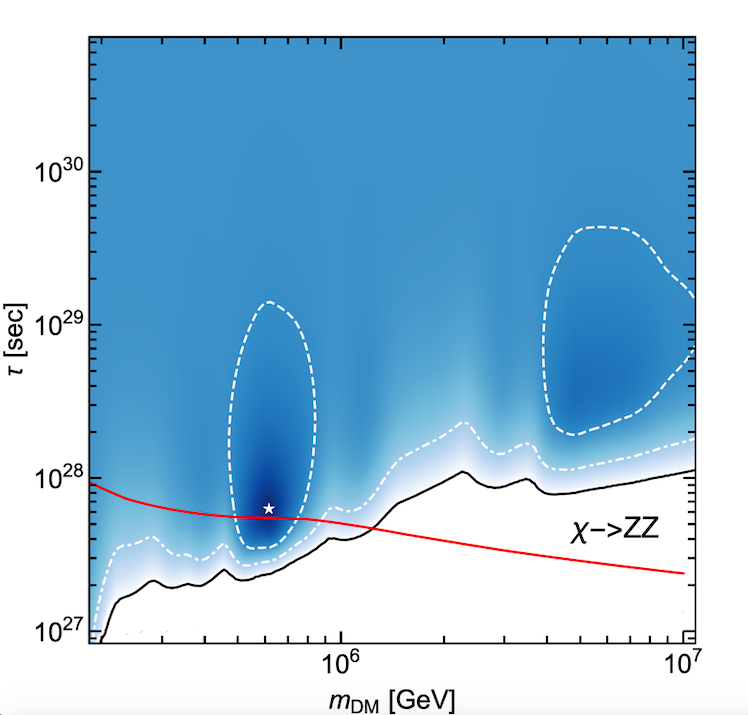}
\caption{Channel $ZZ$}
\end{subfigure}
\begin{subfigure}[b]{0.335\textwidth}
\includegraphics[width=\textwidth]{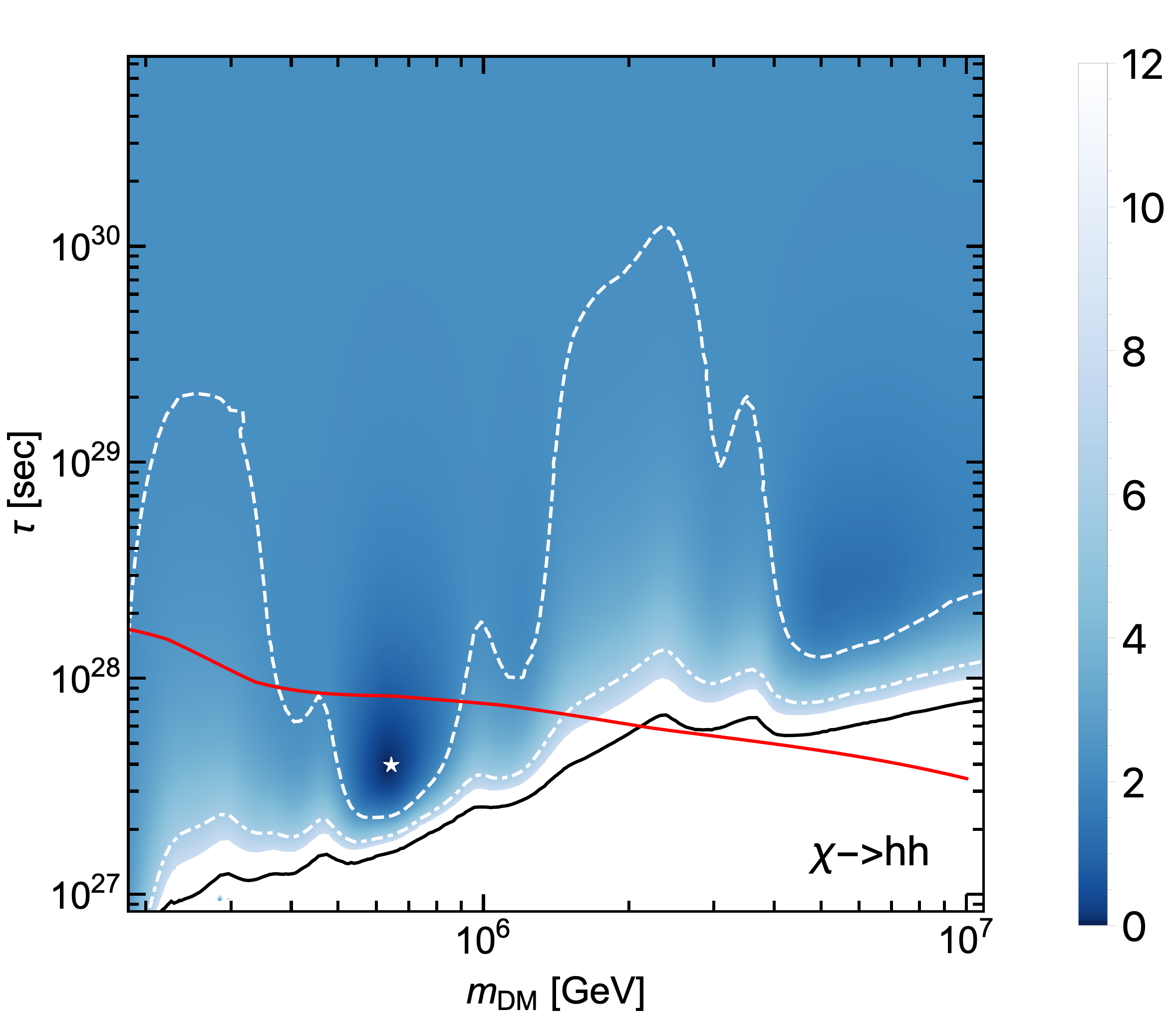}
\caption{Channel $hh$ }
\end{subfigure}
\end{center}
\caption{Likelihood contours for the bosons channels in the $m_{\rm DM}$-$\tau$ plane. The description is the same as in figure~\ref{figtaum}.}\label{figtaum2}
\end{figure}
\begin{figure}[t!]
\begin{center}
\begin{subfigure}[b]{0.3\textwidth}
\includegraphics[width=\textwidth]{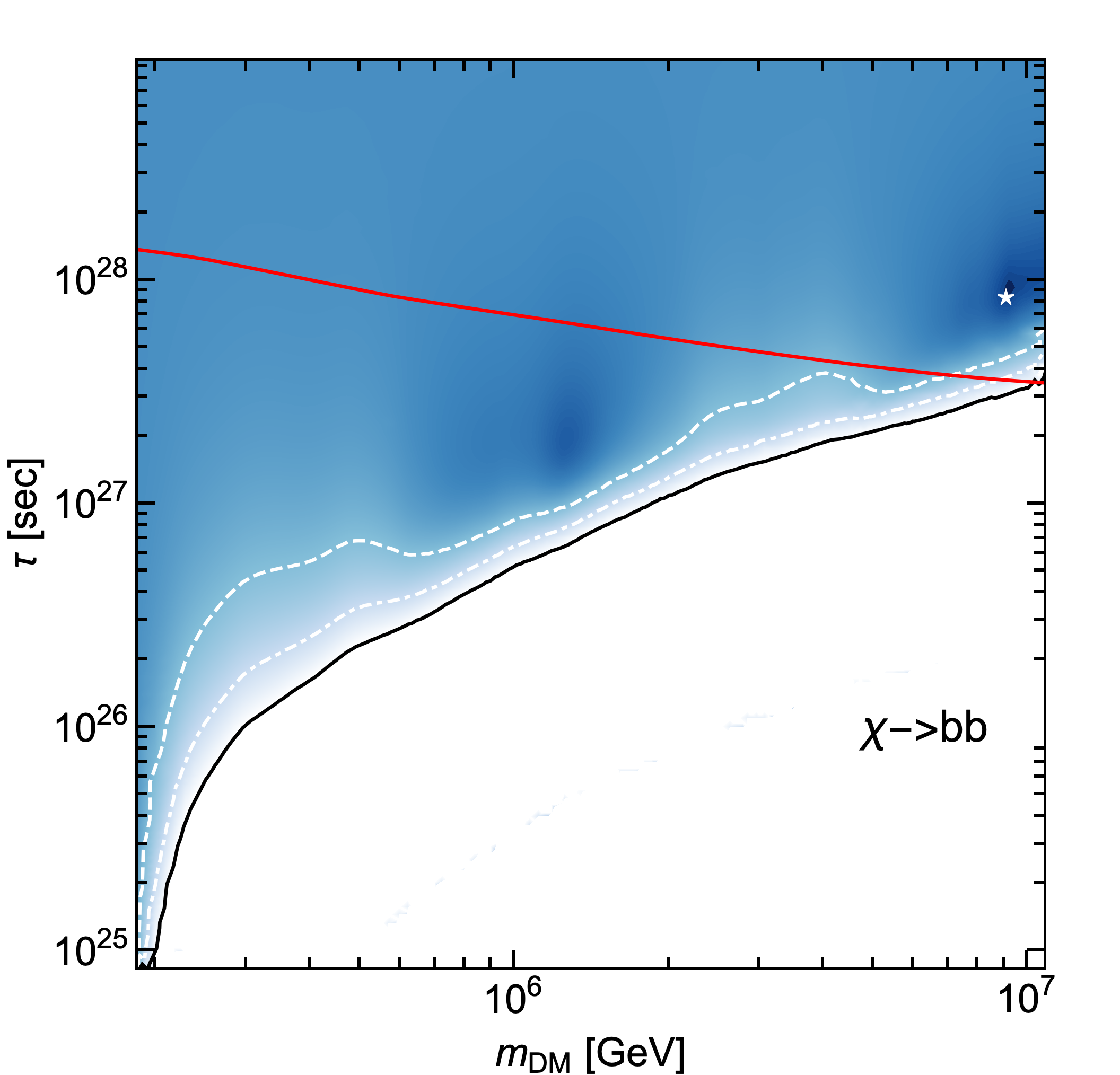}
\caption{Channel $b \overline{b}$}
\end{subfigure}
\begin{subfigure}[b]{0.3\textwidth}
\includegraphics[width=\textwidth]{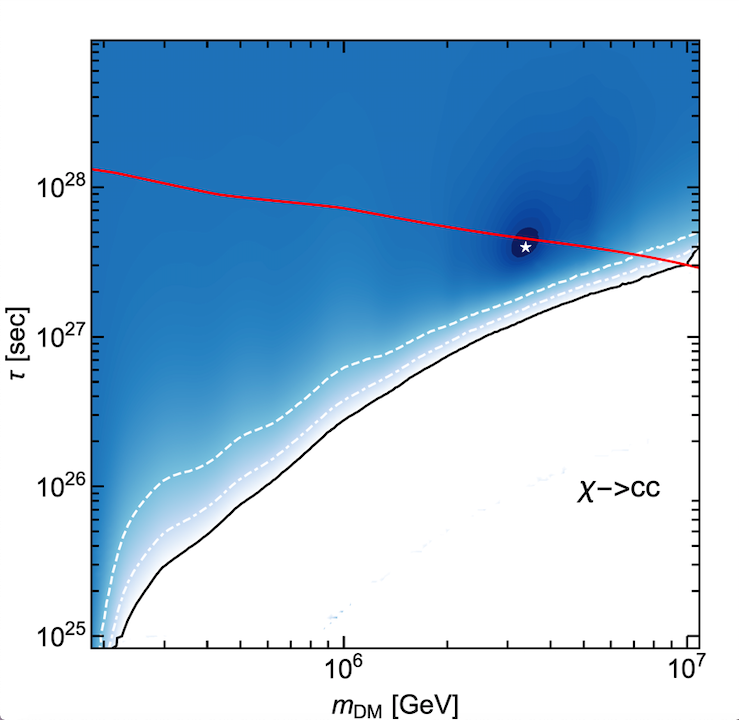}
\caption{Channel $c \overline{c}$}
\end{subfigure}
\begin{subfigure}[b]{0.335\textwidth}
\includegraphics[width=\textwidth]{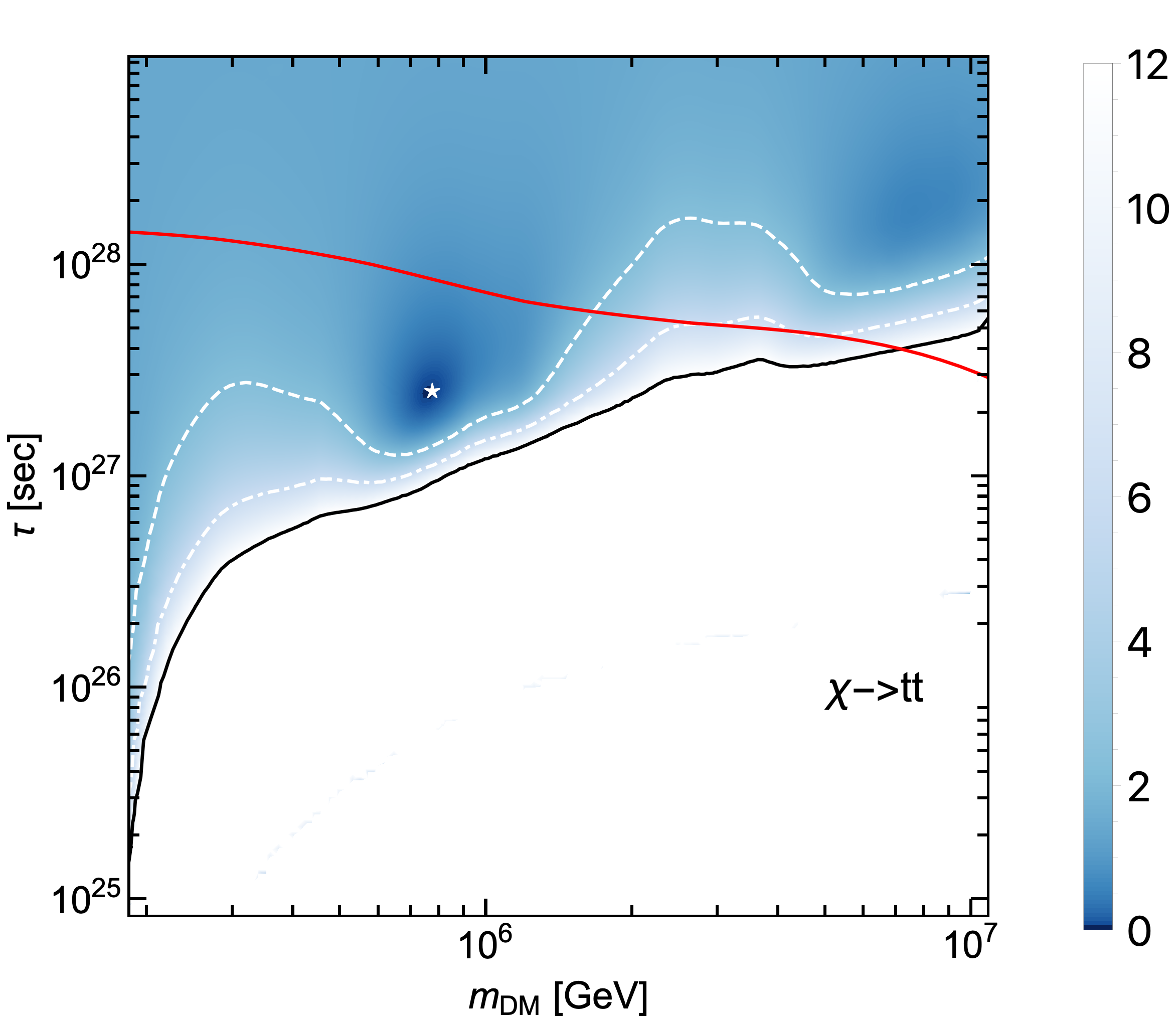}
\caption{Channel $t \overline{t}$}
\end{subfigure}
\begin{subfigure}[b]{0.3\textwidth}
\includegraphics[width=\textwidth]{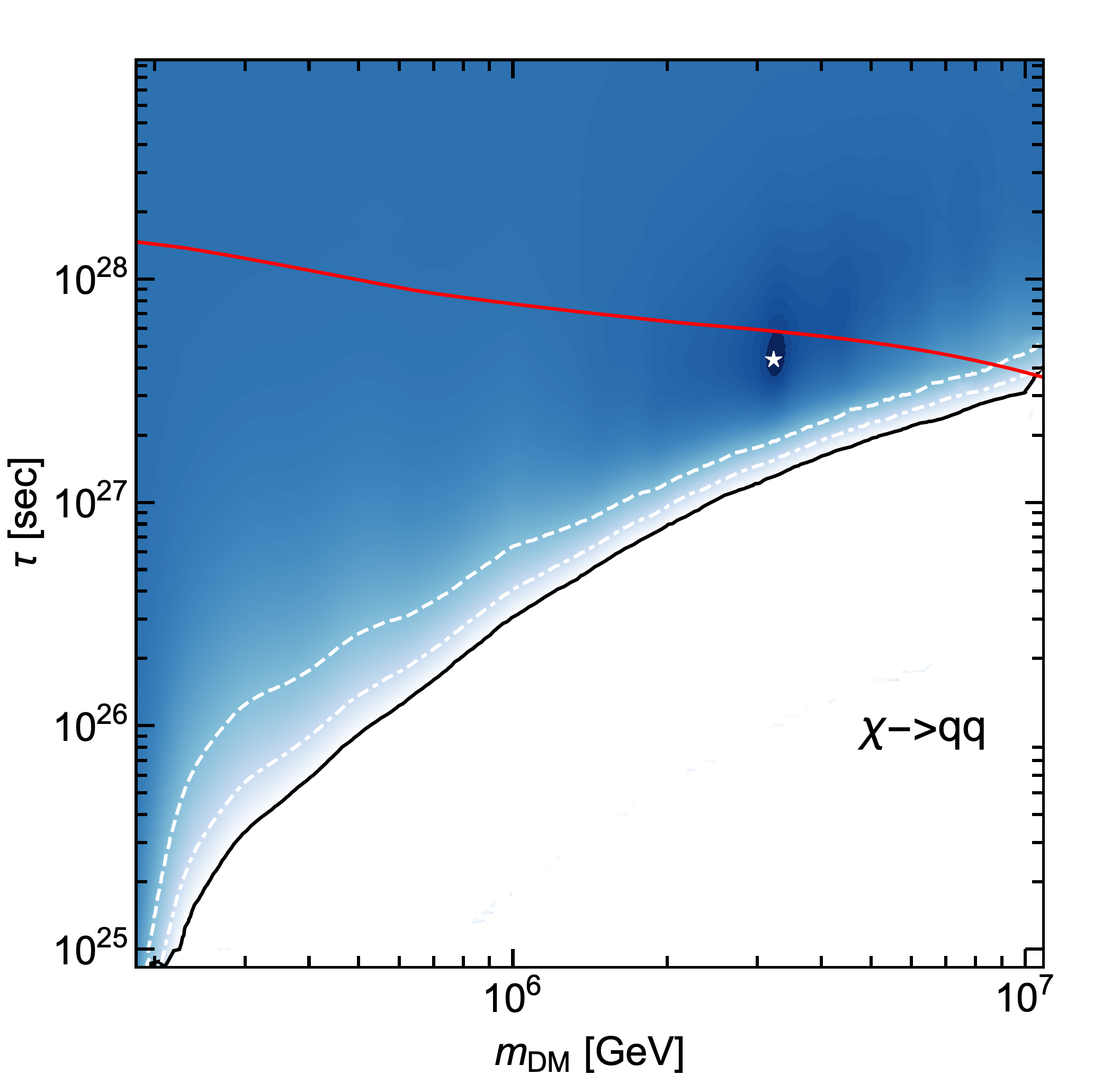}
\caption{Channel $q=u,d,s$ }
\end{subfigure}
\begin{subfigure}[b]{0.3\textwidth}
\includegraphics[width=\textwidth]{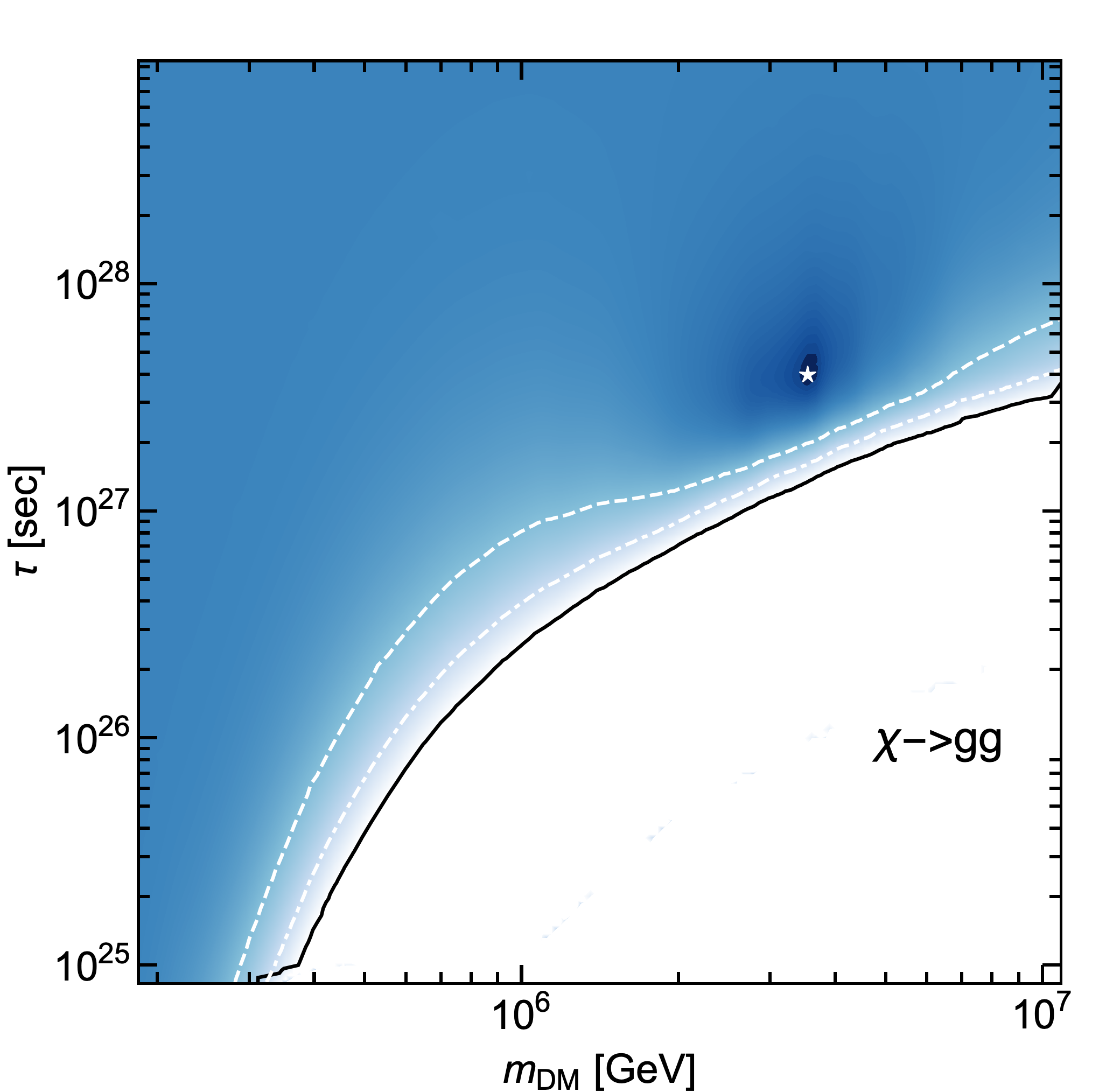}
\caption{Channel $gg$ }
\end{subfigure}
\vskip2.mm
\end{center}
\caption{Likelihood contours for the quark and gluonic channels in the $m_{\rm DM}$-$\tau$ plane. The description is the same as in figure~\ref{figtaum}.}\label{figtaum3}
\end{figure}

For the various channels analyzed, we represent the results in the form of contour plots of the likelihood ratio $\Lambda$ defined above in figures~\ref{figtaum},~\ref{figtaum2} and~\ref{figtaum3} for the various channels, together with the $68\%$, $95\%$ and $99.7\%$ confidence levels. In these plots, the astrophysical parameters are treated as nuisance parameters, and the likelihood is correspondently maximized with respect to them: in particular, $\gamma$ is chosen in a range between $1.5$ and $5$. As anticipated above, these plots highlight a non trivial structure in the likelihood function, which posses different competing maxima. We expect a strong sensitivity in the choice of the dominant maximum on the details of our analysis, such as the adopted functions for the relation between deposited and real energy shown in Table\,(\ref{tab1}).
\begin{table}[h!]
\begin{center}
\begin{tabular}{|c|c|c|c|c|}
\hline
Channel & $\phi_0^{\rm best}(\times 10^{-15} {\rm f.u.})$ &  $\gamma^{\rm best}$& $\tau^{\rm best} (\times 10^{28} \rm{s})$& $m_{\rm DM}^{\rm best} (\rm{TeV})$\\
\hline
$\nu_e\nu_e$ &2.24& 3.33 & 19.10& 4017.35  \\
$\nu_\mu\nu_\mu$ &2.24& 3.33 & 19.10& 4017.35 \\
$\nu_\tau\nu_\tau$ &2.24& 3.33 & 19.10& 4017.35 \\
\hline
$e^+e^-$ & 2.14 & 3.86 & 2.09 & 3846.63 \\
$\mu^+\mu^-$ & 0.66 & 2.64 & 1.91 & 569.17 \\
$\tau^+\tau^-$ & 0.74  & 2.69  & 1.59  & 570.00 \\
\hline
$W^+W^-$ & 0.68 & 2.67 & 0.53 & 620.81 \\
$ZZ$ & 0.72 & 2.69 & 0.63 & 621.00 \\
$hh$ & 0.67 & 2.66 & 0.39 & 645.65 \\
\hline
$b\overline{b}$ & 1.15 & 3.19  & 0.83 & 9168.11  \\
$c\overline{c}$ & 0.78 & 2.78 & 0.40 & 3376.76 \\
$t\overline{t}$ & 0.73 & 2.69 & 0.25 & 776.47 \\
$q\overline{q}$ & 0.88 & 2.81  & 0.44  & 3233.26 \\
$gg$ & 0.77 & 2.74 & 0.40 & 3526.63\\
\hline
\end{tabular}\caption{Summary of the neutrino analysis: the best fit parameters are given for each analyzed channel. The $\phi_0$ is expressed in flux units of f.u.\,$\equiv \rm TeV^{-1} cm^{-2} s^{-1} sr^{-1}$.}\label{tab2}
\end{center}
\end{table}

Qualitatively, the data can be reproduced either by a softer spectrum, with a spectral index nearer to $2$, and a Dark Matter with a mass of order $\sim 100$ $\rm{TeV}$ which can fit the excess in this region, or by a harder spectrum with a DM mass of order $\sim 1$ $\rm{PeV}$, which can fit the high energy data. A major factor in determining which regions of the $m_{\rm DM}-\tau$ plane will be favored by the likelihood ratio is the choice of the interval over which $\gamma$ varies: in fact, according to our qualitative understanding of the solution, restricting $\gamma$ to be near the expected spectral index, which is around $2$, should favor the region of masses of order $\sim 100$ $\rm{TeV}$. 

In figures~\ref{figtaum},~\ref{figtaum2} and~\ref{figtaum3} we also represent, for the channels for which it is available, the exclusion curves obtained in \cite{Cohen:2016uyg} from the Fermi-LAT data with a red line. We also give in Table~\ref{tab2} the best fit values for the astrophysical and Dark Matter parameters. We underline that for the various channels the $2\sigma$ contours are always open, which means that the pure astrophysical spectrum cannot be rejected at more than the $2\sigma$ levels; further, both the hadronic and the neutrinophilic channels are not analyzable with more than the $1\sigma$ confidence level. 

A point of special interest is the comparison between the best fit astrophysical parameters and the astrophysical parameters found by the analysis of the through-going muons. As was mentioned in the introduction, the latter predict a spectral index $\gamma$ of $2.28$. The comparison between the spectral indices can be directly performed by virtue of our expectation that the Dark Matter gives a relatively small contribution to the through-going muons, because of the higher energy range and the more limited angular range which are explored. As we mentioned before, the likelihood maxima with $m_{DM}\sim$ PeV have a $\gamma$ larger than $3$, while those with $m_{DM}\sim 100$ TeV have a $\gamma$ smaller than $3$. Therefore the former case leads to a worse agreement with the through-going muons, while in the latter we have an improvement. However, being the two maxima almost degenerate, we cannot unambiguously say that the Dark Matter deteriorate the agreement. The next section will be dedicated to a partial answer to this question.

\subsection{Astrophysical spectrum properties}

At the end of the previous section we mentioned the importance of the spectral index in estimating the quality of our fit. In fact, as we stated in the introduction, Dark Matter is a possible candidate especially by virtue of the inconsistency of the spectral index found through the through-going muon neutrinos analysis and the one predicted by the HESE fit. 

A rigorous investigation in this respect should be obtained by performing a combined fit on the through-going muon neutrinos data and the HESE data. This is however beyond the scope of the present paper. It is still of interest to take into account the information provided by the through-going muon neutrinos data approximately. This can be done by incorporating, in the likelihood \eqref{multipoisson}, a Gaussian distribution for the astrophysical parameters centered around the values predicted by the IceCube fit with the corresponding uncertainties. In this way, we are introducing a Gaussian approximation to the through-going muon neutrinos likelihood which would be found by fitting them with a pure astrophysical spectrum. The likelihood \eqref{multipoisson} is therefore modified through the multiplication by a factor:
\begin{equation}
\mathcal{L}_{\rm TG} = \exp \left[{-\frac{(\gamma-\bar{\gamma})^2}{2\sigma_{\gamma}^2}-\frac{(\phi-\bar{\phi})^2}{2\sigma_{\phi}^2}} \right]
\end{equation}
where the parameters are extracted by the 10-year through-going muon neutrinos fit to be $\bar{\gamma}=2.28$, $\sigma_{\gamma}=0.08$, $\bar{\phi}=1.44\times 10^{-15} \rm \, TeV^{-1} cm^{-2} s^{-1} sr^{-1}$ and $\sigma_{\phi}=0.25\times 10^{-15} \rm \, TeV^{-1} cm^{-2} s^{-1} sr^{-1}$.

\begin{figure}[t!]
\begin{center}
\begin{subfigure}[b]{0.3\textwidth}
\includegraphics[width=\textwidth]{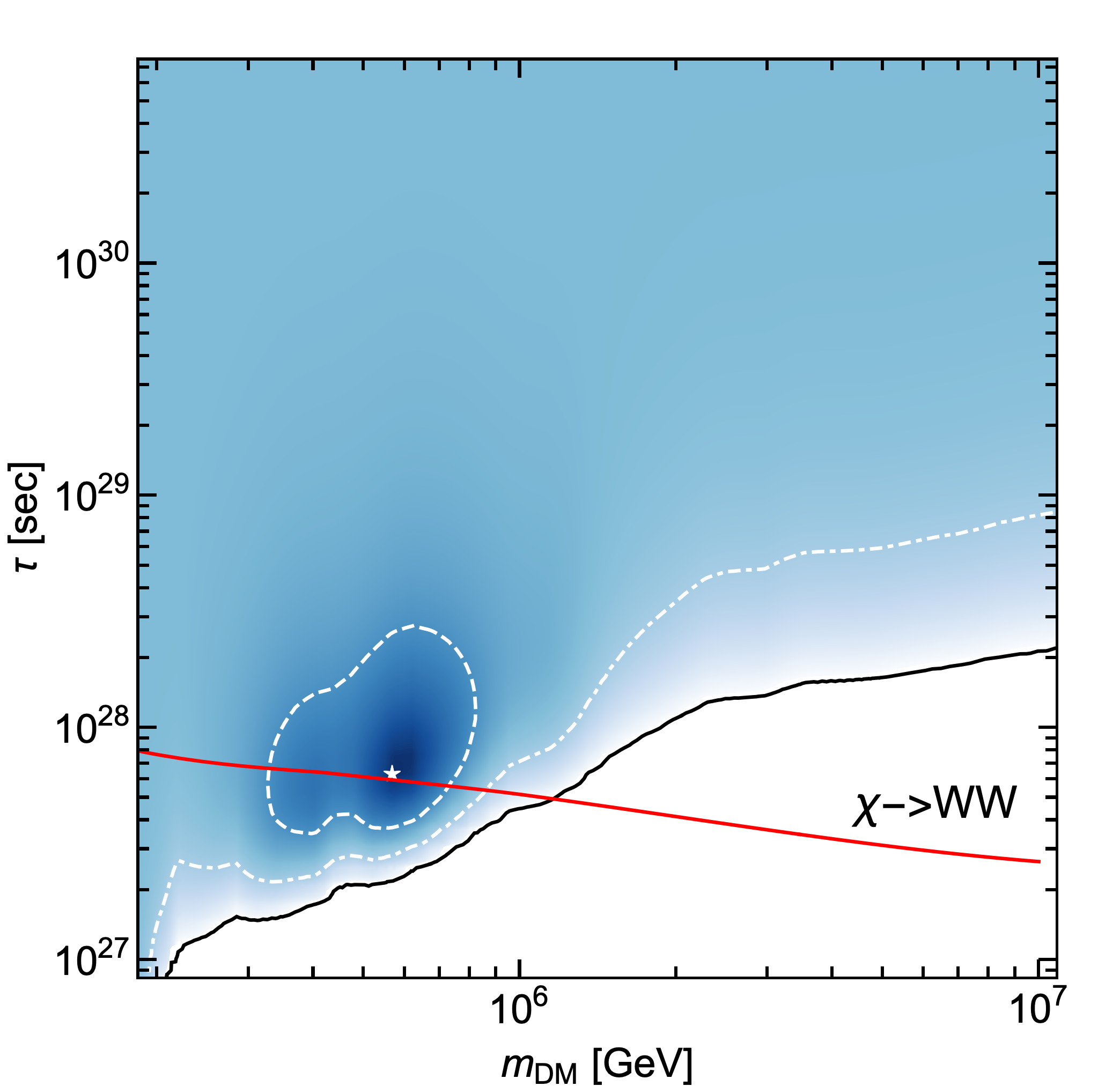}
\caption{Channel $W^+W^-$}
\end{subfigure}
\begin{subfigure}[b]{0.3\textwidth}
\includegraphics[width=\textwidth]{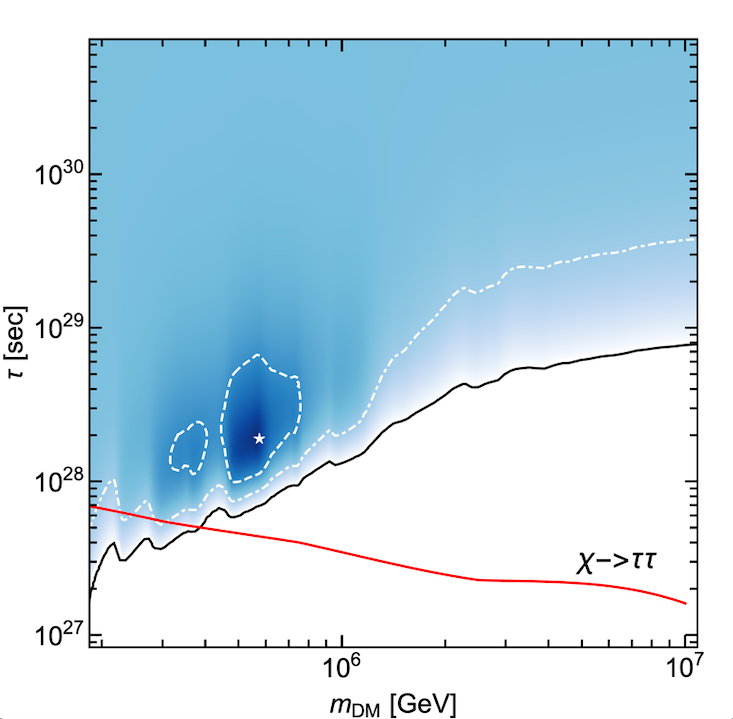}
\caption{Channel $\tau^+\tau^-$}
\end{subfigure}
\begin{subfigure}[b]{0.335\textwidth}
\includegraphics[width=\textwidth]{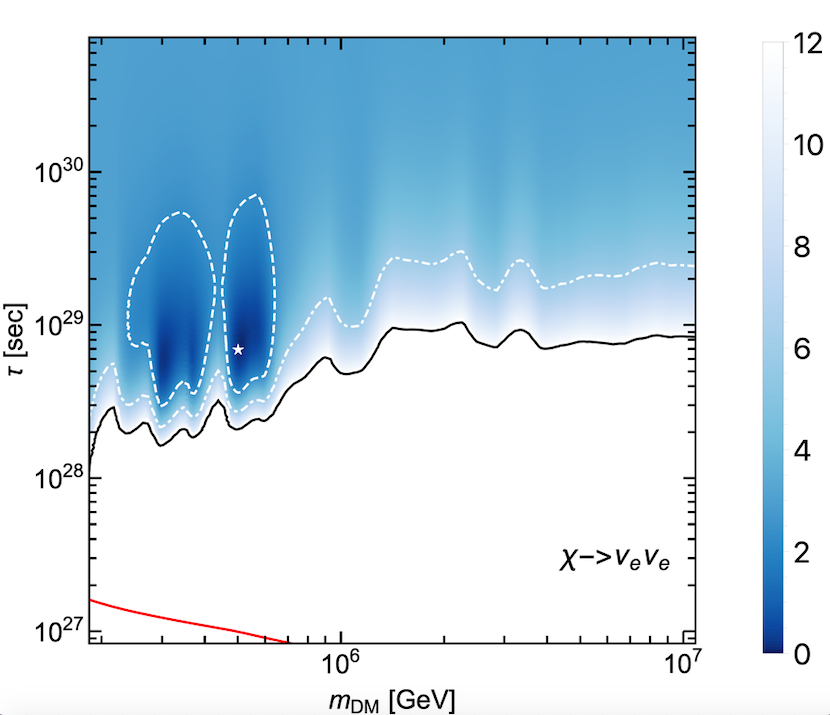}
\caption{Channel $\nu_e\nu_e$ }
\end{subfigure}
\end{center}
\caption{Likelihood contours for a boson, a lepton and a neutrino channel in the $m_{\rm DM}$-$\tau$ plane, with a likelihood modified to take into account the 10-year through-going muon neutrinos.}\label{figmod}
\end{figure}

\begin{figure}[t!]
\begin{center}
\begin{subfigure}[b]{0.3\textwidth}
\includegraphics[width=\textwidth]{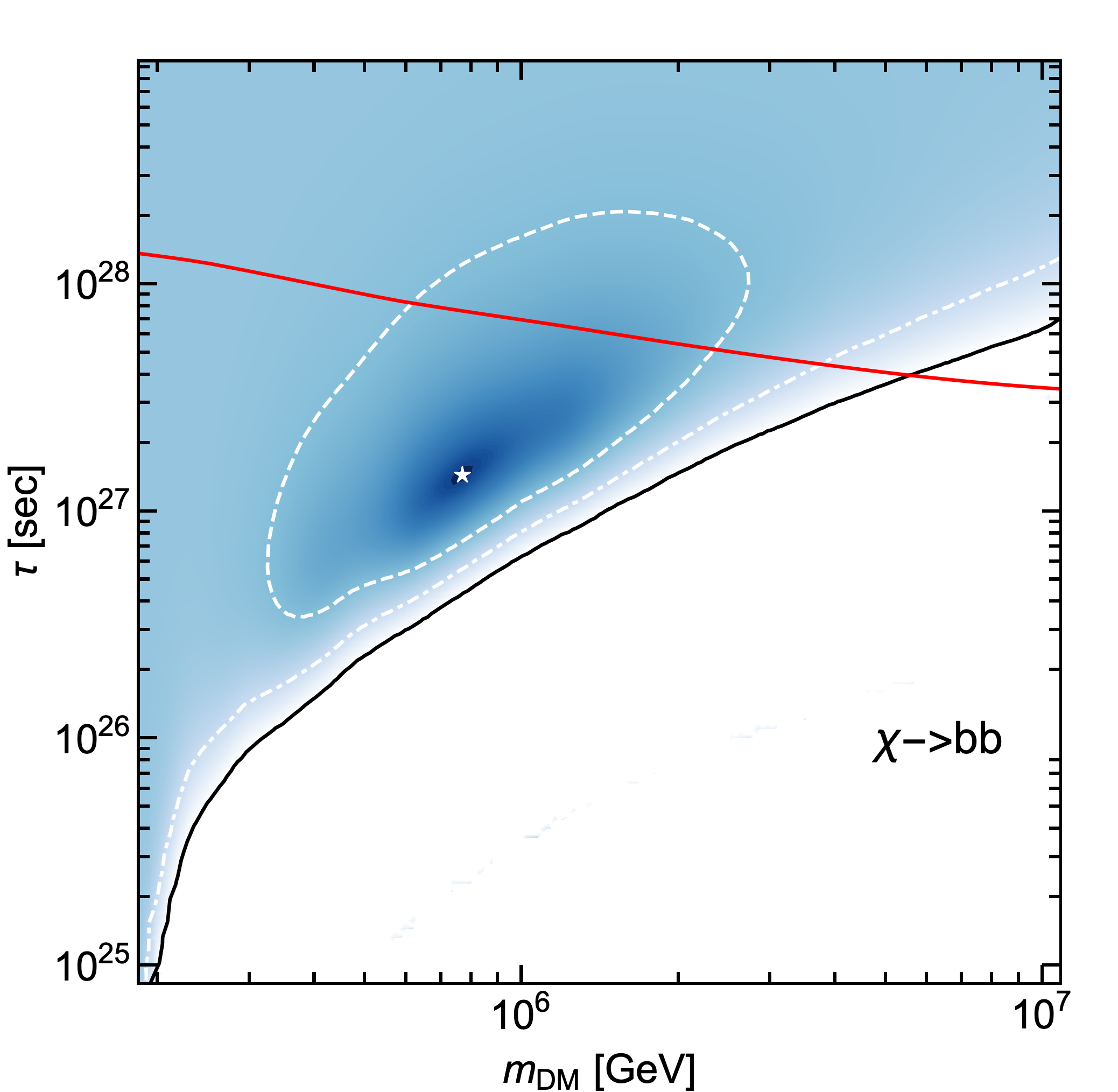}
\caption{Channel $b \overline{b}$}
\end{subfigure}
\begin{subfigure}[b]{0.335\textwidth}
\includegraphics[width=\textwidth]{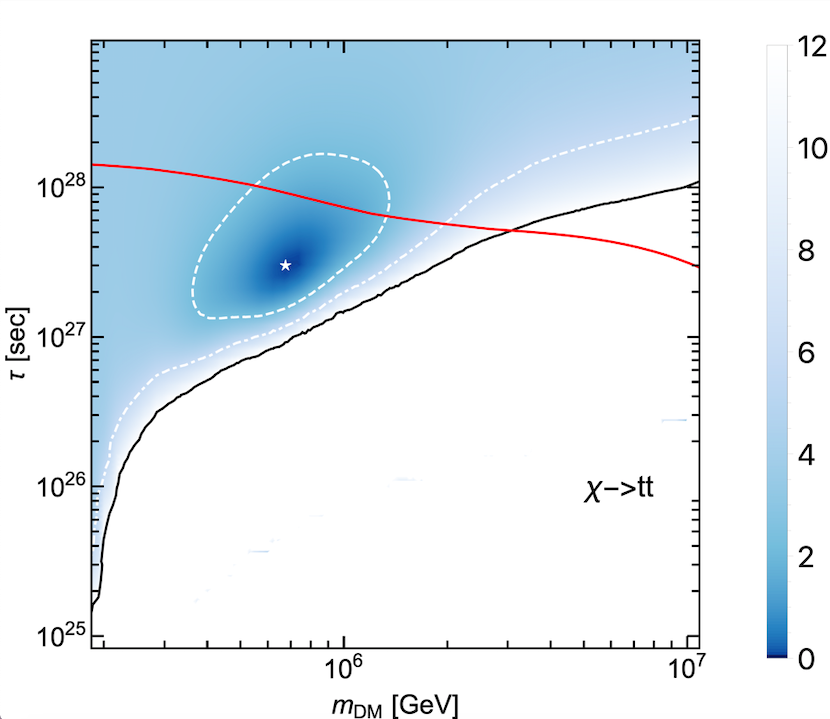}
\caption{Channel $t \overline{t}$}
\end{subfigure}
\vskip2.mm
\end{center}
\caption{Likelihood contours for two hadronic channels in the $m_{\rm DM}$-$\tau$ plane, with a likelihood modified to take into account the 10-year through-going muon neutrinos.}\label{figmod2}
\end{figure}

The incorrectness in our method steals from the fact that we are modeling the HESE data with a double component spectrum, while the through-going muon neutrinos are modeled by a pure astrophysical spectrum. However, due to the higher energy range and the different angular range, which does not involve the Galaxy center where the Dark Matter neutrino production is stronger, we expect the latter to be subdominant for the through-going muon neutrinos. This means that we can get at least a qualitative understanding of the changes which would be introduced in our results by taking into account the through-going muon neutrinos data. The contour plots for some significant channels, which best illustrate the qualitative changes, are represented in Figure \ref{figmod} and Figure \ref{figmod2}.

These plots evidences that the introduction of the through-going muons prior information strongly disfavors the PeV region for the DM masses, so that even those channels which previously had their best fit in that region are moved to the $\sim 100$ TeV region. The most dramatic consequence of this is that the hadronic channels are even more disfavored by the gamma ray constraints. The reason for this is that, as was mentioned above, the PeV region requires spectral indices even larger than $3$ to obtain a consistent fit of the data. An efficient comparison can be made by looking at the best fit parameters obtained in this new analysis: for the decay channels which have been analyzed these results are given in Table \ref{tab3}.

\begin{table}[h!]
\begin{center}
\begin{tabular}{|c|c|c|c|c|}
\hline
Channel & $\phi_0^{\rm best}(\times 10^{-15} {\rm f.u.})$ &  $\gamma^{\rm best}$& $\tau^{\rm best} (\times 10^{28} \rm{s})$& $m_{\rm DM}^{\rm best} (\rm{TeV})$\\
\hline
$\nu_e\nu_e$ &0.88& 2.43 & 6.92& 501.19 \\
$\tau^+\tau^-$ & 0.77  & 2.42  & 1.91  & 569.17 \\
$W^+W^-$ & 0.77 & 2.42 & 0.63 & 575.44 \\
$b\overline{b}$ & 0.86 & 2.43  & 0.14 & 776.25 \\
$t\overline{t}$ & 0.83 & 2.43 & 0.30 & 676.08 \\
\hline
\end{tabular}\caption{Summary of the neutrino analysis with a prior distribution for the through-going muons: the best fit parameters are given for each analyzed channel. The $\phi_0$ is expressed in flux units of f.u.\,$\equiv \rm TeV^{-1} cm^{-2} s^{-1} sr^{-1}$.}\label{tab3}
\end{center}
\end{table}

Another important consequence is that for all channels the $1\sigma$ contours are now closed. This can also be understood on the grounds that the pure astrophysical hypothesis requires, as we know, a spectral index near $3$, which is therefore relatively far from the through-going muon neutrinos value. Therefore, the region of large $\tau$ has a lower likelihood than before and is more strongly disfavored. Even with this information, however, the pure astrophysical spectrum can never be rejected at the $2\sigma$ level.

\begin{figure}[h!]
\begin{center}
\begin{subfigure}[b]{0.45\textwidth}
\includegraphics[width=\textwidth]{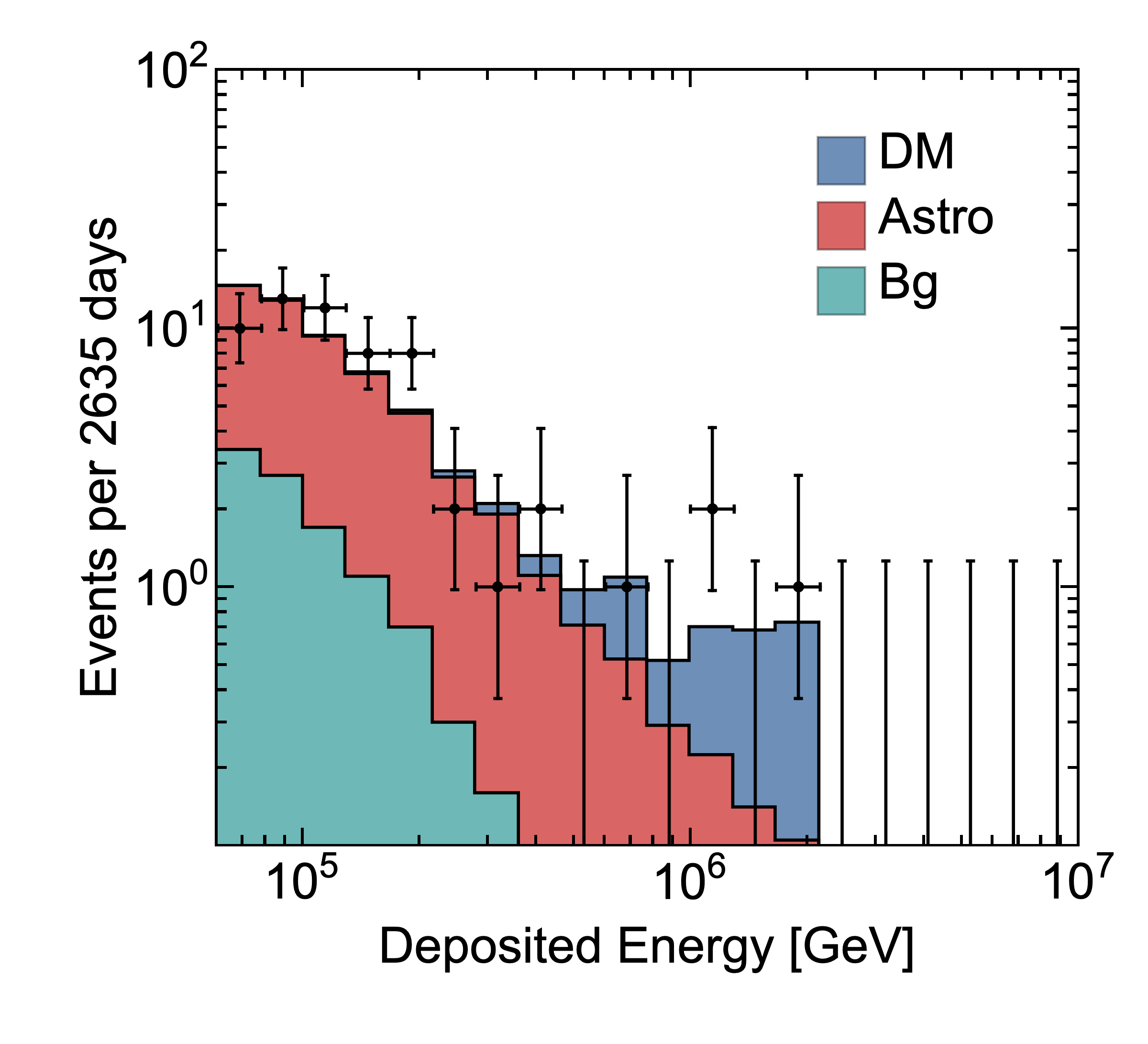}
\caption{Channel $\nu_e$, no prior distribution}
\end{subfigure}
\begin{subfigure}[b]{0.45\textwidth}
\includegraphics[width=\textwidth]{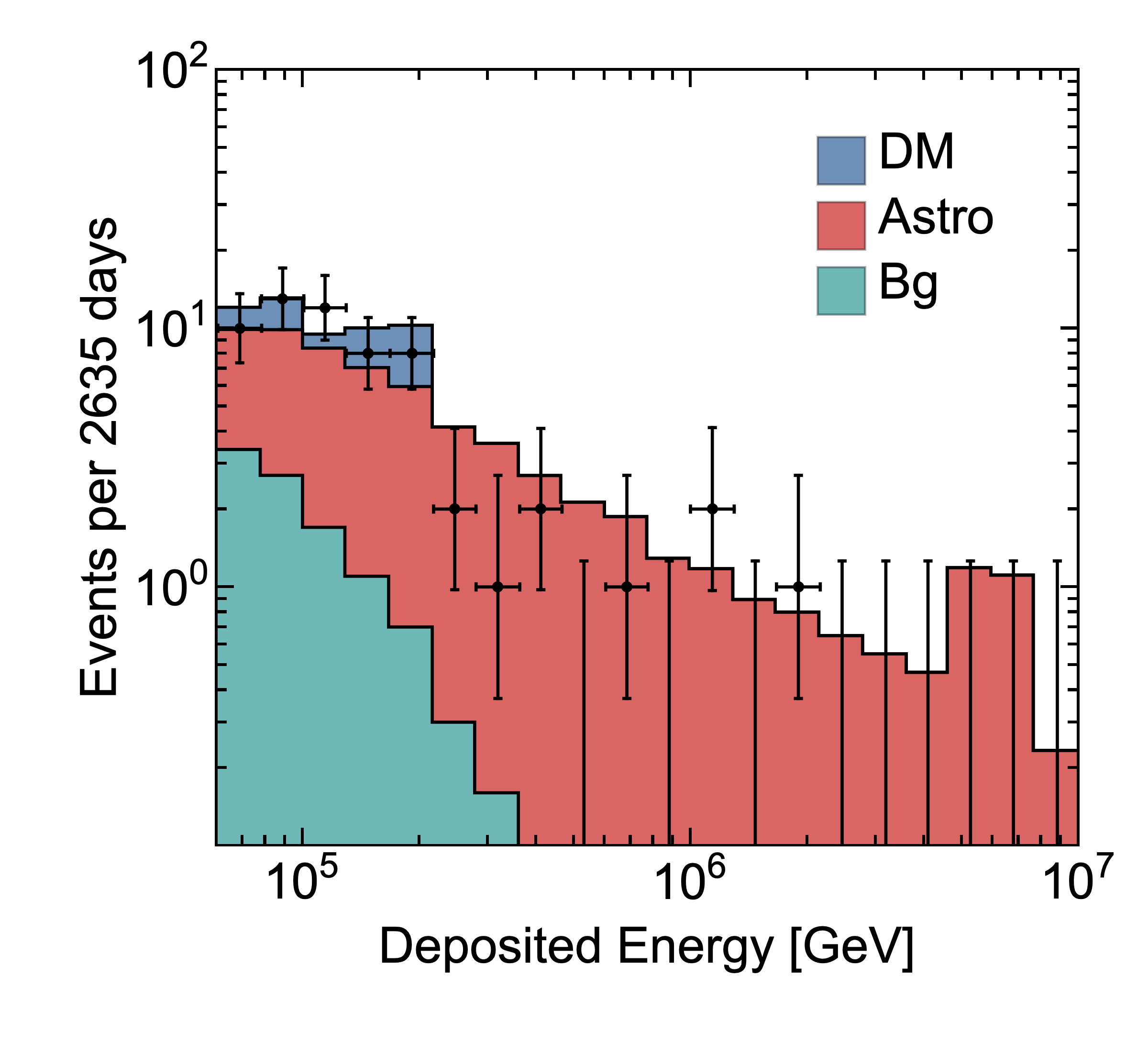}
\caption{Channel $\nu_e$, prior distribution}
\end{subfigure}
\begin{subfigure}[b]{0.45\textwidth}
\includegraphics[width=\textwidth]{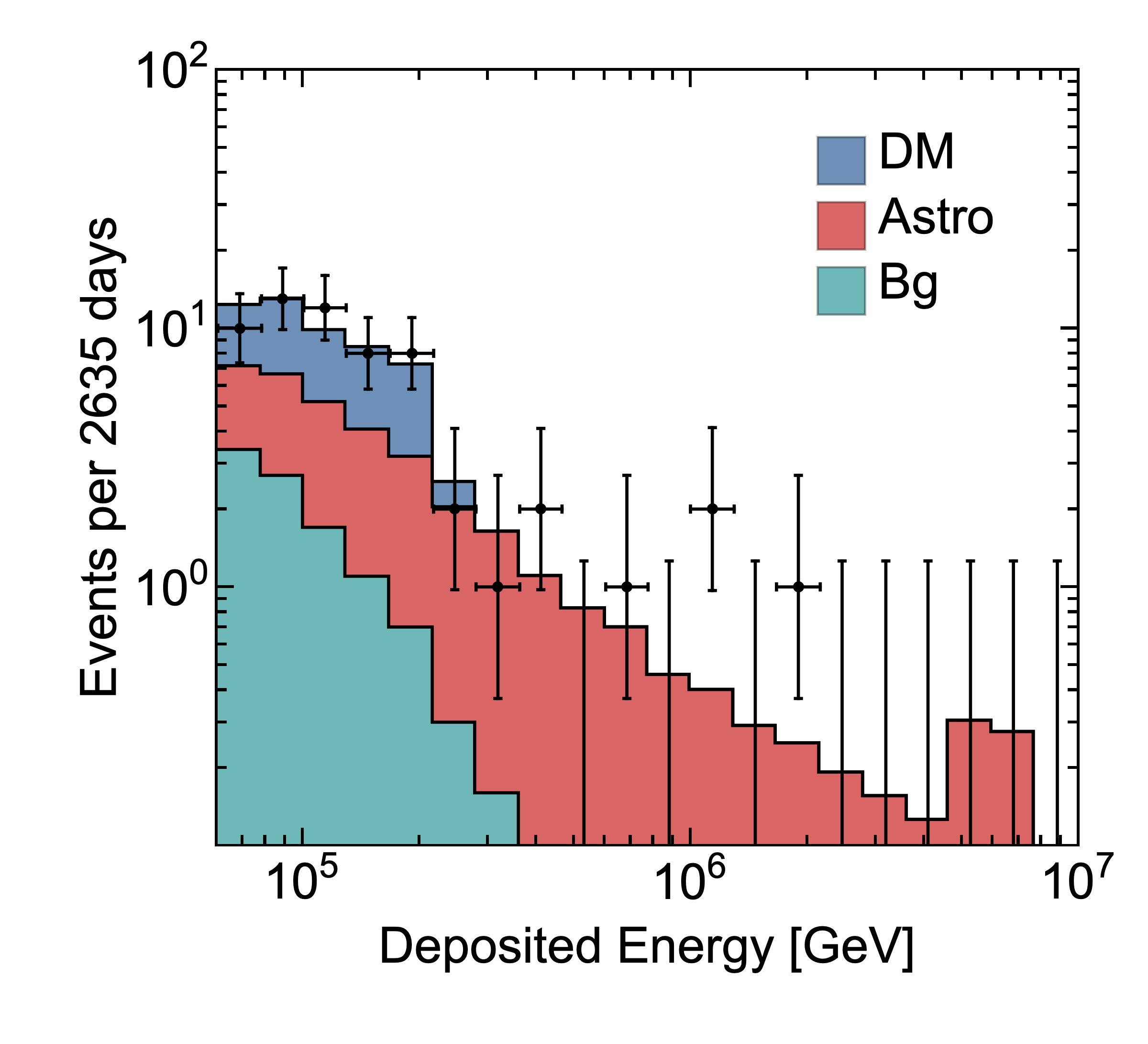}
\caption{Channel $W$, no prior distribution}
\end{subfigure}
\begin{subfigure}[b]{0.45\textwidth}
\includegraphics[width=\textwidth]{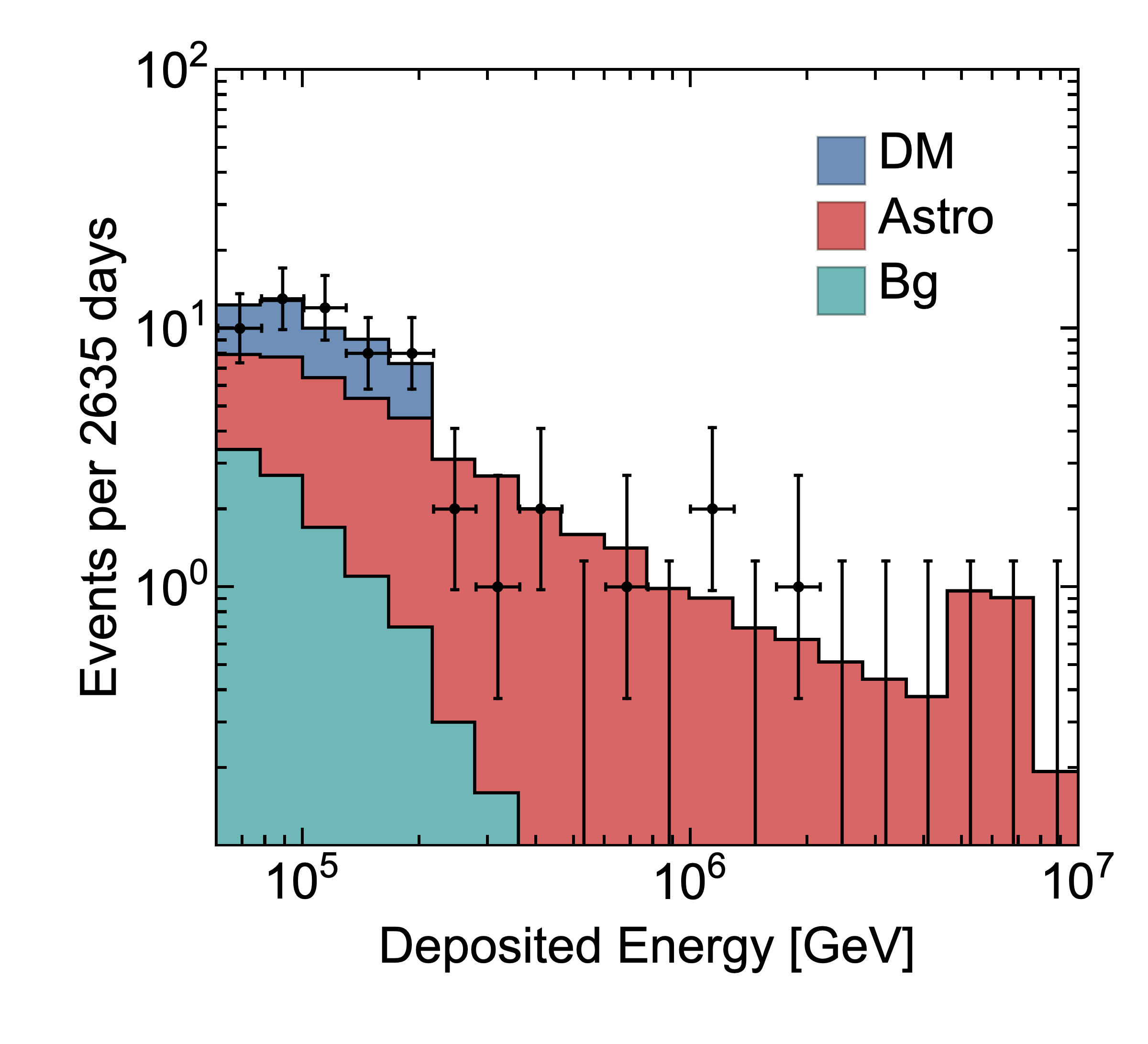}
\caption{Channel $W$, prior distribution}
\end{subfigure}
\vskip2.mm
\end{center}
\caption{Numbers of neutrino events represented as a function of the neutrino energy for the two-component spectrum in the two decay channels of $\chi\rightarrow \nu_e \nu_e$ and $\chi\rightarrow\mu \mu$. The astrophysical and Dark Matter parameters are evaluated at their best fit values both in the case of no prior information and in the case of a gaussian prior distribution taking into account the throughgoing muons information. }\label{fighist}
\end{figure}

The question of the competing maxima, one for a Dark Matter mass of the order of PeV and one at the order of hundreds of TeV, can be made clearer by comparing the relative contribution of the astrophysical and Dark Matter spectrum in either of the two cases above mentioned, both with the introduction of the prior information and without it. We have done so in Figure~\ref{fighist} for two representative cases: in the case $\chi\rightarrow \nu_e \nu_e$ the best fit value of the Dark Matter mass is of order $\sim 1$ PeV, while in the case $\chi\rightarrow WW$ it is of order $\sim 100$ TeV. For both decay channels we represent the histogram for the best fit values in the case with prior information and without it. It can now be seen that in the two channels, for the case without prior information, the Dark Matter excess is relevant in two different regions: in the former only the higher energies are affected, while in the latter the lower energies of order $100$ TeV are influenced. The introduction of the prior distribution moves the Dark Matter contribution in both cases at lower energies, while favoring a slightly harder astrophysical spectrum, as is clear from the presence of more events at higher energies.

\section{High energy gamma from decaying dark matter: background comparison}

As has been detailed in previous sections, a possible way of dealing with the investigation of Dark Matter decay is the analysis of the gamma ray fluxes. There are in fact a large number of experiments detecting high energy gamma rays which might be able to put further constraints on the models which we are analyzing. Some of these experiments generally work through point-like sources investigations: these include MAGIC \cite{Acciari:2018sjn}, HESS \cite{Rinchiuso:2019rrh}, Veritas \cite{Zitzer:2017xlo}, HAWC \cite{Abeysekara:2017jxs}, CARPET-3 \cite{Dzhappuev:2018bnl} and the future CTA \cite{Pierre:2014tra}. A second class of experiments is instead able to detect the diffuse flux: these include Fermi-LAT \cite{Atwood:2009ez}, Pierre Auger \cite{Aab:2015bza}, CASA-MIA \cite{Chantell:1997gs} and KASCADE \cite{Kang:2015gpa}. In the following we will qualitatively investigate their potentialities in detecting the Dark Matter produced photon fluxes for both kinds of experiments. In order for this strategy to be useful, a clear necessity is that the photon fluxes produced by the decay should be distinguishable from the background. We will therefore test whether this requirement is met in the cases of interest for IceCube.

Since most of the high energy experiments mentioned above typically investigate point-like sources, we will make the comparison for the gamma ray fluxes coming from a region of $2^{\circ}\times 2^{\circ}$ around the Galactic Center. We follow the analysis performed in \cite{Ibarra:2015tya}, studying a region with an aperture which is comparable to the field of view of the CTA experiment. The estimation of the background is performed similarly to \cite{Ibarra:2015tya}, taking into account different background components. The detailed procedure is left for the appendix. 

\begin{figure}[t!]
\begin{center}
\includegraphics[width=0.45\textwidth]{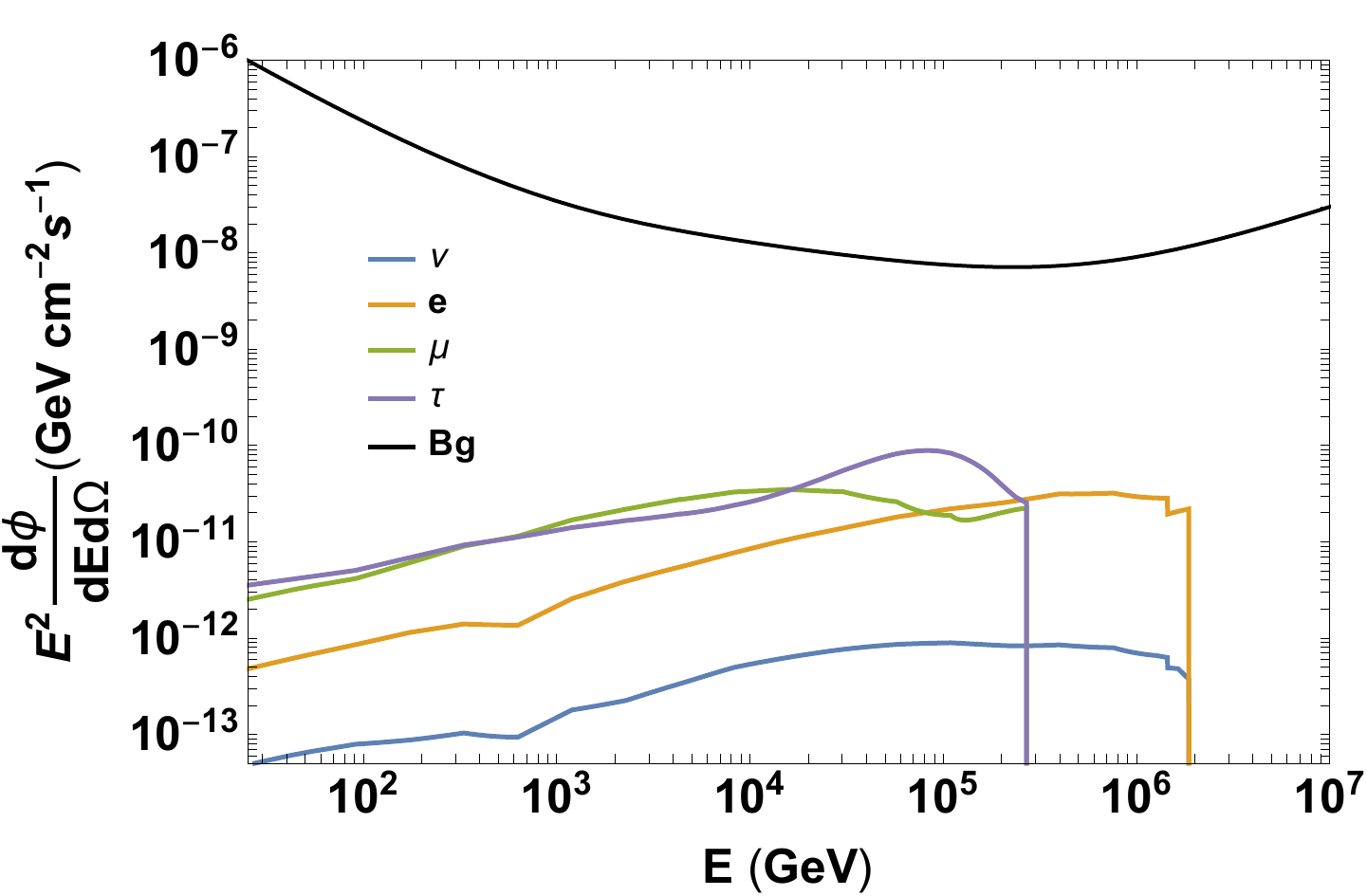}
\hskip4.mm
\includegraphics[width=0.45\textwidth]{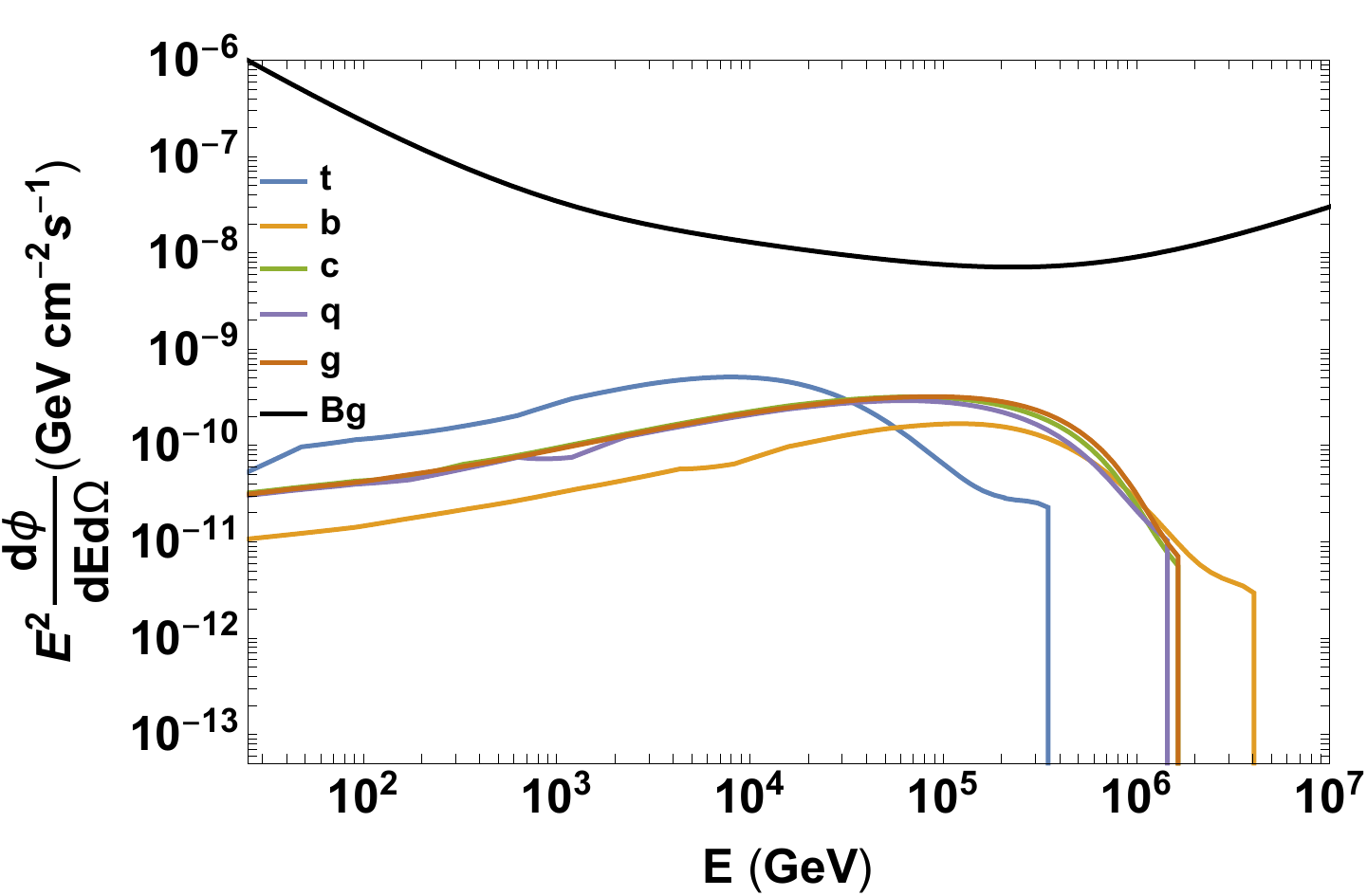}
\includegraphics[width=0.45\textwidth]{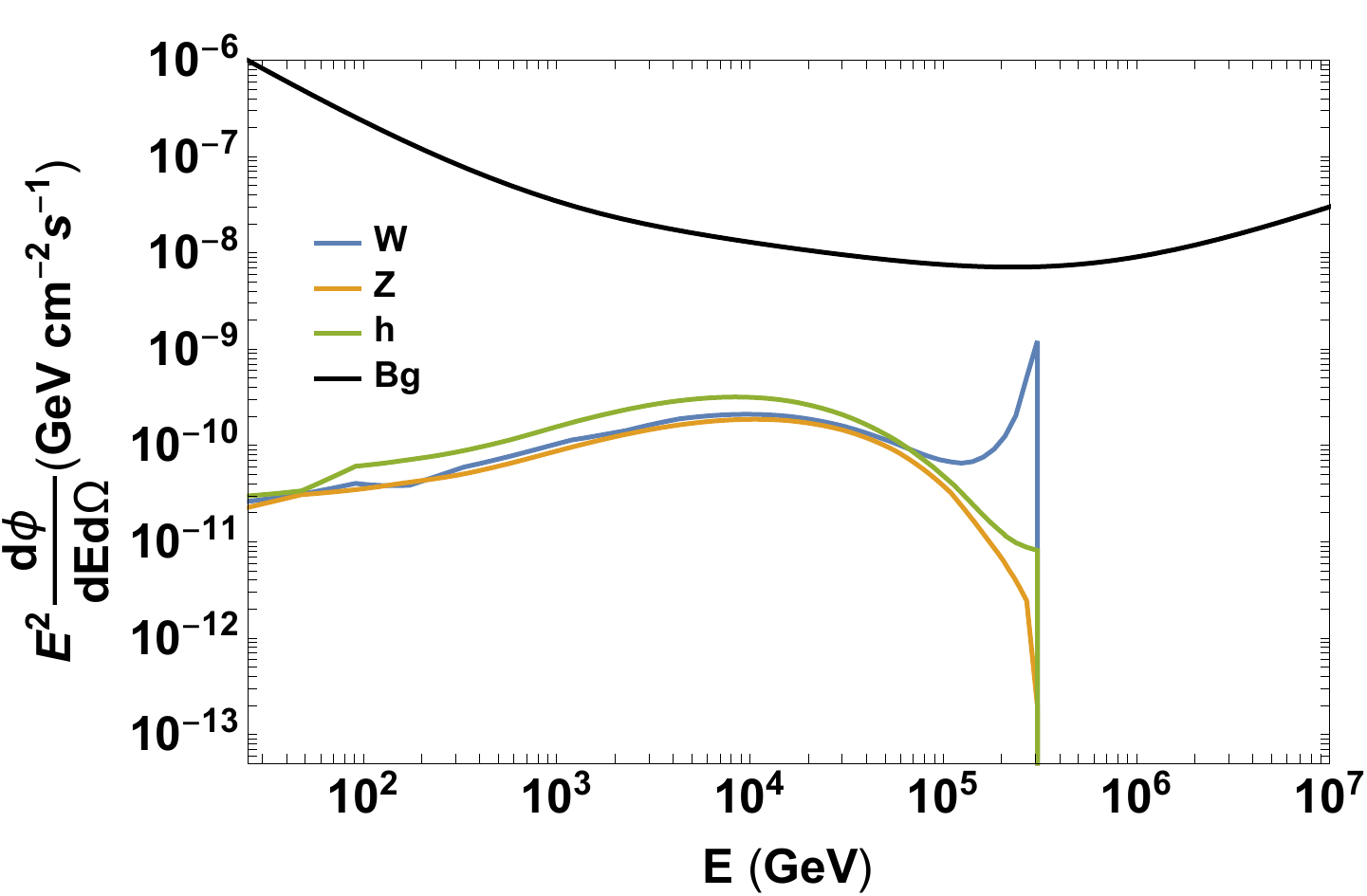}
\end{center}
\caption{Expected photon spectra for all best-fit channels, compared with the expected background flux from cosmic protons, electrons and photons.}\label{fig3}
\end{figure}

In figure~\ref{fig3} we compare the background spectrum with the photon spectra from decaying Dark Matter for all the analyzed channels: the mass and lifetime of the Dark Matter is fixed to the best fit value found in the previous section for each different channel. The DM spectra are lower than the background by at least two orders of magnitude: we therefore reach the conclusions that CTA would not be sensitive to such signals and could therefore not provide further constraints on the region of the parameter space which is of special interest to IceCube. Accordingly, the same can be said for the other high energy experiments such as HESS, VERITAS and MAGIC having a lower sensitivity.

A more detailed analysis can be performed along the guidelines of \cite{Ibarra:2015tya} to determine the significance with which CTA would be able to discriminate the presence of the predicted fluxes. In order to do so, we need to predict, for a given incident flux per unit energy, the number of events expected at CTA. This requires a knowledge of the CTA effective area $A_{eff}(E)$ and reconstruction function $R(E,E')$. The latter is assumed to be Gaussian, with a $68\%$ containment interval provided by the CTA Collaboration, together with the effective area \cite{Ong:2019zyq}. In terms of these elements the expected number of events in an energy bin $\Delta E$ is:
\begin{equation}
n=\Delta t \int_{\Delta E} dE \int dE' R(E,E') A_{eff} (E') \frac{d\Phi}{dE'}(E') 
\end{equation}
where $\Delta t$ is the $50$ hours exposition time.

We divide the energy interval from $25$ GeV to $100$ TeV into $200$ bins per decade and predict the number of background events $b_i$ and of Dark Matter decaying events $a_i$ for each energy bin. This analysis has been separately done for each decay channel, with the DM parameters fixed to their IceCube best fit value.

To foresee the sensitivity of CTA we need to assess the distribution of the likelihood ratio between the case of pure background and the case of background plus signal. This distribution has to be computed in the case of events distributed according to the background plus signal case\footnote{If the events were distributed according to the background only case, we would assess the exclusion p-value.}. The likelihood is assumed to take the multi-Poissonian form \eqref{multipoisson}. If we call $\Lambda$ the likelihood ratio, its distribution in the vicinity of the maximum can be well approximated by a Gaussian with a mean value of:
\begin{equation}
\bar{\Lambda}=\sum_i \left[(a_i+b_i)\ln\left(\frac{a_i+b_i}{b_i}\right)-a_i\right]
\end{equation}
and a standard deviation of:
\begin{equation}
\sigma=\sqrt{\sum_i\left[(a_i+b_i)\ln^2\left(\frac{a_i+b_i}{b_i}\right)\right]}
\end{equation}
To assess the number of standard deviations with which it would be possible to exclude the pure background hypothesis, we notice that in such hypothesis the expected value of the likelihood ratio would be:
\begin{equation}
\bar{\Lambda_{pb}}=\sum_i \left[b_i\ln\left(\frac{a_i+b_i}{b_i}\right)-a_i\right]
\end{equation}
It follows that the number of standard deviations for rejection of the pure background hypothesis would be the ratio:
\begin{equation}\label{nsig}
N_{\sigma}=\frac{\sum_i a_i \ln\left(\frac{a_i+b_i}{b_i}\right)}{\sigma}
\end{equation}
Performing the preceding analysis for each channel, we find the results in Table~\ref{tabsig}, where for each channel we provide the predicted number of standard deviations with which, in the case of signal plus background, it would be possible to reject the hypothesis of pure background. Confirming the intuitive conclusions coming from the comparison with the background, we see that for none of the analyzed channels it would be possible to discriminate the signal. This result is in agreement with \cite{Pierre:2014tra}. We also notice that in \eqref{nsig} we can expand in the small ratio $\frac{a_i}{b_i}$ to obtain the form:
\begin{equation}
N_{\sigma}\simeq \sqrt{\sum_i \frac{a_i^2}{b_i}}
\end{equation}
which shows that the number of standard deviations scales linearly with the signal amplitude for small signals. This means that, for example, the $t\bar{t}$ channel would reach the $5 \sigma$ detection limit for a signal of order $\sim 10$ times larger, which corresponds to a lifetime smaller by a factor of $10$. By comparison with figure~\ref{figtaum3} we see that this value is lower than the Fermi-LAT exclusion limits, which confirms the expectation that analyses of point-like sources are generally not competitive in giving constraints on the Dark Matter fluxes.
\begin{table}[h!]
\begin{center}
\begin{tabular}{|c|c|c|}
\hline
Channel & $N_{\sigma}$ CTA sensitivity  \\
\hline
$\nu\nu$ &0.00066  \\
\hline
$e^+e^-$ & 0.012  \\
$\mu^+\mu^-$ & 0.038  \\
$\tau^+\tau^-$ & 0.046   \\
\hline
$W^+W^-$ & 0.23  \\
$ZZ$ & 0.20  \\
$hh$ & 0.35  \\
\hline
$b\overline{b}$ & 0.11  \\
$c\overline{c}$ & 0.28  \\
$t\overline{t}$ & 0.56  \\
$q\overline{q}$ & 0.25  \\
$gg$ & 0.27  \\
\hline
\end{tabular}\caption{Number of standard deviations with which we expect CTA to detect decaying Dark Matter signals for all channels: the DM parameters are fixed at their IceCube best fit values reported in Table~\ref{tab2}.}\label{tabsig}
\end{center}
\end{table}

A further possibility of investigation of the Dark Matter fluxes, as explained above, is the diffuse flux. We provide for reference the diffuse fluxes expected for all channels with a DM mass and lifetime fixed at their best fit values, together with the data collected by KASCADE \cite{Chantell:1997gs} and CASA-MIA \cite{Schatz:2003aw}. In Figure~\ref{fig4} we show the diffuse fluxes which, as foreseen, are more constrained by the experimental data than the corresponding point-like investigations.


\section{Conclusions}

The multimessenger analysis has recently become a guiding instrument in the investigation of cosmological Dark Matter. Following these guidelines, in this work we have focused on the neutrino and gamma ray production from decaying Dark Matter. The unavoidable model dependence of such study has been reduced to the decay into particle-antiparticle pair from the Standard Model for various decay channels. In particular, we have analyzed the latest IceCube HESE data under the assumption of a dark component superimposed to an astrophysical cumulative spectrum. This constitutes an update of previous analyses, and therefore allows the determination of more stringent bounds in the decaying Dark Matter parameter space. The results for the best fit parameters show that all of the channels favor either a Dark Matter mass around $100$ TeV or a Dark Matter mass in the PeV region: in the former case the spectral index is expected to become nearer to the values of $2.3$, while in the latter it is expected to increase to higher values. The statistical test evidences that for none of the channels the pure astrophysical spectrum can be rejected with more than the $2\sigma$ level, and in some of the channels it cannot be rejected even at the $1\sigma$ level. This implies that the Dark Matter introduction does not substantially improve the data explanation.

Moreover, we have used, for each decay channel, the best fit scenario to the IceCube data to predict the expected gamma ray spectra. The purpose of this examination was to provide an estimation of the gamma rays experiment possibilities in further constraining the model. We have done so both for point-like and diffuse searches. In the former case, we have predicted the expected spectra from Dark Matter for all channels in the region around the Galactic Centre, where these are expected to be higher, with the modeled background from the same region. This analysis has been extended, for the specific case of the Cherenkov Telescope Array, to a statistical investigation of the p-value with which the future experiment should be able to assess the presence of a Dark Matter originated component. The results show that it would not be possible through a point-like analysis to significantly constrain the Dark Matter parameter space beyond the already existing constraints, in that the IceCube best fits are not testable even at the $1\sigma$ level. In the case of the diffuse searches, we have compared the predicted diffuse gamma ray spectra for all channels with the data coming from KASCADE, CASA-MIA and Fermi LAT. While the former two are able to constrain only Dark Matter with masses above the $10$ PeV, the latter is more informative in determining the admissible region of the parameter space, and has in fact already been adopted in the previous analysis of the IceCube data.

\section*{Acknowledgments}
The authors thank Kohta Murase for his useful comments during the 36th International Cosmic-Ray conference in Madison. This work was partially supported by the research grant number 2017W4HA7S  "NAT-NET: Neutrino and Astroparticle Theory Network'' under the program PRIN 2017 funded by the Italian Ministero dell’Istruzione, dell'Università e della Ricerca (MIUR). 
D.F.G.F., G.M., S.M. and O.P. acknowledge partial support by the research
project TAsP (Theoretical Astroparticle Physics) funded by the Instituto
Nazionale di Fisica Nucleare (INFN).

\appendix
\section{Gamma rays background estimation for the Galactic Center}

There are three main components to the gamma rays background at the Galactic Center: one coming from the showers induced by cosmic protons, one coming from the electron and positron induced showers and a gamma ray Galactic background. The component of background due to the showers produced by the cosmic protons is parameterized as \cite{Hoerandel:2002yg}:
\begin{equation}
\frac{d^2\Phi_p}{dEd\Omega}=8.73\times 10^{-6} \left(\frac{E_p}{\rm{TeV}}\right)^{-2.71} \rm{TeV^{-1} cm^{-2} s^{-1} sr^{-1}}
\end{equation}
Due to the different yield of the proton showers, this spectrum is actually calculated for an energy $E'_p = \rho E_p$, where $\rho\simeq 3$. Another background component is the spectrum of the cascades induced by the cosmic electrons. For energies below $\sim 100 \rm{GeV}$ this spectrum is measured by AMS-02 \cite{Aguilar:2014fea} as: \begin{equation}
\frac{d^2\Phi_{e1}}{dEd\Omega}=9.93\times 10^{-9} \left(\frac{E_e}{\rm{TeV}}\right)^{-3.17} \rm{TeV^{-1} cm^{-2} s^{-1} sr^{-1}}
\end{equation}
while above $\sim 30 \rm{GeV}$ is measured by HESS as \cite{Aharonian:2006au}:
\begin{equation}
\frac{d^2\Phi_{e2}}{dEd\Omega}=1.17\times 10^{-8} \left(\frac{E_e}{\rm{TeV}}\right)^{-3.9} \rm{TeV^{-1} cm^{-2} s^{-1} sr^{-1}}
\end{equation}
To ensure a smooth transition between these two spectra, we take
\begin{equation}
\frac{d^2\Phi_{e}}{dEd\Omega}=\left(\frac{d^2\Phi_{e1}}{dEd\Omega}^{-2}+\frac{d^2\Phi_{e2}}{dEd\Omega}^{-2}\right)^{-1/2}
\end{equation}
Finally, the gamma ray background is made up of mainly two components. On the one hand there is the Galactic ridge emission, which is here parameterized as \cite{Aharonian:2006au}:
\begin{equation}
\frac{d^2\Phi_{\gamma 1}}{dEd\Omega}=1.73\times 10^{-8} \left(\frac{E_{\gamma}}{\rm{TeV}}\right)^{-2.29}
\end{equation}
On the other hand, the second component can be identified as a point-like source. Its spectrum is in the range $5-100$ $\rm{GeV}$ \cite{Chernyakova:2011zz}:
\begin{equation}
\frac{d\Phi_{gc1}}{dE}=1.11\times 10^{-12} \left(\frac{E}{\rm{TeV}}\right)^{-2.68} \rm{TeV^{-1} cm^{-2} s^{-1} sr^{-1}}
\end{equation}
and in the range $160$ $\rm{GeV}-30$ $\rm{TeV}$ \cite{Aharonian:2006wh}:
\begin{equation}
\frac{d\Phi_{gc2}}{dE}=2.34\times 10^{-12} \left(\frac{E}{\rm{TeV}}\right)^{-2.25} \rm{TeV^{-1} cm^{-2} s^{-1} sr^{-1}}
\end{equation}
The two spectra are interpolated by the form
\begin{equation}
\frac{d\Phi_{gc}}{dE}=\left(\frac{d\Phi_{gc1}}{dE}^5+\frac{d\Phi_{gc2}}{dE}^5\right)^{1/5}
\end{equation}
In the overall background flux the different efficiency in the detection of cosmic protons is taken into account by multiplying the corresponding flux by a factor $\epsilon_{p}=0.001+0.002\left(\frac{E_p}{20 \rm{TeV}}\right)^{1.4}$.


\begin{thebibliography}{109}%
\makeatletter
\providecommand \@ifxundefined [1]{%
 \@ifx{#1\undefined}
}%
\providecommand \@ifnum [1]{%
 \ifnum #1\expandafter \@firstoftwo
 \else \expandafter \@secondoftwo
 \fi
}%
\providecommand \@ifx [1]{%
 \ifx #1\expandafter \@firstoftwo
 \else \expandafter \@secondoftwo
 \fi
}%
\providecommand \natexlab [1]{#1}%
\providecommand \enquote  [1]{``#1''}%
\providecommand \bibnamefont  [1]{#1}%
\providecommand \bibfnamefont [1]{#1}%
\providecommand \citenamefont [1]{#1}%
\providecommand \href@noop [0]{\@secondoftwo}%
\providecommand \href [0]{\begingroup \@sanitize@url \@href}%
\providecommand \@href[1]{\@@startlink{#1}\@@href}%
\providecommand \@@href[1]{\endgroup#1\@@endlink}%
\providecommand \@sanitize@url [0]{\catcode `\\12\catcode `\$12\catcode
  `\&12\catcode `\#12\catcode `\^12\catcode `\_12\catcode `\%12\relax}%
\providecommand \@@startlink[1]{}%
\providecommand \@@endlink[0]{}%
\providecommand \url  [0]{\begingroup\@sanitize@url \@url }%
\providecommand \@url [1]{\endgroup\@href {#1}{\urlprefix }}%
\providecommand \urlprefix  [0]{URL }%
\providecommand \Eprint [0]{\href }%
\providecommand \doibase [0]{http://dx.doi.org/}%
\providecommand \selectlanguage [0]{\@gobble}%
\providecommand \bibinfo  [0]{\@secondoftwo}%
\providecommand \bibfield  [0]{\@secondoftwo}%
\providecommand \translation [1]{[#1]}%
\providecommand \BibitemOpen [0]{}%
\providecommand \bibitemStop [0]{}%
\providecommand \bibitemNoStop [0]{.\EOS\space}%
\providecommand \EOS [0]{\spacefactor3000\relax}%
\providecommand \BibitemShut  [1]{\csname bibitem#1\endcsname}%
\let\auto@bib@innerbib\@empty
\bibitem [{\citenamefont {Aartsen}\ \emph {et~al.}(2013)\citenamefont {Aartsen}
  \emph {et~al.}}]{Aartsen:2013jdh}%
  \BibitemOpen
  \bibfield  {author} {\bibinfo {author} {\bibfnamefont {M.~G.}\ \bibnamefont
  {Aartsen}} \emph {et~al.} (\bibinfo {collaboration} {IceCube}),\ }\href
  {\doibase 10.1126/science.1242856} {\bibfield  {journal} {\bibinfo  {journal}
  {Science}\ }\textbf {\bibinfo {volume} {342}},\ \bibinfo {pages} {1242856}
  (\bibinfo {year} {2013})},\ \Eprint {http://arxiv.org/abs/1311.5238}
  {arXiv:1311.5238 [astro-ph.HE]} \BibitemShut {NoStop}%
\bibitem [{\citenamefont {Aartsen}\ \emph {et~al.}(2014)\citenamefont {Aartsen}
  \emph {et~al.}}]{Aartsen:2014gkd}%
  \BibitemOpen
  \bibfield  {author} {\bibinfo {author} {\bibfnamefont {M.~G.}\ \bibnamefont
  {Aartsen}} \emph {et~al.} (\bibinfo {collaboration} {IceCube}),\ }\href
  {\doibase 10.1103/PhysRevLett.113.101101} {\bibfield  {journal} {\bibinfo
  {journal} {Phys. Rev. Lett.}\ }\textbf {\bibinfo {volume} {113}},\ \bibinfo
  {pages} {101101} (\bibinfo {year} {2014})},\ \Eprint
  {http://arxiv.org/abs/1405.5303} {arXiv:1405.5303 [astro-ph.HE]} \BibitemShut
  {NoStop}%
\bibitem [{\citenamefont {Aartsen}\ \emph
  {et~al.}(2017{\natexlab{a}})\citenamefont {Aartsen} \emph
  {et~al.}}]{Aartsen:2017mau}%
  \BibitemOpen
  \bibfield  {author} {\bibinfo {author} {\bibfnamefont {M.~G.}\ \bibnamefont
  {Aartsen}} \emph {et~al.} (\bibinfo {collaboration} {IceCube}),\ }\href@noop
  {} {\  (\bibinfo {year} {2017}{\natexlab{a}})},\ \Eprint
  {http://arxiv.org/abs/1710.01191} {arXiv:1710.01191 [astro-ph.HE]}
  \BibitemShut {NoStop}%
\bibitem [{\citenamefont {Loeb}\ and\ \citenamefont
  {Waxman}(2006)}]{Loeb:2006tw}%
  \BibitemOpen
  \bibfield  {author} {\bibinfo {author} {\bibfnamefont {A.}~\bibnamefont
  {Loeb}}\ and\ \bibinfo {author} {\bibfnamefont {E.}~\bibnamefont {Waxman}},\
  }\href {\doibase 10.1088/1475-7516/2006/05/003} {\bibfield  {journal}
  {\bibinfo  {journal} {JCAP}\ }\textbf {\bibinfo {volume} {0605}},\ \bibinfo
  {pages} {003} (\bibinfo {year} {2006})},\ \Eprint
  {http://arxiv.org/abs/astro-ph/0601695} {arXiv:astro-ph/0601695 [astro-ph]}
  \BibitemShut {NoStop}%
\bibitem [{\citenamefont {Murase}\ \emph {et~al.}(2013)\citenamefont {Murase},
  \citenamefont {Ahlers},\ and\ \citenamefont {Lacki}}]{Murase:2013rfa}%
  \BibitemOpen
  \bibfield  {author} {\bibinfo {author} {\bibfnamefont {K.}~\bibnamefont
  {Murase}}, \bibinfo {author} {\bibfnamefont {M.}~\bibnamefont {Ahlers}}, \
  and\ \bibinfo {author} {\bibfnamefont {B.~C.}\ \bibnamefont {Lacki}},\ }\href
  {\doibase 10.1103/PhysRevD.88.121301} {\bibfield  {journal} {\bibinfo
  {journal} {Phys. Rev.}\ }\textbf {\bibinfo {volume} {D88}},\ \bibinfo {pages}
  {121301} (\bibinfo {year} {2013})},\ \Eprint {http://arxiv.org/abs/1306.3417}
  {arXiv:1306.3417 [astro-ph.HE]} \BibitemShut {NoStop}%
\bibitem [{\citenamefont {Tamborra}\ \emph {et~al.}(2014)\citenamefont
  {Tamborra}, \citenamefont {Ando},\ and\ \citenamefont
  {Murase}}]{Tamborra:2014xia}%
  \BibitemOpen
  \bibfield  {author} {\bibinfo {author} {\bibfnamefont {I.}~\bibnamefont
  {Tamborra}}, \bibinfo {author} {\bibfnamefont {S.}~\bibnamefont {Ando}}, \
  and\ \bibinfo {author} {\bibfnamefont {K.}~\bibnamefont {Murase}},\ }\href
  {\doibase 10.1088/1475-7516/2014/09/043} {\bibfield  {journal} {\bibinfo
  {journal} {JCAP}\ }\textbf {\bibinfo {volume} {1409}},\ \bibinfo {pages}
  {043} (\bibinfo {year} {2014})},\ \Eprint {http://arxiv.org/abs/1404.1189}
  {arXiv:1404.1189 [astro-ph.HE]} \BibitemShut {NoStop}%
\bibitem [{\citenamefont {Bechtol}\ \emph {et~al.}(2017)\citenamefont
  {Bechtol}, \citenamefont {Ahlers}, \citenamefont {Di~Mauro}, \citenamefont
  {Ajello},\ and\ \citenamefont {Vandenbroucke}}]{Bechtol:2015uqb}%
  \BibitemOpen
  \bibfield  {author} {\bibinfo {author} {\bibfnamefont {K.}~\bibnamefont
  {Bechtol}}, \bibinfo {author} {\bibfnamefont {M.}~\bibnamefont {Ahlers}},
  \bibinfo {author} {\bibfnamefont {M.}~\bibnamefont {Di~Mauro}}, \bibinfo
  {author} {\bibfnamefont {M.}~\bibnamefont {Ajello}}, \ and\ \bibinfo {author}
  {\bibfnamefont {J.}~\bibnamefont {Vandenbroucke}},\ }\href {\doibase
  10.3847/1538-4357/836/1/47} {\bibfield  {journal} {\bibinfo  {journal}
  {Astrophys. J.}\ }\textbf {\bibinfo {volume} {836}},\ \bibinfo {pages} {47}
  (\bibinfo {year} {2017})},\ \Eprint {http://arxiv.org/abs/1511.00688}
  {arXiv:1511.00688 [astro-ph.HE]} \BibitemShut {NoStop}%
\bibitem [{\citenamefont {Winter}(2013)}]{Winter:2013cla}%
  \BibitemOpen
  \bibfield  {author} {\bibinfo {author} {\bibfnamefont {W.}~\bibnamefont
  {Winter}},\ }\href {\doibase 10.1103/PhysRevD.88.083007} {\bibfield
  {journal} {\bibinfo  {journal} {Phys. Rev.}\ }\textbf {\bibinfo {volume}
  {D88}},\ \bibinfo {pages} {083007} (\bibinfo {year} {2013})},\ \Eprint
  {http://arxiv.org/abs/1307.2793} {arXiv:1307.2793 [astro-ph.HE]} \BibitemShut
  {NoStop}%
\bibitem [{\citenamefont {Murase}\ \emph {et~al.}(2016)\citenamefont {Murase},
  \citenamefont {Guetta},\ and\ \citenamefont {Ahlers}}]{Murase:2015xka}%
  \BibitemOpen
  \bibfield  {author} {\bibinfo {author} {\bibfnamefont {K.}~\bibnamefont
  {Murase}}, \bibinfo {author} {\bibfnamefont {D.}~\bibnamefont {Guetta}}, \
  and\ \bibinfo {author} {\bibfnamefont {M.}~\bibnamefont {Ahlers}},\ }\href
  {\doibase 10.1103/PhysRevLett.116.071101} {\bibfield  {journal} {\bibinfo
  {journal} {Phys. Rev. Lett.}\ }\textbf {\bibinfo {volume} {116}},\ \bibinfo
  {pages} {071101} (\bibinfo {year} {2016})},\ \Eprint
  {http://arxiv.org/abs/1509.00805} {arXiv:1509.00805 [astro-ph.HE]}
  \BibitemShut {NoStop}%
\bibitem [{\citenamefont {Aartsen}\ \emph
  {et~al.}(2018{\natexlab{a}})\citenamefont {Aartsen} \emph
  {et~al.}}]{IceCube:2018cha}%
  \BibitemOpen
  \bibfield  {author} {\bibinfo {author} {\bibfnamefont {M.~G.}\ \bibnamefont
  {Aartsen}} \emph {et~al.} (\bibinfo {collaboration} {IceCube}),\ }\href
  {\doibase 10.1126/science.aat2890} {\bibfield  {journal} {\bibinfo  {journal}
  {Science}\ }\textbf {\bibinfo {volume} {361}},\ \bibinfo {pages} {147}
  (\bibinfo {year} {2018}{\natexlab{a}})},\ \Eprint
  {http://arxiv.org/abs/1807.08794} {arXiv:1807.08794 [astro-ph.HE]}
  \BibitemShut {NoStop}%
\bibitem [{\citenamefont {Aartsen}\ \emph
  {et~al.}(2018{\natexlab{b}})\citenamefont {Aartsen} \emph
  {et~al.}}]{IceCube:2018dnn}%
  \BibitemOpen
  \bibfield  {author} {\bibinfo {author} {\bibfnamefont {M.~G.}\ \bibnamefont
  {Aartsen}} \emph {et~al.} (\bibinfo {collaboration} {IceCube, Fermi-LAT,
  MAGIC, AGILE, ASAS-SN, HAWC, H.E.S.S., INTEGRAL, Kanata, Kiso, Kapteyn,
  Liverpool Telescope, Subaru, Swift NuSTAR, VERITAS, VLA/17B-403}),\ }\href
  {\doibase 10.1126/science.aat1378} {\bibfield  {journal} {\bibinfo  {journal}
  {Science}\ }\textbf {\bibinfo {volume} {361}},\ \bibinfo {pages} {eaat1378}
  (\bibinfo {year} {2018}{\natexlab{b}})},\ \Eprint
  {http://arxiv.org/abs/1807.08816} {arXiv:1807.08816 [astro-ph.HE]}
  \BibitemShut {NoStop}%
\bibitem [{\citenamefont {Ansoldi}\ \emph {et~al.}(2018)\citenamefont {Ansoldi}
  \emph {et~al.}}]{Ahnen:2018mvi}%
  \BibitemOpen
  \bibfield  {author} {\bibinfo {author} {\bibfnamefont {S.}~\bibnamefont
  {Ansoldi}} \emph {et~al.} (\bibinfo {collaboration} {MAGIC}),\ }\href
  {\doibase 10.3847/2041-8213/aad083} {\bibfield  {journal} {\bibinfo
  {journal} {Astrophys. J. Lett.}\ } (\bibinfo {year} {2018}),\
  10.3847/2041-8213/aad083},\ \bibinfo {note} {[Astrophys. J.863,L10(2018)]},\
  \Eprint {http://arxiv.org/abs/1807.04300} {arXiv:1807.04300 [astro-ph.HE]}
  \BibitemShut {NoStop}%
\bibitem [{\citenamefont {Murase}\ \emph {et~al.}(2018)\citenamefont {Murase},
  \citenamefont {Oikonomou},\ and\ \citenamefont
  {Petropoulou}}]{Murase:2018iyl}%
  \BibitemOpen
  \bibfield  {author} {\bibinfo {author} {\bibfnamefont {K.}~\bibnamefont
  {Murase}}, \bibinfo {author} {\bibfnamefont {F.}~\bibnamefont {Oikonomou}}, \
  and\ \bibinfo {author} {\bibfnamefont {M.}~\bibnamefont {Petropoulou}},\
  }\href {\doibase 10.3847/1538-4357/aada00} {\bibfield  {journal} {\bibinfo
  {journal} {Astrophys. J.}\ }\textbf {\bibinfo {volume} {865}},\ \bibinfo
  {pages} {124} (\bibinfo {year} {2018})},\ \Eprint
  {http://arxiv.org/abs/1807.04748} {arXiv:1807.04748 [astro-ph.HE]}
  \BibitemShut {NoStop}%
\bibitem [{\citenamefont {Adrian-Martinez}\ \emph {et~al.}(2016)\citenamefont
  {Adrian-Martinez} \emph {et~al.}}]{Adrian-Martinez:2015ver}%
  \BibitemOpen
  \bibfield  {author} {\bibinfo {author} {\bibfnamefont {S.}~\bibnamefont
  {Adrian-Martinez}} \emph {et~al.} (\bibinfo {collaboration} {ANTARES,
  IceCube}),\ }\href {\doibase 10.3847/0004-637X/823/1/65} {\bibfield
  {journal} {\bibinfo  {journal} {Astrophys. J.}\ }\textbf {\bibinfo {volume}
  {823}},\ \bibinfo {pages} {65} (\bibinfo {year} {2016})},\ \Eprint
  {http://arxiv.org/abs/1511.02149} {arXiv:1511.02149 [hep-ex]} \BibitemShut
  {NoStop}%
\bibitem [{\citenamefont {Aartsen}\ \emph
  {et~al.}(2017{\natexlab{b}})\citenamefont {Aartsen} \emph
  {et~al.}}]{Aartsen:2016oji}%
  \BibitemOpen
  \bibfield  {author} {\bibinfo {author} {\bibfnamefont {M.~G.}\ \bibnamefont
  {Aartsen}} \emph {et~al.} (\bibinfo {collaboration} {IceCube}),\ }\href
  {\doibase 10.3847/1538-4357/835/2/151} {\bibfield  {journal} {\bibinfo
  {journal} {Astrophys. J.}\ }\textbf {\bibinfo {volume} {835}},\ \bibinfo
  {pages} {151} (\bibinfo {year} {2017}{\natexlab{b}})},\ \Eprint
  {http://arxiv.org/abs/1609.04981} {arXiv:1609.04981 [astro-ph.HE]}
  \BibitemShut {NoStop}%
\bibitem [{\citenamefont {Aartsen}\ \emph
  {et~al.}(2019{\natexlab{a}})\citenamefont {Aartsen} \emph
  {et~al.}}]{Aartsen:2018fpd}%
  \BibitemOpen
  \bibfield  {author} {\bibinfo {author} {\bibfnamefont {M.~G.}\ \bibnamefont
  {Aartsen}} \emph {et~al.} (\bibinfo {collaboration} {IceCube}),\ }\href
  {\doibase 10.1103/PhysRevLett.122.051102} {\bibfield  {journal} {\bibinfo
  {journal} {Phys. Rev. Lett.}\ }\textbf {\bibinfo {volume} {122}},\ \bibinfo
  {pages} {051102} (\bibinfo {year} {2019}{\natexlab{a}})},\ \Eprint
  {http://arxiv.org/abs/1807.11492} {arXiv:1807.11492 [astro-ph.HE]}
  \BibitemShut {NoStop}%
\bibitem [{\citenamefont {Aartsen}\ \emph
  {et~al.}(2019{\natexlab{b}})\citenamefont {Aartsen} \emph
  {et~al.}}]{Aartsen:2018ywr}%
  \BibitemOpen
  \bibfield  {author} {\bibinfo {author} {\bibfnamefont {M.~G.}\ \bibnamefont
  {Aartsen}} \emph {et~al.} (\bibinfo {collaboration} {IceCube}),\ }\href
  {\doibase 10.1140/epjc/s10052-019-6680-0} {\bibfield  {journal} {\bibinfo
  {journal} {Eur. Phys. J.}\ }\textbf {\bibinfo {volume} {C79}},\ \bibinfo
  {pages} {234} (\bibinfo {year} {2019}{\natexlab{b}})},\ \Eprint
  {http://arxiv.org/abs/1811.07979} {arXiv:1811.07979 [hep-ph]} \BibitemShut
  {NoStop}%
\bibitem [{\citenamefont {Murase}\ and\ \citenamefont
  {Waxman}(2016)}]{Murase:2016gly}%
  \BibitemOpen
  \bibfield  {author} {\bibinfo {author} {\bibfnamefont {K.}~\bibnamefont
  {Murase}}\ and\ \bibinfo {author} {\bibfnamefont {E.}~\bibnamefont
  {Waxman}},\ }\href {\doibase 10.1103/PhysRevD.94.103006} {\bibfield
  {journal} {\bibinfo  {journal} {Phys. Rev.}\ }\textbf {\bibinfo {volume}
  {D94}},\ \bibinfo {pages} {103006} (\bibinfo {year} {2016})},\ \Eprint
  {http://arxiv.org/abs/1607.01601} {arXiv:1607.01601 [astro-ph.HE]}
  \BibitemShut {NoStop}%
\bibitem [{\citenamefont {Aartsen}\ \emph
  {et~al.}(2015{\natexlab{a}})\citenamefont {Aartsen} \emph
  {et~al.}}]{Aartsen:2014ivk}%
  \BibitemOpen
  \bibfield  {author} {\bibinfo {author} {\bibfnamefont {M.~G.}\ \bibnamefont
  {Aartsen}} \emph {et~al.} (\bibinfo {collaboration} {IceCube}),\ }\href
  {\doibase 10.1016/j.astropartphys.2015.01.001} {\bibfield  {journal}
  {\bibinfo  {journal} {Astropart. Phys.}\ }\textbf {\bibinfo {volume} {66}},\
  \bibinfo {pages} {39} (\bibinfo {year} {2015}{\natexlab{a}})},\ \Eprint
  {http://arxiv.org/abs/1408.0634} {arXiv:1408.0634 [astro-ph.HE]} \BibitemShut
  {NoStop}%
\bibitem [{\citenamefont {Ando}\ \emph {et~al.}(2017)\citenamefont {Ando},
  \citenamefont {Feyereisen},\ and\ \citenamefont {Fornasa}}]{Ando:2017xcb}%
  \BibitemOpen
  \bibfield  {author} {\bibinfo {author} {\bibfnamefont {S.}~\bibnamefont
  {Ando}}, \bibinfo {author} {\bibfnamefont {M.~R.}\ \bibnamefont
  {Feyereisen}}, \ and\ \bibinfo {author} {\bibfnamefont {M.}~\bibnamefont
  {Fornasa}},\ }\href {\doibase 10.1103/PhysRevD.95.103003} {\bibfield
  {journal} {\bibinfo  {journal} {Phys. Rev.}\ }\textbf {\bibinfo {volume}
  {D95}},\ \bibinfo {pages} {103003} (\bibinfo {year} {2017})},\ \Eprint
  {http://arxiv.org/abs/1701.02165} {arXiv:1701.02165 [astro-ph.HE]}
  \BibitemShut {NoStop}%
\bibitem [{\citenamefont {Mertsch}\ \emph {et~al.}(2017)\citenamefont
  {Mertsch}, \citenamefont {Rameez},\ and\ \citenamefont
  {Tamborra}}]{Mertsch:2016hcd}%
  \BibitemOpen
  \bibfield  {author} {\bibinfo {author} {\bibfnamefont {P.}~\bibnamefont
  {Mertsch}}, \bibinfo {author} {\bibfnamefont {M.}~\bibnamefont {Rameez}}, \
  and\ \bibinfo {author} {\bibfnamefont {I.}~\bibnamefont {Tamborra}},\ }\href
  {\doibase 10.1088/1475-7516/2017/03/011} {\bibfield  {journal} {\bibinfo
  {journal} {JCAP}\ }\textbf {\bibinfo {volume} {1703}},\ \bibinfo {pages}
  {011} (\bibinfo {year} {2017})},\ \Eprint {http://arxiv.org/abs/1612.07311}
  {arXiv:1612.07311 [astro-ph.HE]} \BibitemShut {NoStop}%
\bibitem [{\citenamefont {Dekker}\ and\ \citenamefont
  {Ando}(2019)}]{Dekker:2018cqu}%
  \BibitemOpen
  \bibfield  {author} {\bibinfo {author} {\bibfnamefont {A.}~\bibnamefont
  {Dekker}}\ and\ \bibinfo {author} {\bibfnamefont {S.}~\bibnamefont {Ando}},\
  }\href {\doibase 10.1088/1475-7516/2019/02/002} {\bibfield  {journal}
  {\bibinfo  {journal} {JCAP}\ }\textbf {\bibinfo {volume} {1902}},\ \bibinfo
  {pages} {002} (\bibinfo {year} {2019})},\ \Eprint
  {http://arxiv.org/abs/1811.02576} {arXiv:1811.02576 [astro-ph.HE]}
  \BibitemShut {NoStop}%
\bibitem [{\citenamefont {Senno}\ \emph {et~al.}(2016)\citenamefont {Senno},
  \citenamefont {Murase},\ and\ \citenamefont {Meszaros}}]{Senno:2015tsn}%
  \BibitemOpen
  \bibfield  {author} {\bibinfo {author} {\bibfnamefont {N.}~\bibnamefont
  {Senno}}, \bibinfo {author} {\bibfnamefont {K.}~\bibnamefont {Murase}}, \
  and\ \bibinfo {author} {\bibfnamefont {P.}~\bibnamefont {Meszaros}},\ }\href
  {\doibase 10.1103/PhysRevD.93.083003} {\bibfield  {journal} {\bibinfo
  {journal} {Phys. Rev.}\ }\textbf {\bibinfo {volume} {D93}},\ \bibinfo {pages}
  {083003} (\bibinfo {year} {2016})},\ \Eprint
  {http://arxiv.org/abs/1512.08513} {arXiv:1512.08513 [astro-ph.HE]}
  \BibitemShut {NoStop}%
\bibitem [{\citenamefont {Stettner}(2019)}]{ICRC_TG}%
  \BibitemOpen
  \bibfield  {author} {\bibinfo {author} {\bibfnamefont {J.}~\bibnamefont
  {Stettner}} (\bibinfo {collaboration} {IceCube}),\ }\bibfield  {booktitle}
  {\emph {\bibinfo {booktitle} {{Measurement of the diffuse astrophysical muon-
  neutrino spectrum with ten years of IceCube data}}},\ }\href@noop {}
  {\bibfield  {journal} {\bibinfo  {journal} {PoS (ICRC2019)1017}\ } (\bibinfo
  {year} {2019})}\BibitemShut {NoStop}%
\bibitem [{\citenamefont {Waxman}\ and\ \citenamefont
  {Bahcall}(1999)}]{Waxman:1998yy}%
  \BibitemOpen
  \bibfield  {author} {\bibinfo {author} {\bibfnamefont {E.}~\bibnamefont
  {Waxman}}\ and\ \bibinfo {author} {\bibfnamefont {J.~N.}\ \bibnamefont
  {Bahcall}},\ }\href {\doibase 10.1103/PhysRevD.59.023002} {\bibfield
  {journal} {\bibinfo  {journal} {Phys. Rev.}\ }\textbf {\bibinfo {volume}
  {D59}},\ \bibinfo {pages} {023002} (\bibinfo {year} {1999})},\ \Eprint
  {http://arxiv.org/abs/hep-ph/9807282} {arXiv:hep-ph/9807282 [hep-ph]}
  \BibitemShut {NoStop}%
\bibitem [{\citenamefont {Schneider}(2019)}]{ICRC_HESE}%
  \BibitemOpen
  \bibfield  {author} {\bibinfo {author} {\bibfnamefont {A.}~\bibnamefont
  {Schneider}} (\bibinfo {collaboration} {IceCube}),\ }\bibfield  {booktitle}
  {\emph {\bibinfo {booktitle} {{Characterization of the Astrophysical Diffuse
  Neutrino Flux with High-Energy Starting Events and Prospects for Future
  Measurements with IceCube}}},\ }\href@noop {} {\bibfield  {journal} {\bibinfo
   {journal} {PoS (ICRC2019)1004}\ } (\bibinfo {year} {2019})}\BibitemShut
  {NoStop}%
\bibitem [{\citenamefont {Aartsen}\ \emph
  {et~al.}(2015{\natexlab{b}})\citenamefont {Aartsen} \emph
  {et~al.}}]{Aartsen:2015knd}%
  \BibitemOpen
  \bibfield  {author} {\bibinfo {author} {\bibfnamefont {M.~G.}\ \bibnamefont
  {Aartsen}} \emph {et~al.} (\bibinfo {collaboration} {IceCube}),\ }\href
  {\doibase 10.1088/0004-637X/809/1/98} {\bibfield  {journal} {\bibinfo
  {journal} {Astrophys. J.}\ }\textbf {\bibinfo {volume} {809}},\ \bibinfo
  {pages} {98} (\bibinfo {year} {2015}{\natexlab{b}})},\ \Eprint
  {http://arxiv.org/abs/1507.03991} {arXiv:1507.03991 [astro-ph.HE]}
  \BibitemShut {NoStop}%
\bibitem [{\citenamefont {Chen}\ \emph {et~al.}(2015)\citenamefont {Chen},
  \citenamefont {Bhupal~Dev},\ and\ \citenamefont {Soni}}]{Chen:2014gxa}%
  \BibitemOpen
  \bibfield  {author} {\bibinfo {author} {\bibfnamefont {C.-Y.}\ \bibnamefont
  {Chen}}, \bibinfo {author} {\bibfnamefont {P.~S.}\ \bibnamefont
  {Bhupal~Dev}}, \ and\ \bibinfo {author} {\bibfnamefont {A.}~\bibnamefont
  {Soni}},\ }\href {\doibase 10.1103/PhysRevD.92.073001} {\bibfield  {journal}
  {\bibinfo  {journal} {Phys. Rev.}\ }\textbf {\bibinfo {volume} {D92}},\
  \bibinfo {pages} {073001} (\bibinfo {year} {2015})},\ \Eprint
  {http://arxiv.org/abs/1411.5658} {arXiv:1411.5658 [hep-ph]} \BibitemShut
  {NoStop}%
\bibitem [{\citenamefont {Palladino}\ and\ \citenamefont
  {Vissani}(2016)}]{Palladino:2016zoe}%
  \BibitemOpen
  \bibfield  {author} {\bibinfo {author} {\bibfnamefont {A.}~\bibnamefont
  {Palladino}}\ and\ \bibinfo {author} {\bibfnamefont {F.}~\bibnamefont
  {Vissani}},\ }\href {\doibase 10.3847/0004-637X/826/2/185} {\bibfield
  {journal} {\bibinfo  {journal} {Astrophys. J.}\ }\textbf {\bibinfo {volume}
  {826}},\ \bibinfo {pages} {185} (\bibinfo {year} {2016})},\ \Eprint
  {http://arxiv.org/abs/1601.06678} {arXiv:1601.06678 [astro-ph.HE]}
  \BibitemShut {NoStop}%
\bibitem [{\citenamefont {Vincent}\ \emph {et~al.}(2016)\citenamefont
  {Vincent}, \citenamefont {Palomares-Ruiz},\ and\ \citenamefont
  {Mena}}]{Vincent:2016nut}%
  \BibitemOpen
  \bibfield  {author} {\bibinfo {author} {\bibfnamefont {A.~C.}\ \bibnamefont
  {Vincent}}, \bibinfo {author} {\bibfnamefont {S.}~\bibnamefont
  {Palomares-Ruiz}}, \ and\ \bibinfo {author} {\bibfnamefont {O.}~\bibnamefont
  {Mena}},\ }\href {\doibase 10.1103/PhysRevD.94.023009} {\bibfield  {journal}
  {\bibinfo  {journal} {Phys. Rev.}\ }\textbf {\bibinfo {volume} {D94}},\
  \bibinfo {pages} {023009} (\bibinfo {year} {2016})},\ \Eprint
  {http://arxiv.org/abs/1605.01556} {arXiv:1605.01556 [astro-ph.HE]}
  \BibitemShut {NoStop}%
\bibitem [{\citenamefont {Palladino}\ \emph {et~al.}(2016)\citenamefont
  {Palladino}, \citenamefont {Spurio},\ and\ \citenamefont
  {Vissani}}]{Palladino:2016xsy}%
  \BibitemOpen
  \bibfield  {author} {\bibinfo {author} {\bibfnamefont {A.}~\bibnamefont
  {Palladino}}, \bibinfo {author} {\bibfnamefont {M.}~\bibnamefont {Spurio}}, \
  and\ \bibinfo {author} {\bibfnamefont {F.}~\bibnamefont {Vissani}},\ }\href
  {\doibase 10.1088/1475-7516/2016/12/045} {\bibfield  {journal} {\bibinfo
  {journal} {JCAP}\ }\textbf {\bibinfo {volume} {1612}},\ \bibinfo {pages}
  {045} (\bibinfo {year} {2016})},\ \Eprint {http://arxiv.org/abs/1610.07015}
  {arXiv:1610.07015 [astro-ph.HE]} \BibitemShut {NoStop}%
\bibitem [{\citenamefont {Anchordoqui}\ \emph {et~al.}(2017)\citenamefont
  {Anchordoqui}, \citenamefont {Block}, \citenamefont {Durand}, \citenamefont
  {Ha}, \citenamefont {Soriano},\ and\ \citenamefont
  {Weiler}}]{Anchordoqui:2016ewn}%
  \BibitemOpen
  \bibfield  {author} {\bibinfo {author} {\bibfnamefont {L.~A.}\ \bibnamefont
  {Anchordoqui}}, \bibinfo {author} {\bibfnamefont {M.~M.}\ \bibnamefont
  {Block}}, \bibinfo {author} {\bibfnamefont {L.}~\bibnamefont {Durand}},
  \bibinfo {author} {\bibfnamefont {P.}~\bibnamefont {Ha}}, \bibinfo {author}
  {\bibfnamefont {J.~F.}\ \bibnamefont {Soriano}}, \ and\ \bibinfo {author}
  {\bibfnamefont {T.~J.}\ \bibnamefont {Weiler}},\ }\href {\doibase
  10.1103/PhysRevD.95.083009} {\bibfield  {journal} {\bibinfo  {journal} {Phys.
  Rev.}\ }\textbf {\bibinfo {volume} {D95}},\ \bibinfo {pages} {083009}
  (\bibinfo {year} {2017})},\ \Eprint {http://arxiv.org/abs/1611.07905}
  {arXiv:1611.07905 [astro-ph.HE]} \BibitemShut {NoStop}%
\bibitem [{\citenamefont {Palladino}\ and\ \citenamefont
  {Winter}(2018)}]{Palladino:2018evm}%
  \BibitemOpen
  \bibfield  {author} {\bibinfo {author} {\bibfnamefont {A.}~\bibnamefont
  {Palladino}}\ and\ \bibinfo {author} {\bibfnamefont {W.}~\bibnamefont
  {Winter}},\ }\href {\doibase 10.3204/PUBDB-2018-01376,
  10.1051/0004-6361/201832731} {\bibfield  {journal} {\bibinfo  {journal}
  {Astron. Astrophys.}\ }\textbf {\bibinfo {volume} {615}},\ \bibinfo {pages}
  {A168} (\bibinfo {year} {2018})},\ \Eprint {http://arxiv.org/abs/1801.07277}
  {arXiv:1801.07277 [astro-ph.HE]} \BibitemShut {NoStop}%
\bibitem [{\citenamefont {Sui}\ and\ \citenamefont
  {Bhupal~Dev}(2018)}]{Sui:2018bbh}%
  \BibitemOpen
  \bibfield  {author} {\bibinfo {author} {\bibfnamefont {Y.}~\bibnamefont
  {Sui}}\ and\ \bibinfo {author} {\bibfnamefont {P.~S.}\ \bibnamefont
  {Bhupal~Dev}},\ }\href {\doibase 10.1088/1475-7516/2018/07/020} {\bibfield
  {journal} {\bibinfo  {journal} {JCAP}\ }\textbf {\bibinfo {volume} {1807}},\
  \bibinfo {pages} {020} (\bibinfo {year} {2018})},\ \Eprint
  {http://arxiv.org/abs/1804.04919} {arXiv:1804.04919 [hep-ph]} \BibitemShut
  {NoStop}%
\bibitem [{\citenamefont {Chianese}\ \emph
  {et~al.}(2017{\natexlab{a}})\citenamefont {Chianese}, \citenamefont {Mele},
  \citenamefont {Miele}, \citenamefont {Migliozzi},\ and\ \citenamefont
  {Morisi}}]{Chianese:2017jfa}%
  \BibitemOpen
  \bibfield  {author} {\bibinfo {author} {\bibfnamefont {M.}~\bibnamefont
  {Chianese}}, \bibinfo {author} {\bibfnamefont {R.}~\bibnamefont {Mele}},
  \bibinfo {author} {\bibfnamefont {G.}~\bibnamefont {Miele}}, \bibinfo
  {author} {\bibfnamefont {P.}~\bibnamefont {Migliozzi}}, \ and\ \bibinfo
  {author} {\bibfnamefont {S.}~\bibnamefont {Morisi}},\ }\href {\doibase
  10.3847/1538-4357/aa97e6} {\bibfield  {journal} {\bibinfo  {journal}
  {Astrophys. J.}\ }\textbf {\bibinfo {volume} {851}},\ \bibinfo {pages} {36}
  (\bibinfo {year} {2017}{\natexlab{a}})},\ \Eprint
  {http://arxiv.org/abs/1707.05168} {arXiv:1707.05168 [hep-ph]} \BibitemShut
  {NoStop}%
\bibitem [{\citenamefont {Chianese}\ \emph {et~al.}(2016)\citenamefont
  {Chianese}, \citenamefont {Miele}, \citenamefont {Morisi},\ and\
  \citenamefont {Vitagliano}}]{Chianese:2016opp}%
  \BibitemOpen
  \bibfield  {author} {\bibinfo {author} {\bibfnamefont {M.}~\bibnamefont
  {Chianese}}, \bibinfo {author} {\bibfnamefont {G.}~\bibnamefont {Miele}},
  \bibinfo {author} {\bibfnamefont {S.}~\bibnamefont {Morisi}}, \ and\ \bibinfo
  {author} {\bibfnamefont {E.}~\bibnamefont {Vitagliano}},\ }\href {\doibase
  10.1016/j.physletb.2016.03.084} {\bibfield  {journal} {\bibinfo  {journal}
  {Phys. Lett.}\ }\textbf {\bibinfo {volume} {B757}},\ \bibinfo {pages} {251}
  (\bibinfo {year} {2016})},\ \Eprint {http://arxiv.org/abs/1601.02934}
  {arXiv:1601.02934 [hep-ph]} \BibitemShut {NoStop}%
\bibitem [{\citenamefont {Chianese}\ \emph
  {et~al.}(2017{\natexlab{b}})\citenamefont {Chianese}, \citenamefont {Miele},\
  and\ \citenamefont {Morisi}}]{Chianese:2016kpu}%
  \BibitemOpen
  \bibfield  {author} {\bibinfo {author} {\bibfnamefont {M.}~\bibnamefont
  {Chianese}}, \bibinfo {author} {\bibfnamefont {G.}~\bibnamefont {Miele}}, \
  and\ \bibinfo {author} {\bibfnamefont {S.}~\bibnamefont {Morisi}},\ }\href
  {\doibase 10.1088/1475-7516/2017/01/007} {\bibfield  {journal} {\bibinfo
  {journal} {JCAP}\ }\textbf {\bibinfo {volume} {1701}},\ \bibinfo {pages}
  {007} (\bibinfo {year} {2017}{\natexlab{b}})},\ \Eprint
  {http://arxiv.org/abs/1610.04612} {arXiv:1610.04612 [hep-ph]} \BibitemShut
  {NoStop}%
\bibitem [{\citenamefont {Chianese}\ \emph
  {et~al.}(2017{\natexlab{c}})\citenamefont {Chianese}, \citenamefont {Miele},\
  and\ \citenamefont {Morisi}}]{Chianese:2017nwe}%
  \BibitemOpen
  \bibfield  {author} {\bibinfo {author} {\bibfnamefont {M.}~\bibnamefont
  {Chianese}}, \bibinfo {author} {\bibfnamefont {G.}~\bibnamefont {Miele}}, \
  and\ \bibinfo {author} {\bibfnamefont {S.}~\bibnamefont {Morisi}},\ }\href
  {\doibase 10.1016/j.physletb.2017.09.016} {\bibfield  {journal} {\bibinfo
  {journal} {Phys. Lett.}\ }\textbf {\bibinfo {volume} {B773}},\ \bibinfo
  {pages} {591} (\bibinfo {year} {2017}{\natexlab{c}})},\ \Eprint
  {http://arxiv.org/abs/1707.05241} {arXiv:1707.05241 [hep-ph]} \BibitemShut
  {NoStop}%
\bibitem [{\citenamefont {Anisimov}\ and\ \citenamefont
  {Di~Bari}(2009)}]{Anisimov:2008gg}%
  \BibitemOpen
  \bibfield  {author} {\bibinfo {author} {\bibfnamefont {A.}~\bibnamefont
  {Anisimov}}\ and\ \bibinfo {author} {\bibfnamefont {P.}~\bibnamefont
  {Di~Bari}},\ }\href {\doibase 10.1103/PhysRevD.80.073017} {\bibfield
  {journal} {\bibinfo  {journal} {Phys. Rev.}\ }\textbf {\bibinfo {volume}
  {D80}},\ \bibinfo {pages} {073017} (\bibinfo {year} {2009})},\ \Eprint
  {http://arxiv.org/abs/0812.5085} {arXiv:0812.5085 [hep-ph]} \BibitemShut
  {NoStop}%
\bibitem [{\citenamefont {Feldstein}\ \emph {et~al.}(2013)\citenamefont
  {Feldstein}, \citenamefont {Kusenko}, \citenamefont {Matsumoto},\ and\
  \citenamefont {Yanagida}}]{Feldstein:2013kka}%
  \BibitemOpen
  \bibfield  {author} {\bibinfo {author} {\bibfnamefont {B.}~\bibnamefont
  {Feldstein}}, \bibinfo {author} {\bibfnamefont {A.}~\bibnamefont {Kusenko}},
  \bibinfo {author} {\bibfnamefont {S.}~\bibnamefont {Matsumoto}}, \ and\
  \bibinfo {author} {\bibfnamefont {T.~T.}\ \bibnamefont {Yanagida}},\ }\href
  {\doibase 10.1103/PhysRevD.88.015004} {\bibfield  {journal} {\bibinfo
  {journal} {Phys. Rev.}\ }\textbf {\bibinfo {volume} {D88}},\ \bibinfo {pages}
  {015004} (\bibinfo {year} {2013})},\ \Eprint {http://arxiv.org/abs/1303.7320}
  {arXiv:1303.7320 [hep-ph]} \BibitemShut {NoStop}%
\bibitem [{\citenamefont {Esmaili}\ and\ \citenamefont
  {Serpico}(2013)}]{Esmaili:2013gha}%
  \BibitemOpen
  \bibfield  {author} {\bibinfo {author} {\bibfnamefont {A.}~\bibnamefont
  {Esmaili}}\ and\ \bibinfo {author} {\bibfnamefont {P.~D.}\ \bibnamefont
  {Serpico}},\ }\href {\doibase 10.1088/1475-7516/2013/11/054} {\bibfield
  {journal} {\bibinfo  {journal} {JCAP}\ }\textbf {\bibinfo {volume} {1311}},\
  \bibinfo {pages} {054} (\bibinfo {year} {2013})},\ \Eprint
  {http://arxiv.org/abs/1308.1105} {arXiv:1308.1105 [hep-ph]} \BibitemShut
  {NoStop}%
\bibitem [{\citenamefont {Bai}\ \emph {et~al.}(2016)\citenamefont {Bai},
  \citenamefont {Lu},\ and\ \citenamefont {Salvado}}]{Bai:2013nga}%
  \BibitemOpen
  \bibfield  {author} {\bibinfo {author} {\bibfnamefont {Y.}~\bibnamefont
  {Bai}}, \bibinfo {author} {\bibfnamefont {R.}~\bibnamefont {Lu}}, \ and\
  \bibinfo {author} {\bibfnamefont {J.}~\bibnamefont {Salvado}},\ }\href
  {\doibase 10.1007/JHEP01(2016)161} {\bibfield  {journal} {\bibinfo  {journal}
  {JHEP}\ }\textbf {\bibinfo {volume} {01}},\ \bibinfo {pages} {161} (\bibinfo
  {year} {2016})},\ \Eprint {http://arxiv.org/abs/1311.5864} {arXiv:1311.5864
  [hep-ph]} \BibitemShut {NoStop}%
\bibitem [{\citenamefont {Ema}\ \emph {et~al.}(2014{\natexlab{a}})\citenamefont
  {Ema}, \citenamefont {Jinno},\ and\ \citenamefont {Moroi}}]{Ema:2013nda}%
  \BibitemOpen
  \bibfield  {author} {\bibinfo {author} {\bibfnamefont {Y.}~\bibnamefont
  {Ema}}, \bibinfo {author} {\bibfnamefont {R.}~\bibnamefont {Jinno}}, \ and\
  \bibinfo {author} {\bibfnamefont {T.}~\bibnamefont {Moroi}},\ }\href
  {\doibase 10.1016/j.physletb.2014.04.021} {\bibfield  {journal} {\bibinfo
  {journal} {Phys. Lett.}\ }\textbf {\bibinfo {volume} {B733}},\ \bibinfo
  {pages} {120} (\bibinfo {year} {2014}{\natexlab{a}})},\ \Eprint
  {http://arxiv.org/abs/1312.3501} {arXiv:1312.3501 [hep-ph]} \BibitemShut
  {NoStop}%
\bibitem [{\citenamefont {Esmaili}\ \emph {et~al.}(2014)\citenamefont
  {Esmaili}, \citenamefont {Kang},\ and\ \citenamefont
  {Serpico}}]{Esmaili:2014rma}%
  \BibitemOpen
  \bibfield  {author} {\bibinfo {author} {\bibfnamefont {A.}~\bibnamefont
  {Esmaili}}, \bibinfo {author} {\bibfnamefont {S.~K.}\ \bibnamefont {Kang}}, \
  and\ \bibinfo {author} {\bibfnamefont {P.~D.}\ \bibnamefont {Serpico}},\
  }\href {\doibase 10.1088/1475-7516/2014/12/054} {\bibfield  {journal}
  {\bibinfo  {journal} {JCAP}\ }\textbf {\bibinfo {volume} {1412}},\ \bibinfo
  {pages} {054} (\bibinfo {year} {2014})},\ \Eprint
  {http://arxiv.org/abs/1410.5979} {arXiv:1410.5979 [hep-ph]} \BibitemShut
  {NoStop}%
\bibitem [{\citenamefont {Bhattacharya}\ \emph {et~al.}(2014)\citenamefont
  {Bhattacharya}, \citenamefont {Reno},\ and\ \citenamefont
  {Sarcevic}}]{Bhattacharya:2014vwa}%
  \BibitemOpen
  \bibfield  {author} {\bibinfo {author} {\bibfnamefont {A.}~\bibnamefont
  {Bhattacharya}}, \bibinfo {author} {\bibfnamefont {M.~H.}\ \bibnamefont
  {Reno}}, \ and\ \bibinfo {author} {\bibfnamefont {I.}~\bibnamefont
  {Sarcevic}},\ }\href {\doibase 10.1007/JHEP06(2014)110} {\bibfield  {journal}
  {\bibinfo  {journal} {JHEP}\ }\textbf {\bibinfo {volume} {06}},\ \bibinfo
  {pages} {110} (\bibinfo {year} {2014})},\ \Eprint
  {http://arxiv.org/abs/1403.1862} {arXiv:1403.1862 [hep-ph]} \BibitemShut
  {NoStop}%
\bibitem [{\citenamefont {Higaki}\ \emph {et~al.}(2014)\citenamefont {Higaki},
  \citenamefont {Kitano},\ and\ \citenamefont {Sato}}]{Higaki:2014dwa}%
  \BibitemOpen
  \bibfield  {author} {\bibinfo {author} {\bibfnamefont {T.}~\bibnamefont
  {Higaki}}, \bibinfo {author} {\bibfnamefont {R.}~\bibnamefont {Kitano}}, \
  and\ \bibinfo {author} {\bibfnamefont {R.}~\bibnamefont {Sato}},\ }\href
  {\doibase 10.1007/JHEP07(2014)044} {\bibfield  {journal} {\bibinfo  {journal}
  {JHEP}\ }\textbf {\bibinfo {volume} {07}},\ \bibinfo {pages} {044} (\bibinfo
  {year} {2014})},\ \Eprint {http://arxiv.org/abs/1405.0013} {arXiv:1405.0013
  [hep-ph]} \BibitemShut {NoStop}%
\bibitem [{\citenamefont {Rott}\ \emph {et~al.}(2015)\citenamefont {Rott},
  \citenamefont {Kohri},\ and\ \citenamefont {Park}}]{Rott:2014kfa}%
  \BibitemOpen
  \bibfield  {author} {\bibinfo {author} {\bibfnamefont {C.}~\bibnamefont
  {Rott}}, \bibinfo {author} {\bibfnamefont {K.}~\bibnamefont {Kohri}}, \ and\
  \bibinfo {author} {\bibfnamefont {S.~C.}\ \bibnamefont {Park}},\ }\href
  {\doibase 10.1103/PhysRevD.92.023529} {\bibfield  {journal} {\bibinfo
  {journal} {Phys. Rev.}\ }\textbf {\bibinfo {volume} {D92}},\ \bibinfo {pages}
  {023529} (\bibinfo {year} {2015})},\ \Eprint {http://arxiv.org/abs/1408.4575}
  {arXiv:1408.4575 [hep-ph]} \BibitemShut {NoStop}%
\bibitem [{\citenamefont {Ema}\ \emph {et~al.}(2014{\natexlab{b}})\citenamefont
  {Ema}, \citenamefont {Jinno},\ and\ \citenamefont {Moroi}}]{Ema:2014ufa}%
  \BibitemOpen
  \bibfield  {author} {\bibinfo {author} {\bibfnamefont {Y.}~\bibnamefont
  {Ema}}, \bibinfo {author} {\bibfnamefont {R.}~\bibnamefont {Jinno}}, \ and\
  \bibinfo {author} {\bibfnamefont {T.}~\bibnamefont {Moroi}},\ }\href
  {\doibase 10.1007/JHEP10(2014)150} {\bibfield  {journal} {\bibinfo  {journal}
  {JHEP}\ }\textbf {\bibinfo {volume} {10}},\ \bibinfo {pages} {150} (\bibinfo
  {year} {2014}{\natexlab{b}})},\ \Eprint {http://arxiv.org/abs/1408.1745}
  {arXiv:1408.1745 [hep-ph]} \BibitemShut {NoStop}%
\bibitem [{\citenamefont {Murase}\ \emph {et~al.}(2015)\citenamefont {Murase},
  \citenamefont {Laha}, \citenamefont {Ando},\ and\ \citenamefont
  {Ahlers}}]{Murase:2015gea}%
  \BibitemOpen
  \bibfield  {author} {\bibinfo {author} {\bibfnamefont {K.}~\bibnamefont
  {Murase}}, \bibinfo {author} {\bibfnamefont {R.}~\bibnamefont {Laha}},
  \bibinfo {author} {\bibfnamefont {S.}~\bibnamefont {Ando}}, \ and\ \bibinfo
  {author} {\bibfnamefont {M.}~\bibnamefont {Ahlers}},\ }\href {\doibase
  10.1103/PhysRevLett.115.071301} {\bibfield  {journal} {\bibinfo  {journal}
  {Phys. Rev. Lett.}\ }\textbf {\bibinfo {volume} {115}},\ \bibinfo {pages}
  {071301} (\bibinfo {year} {2015})},\ \Eprint
  {http://arxiv.org/abs/1503.04663} {arXiv:1503.04663 [hep-ph]} \BibitemShut
  {NoStop}%
\bibitem [{\citenamefont {Dudas}\ \emph {et~al.}(2015)\citenamefont {Dudas},
  \citenamefont {Mambrini},\ and\ \citenamefont {Olive}}]{Dudas:2014bca}%
  \BibitemOpen
  \bibfield  {author} {\bibinfo {author} {\bibfnamefont {E.}~\bibnamefont
  {Dudas}}, \bibinfo {author} {\bibfnamefont {Y.}~\bibnamefont {Mambrini}}, \
  and\ \bibinfo {author} {\bibfnamefont {K.~A.}\ \bibnamefont {Olive}},\ }\href
  {\doibase 10.1103/PhysRevD.91.075001} {\bibfield  {journal} {\bibinfo
  {journal} {Phys. Rev.}\ }\textbf {\bibinfo {volume} {D91}},\ \bibinfo {pages}
  {075001} (\bibinfo {year} {2015})},\ \Eprint {http://arxiv.org/abs/1412.3459}
  {arXiv:1412.3459 [hep-ph]} \BibitemShut {NoStop}%
\bibitem [{\citenamefont {Fong}\ \emph {et~al.}(2015)\citenamefont {Fong},
  \citenamefont {Minakata}, \citenamefont {Panes},\ and\ \citenamefont
  {Zukanovich~Funchal}}]{Fong:2014bsa}%
  \BibitemOpen
  \bibfield  {author} {\bibinfo {author} {\bibfnamefont {C.~S.}\ \bibnamefont
  {Fong}}, \bibinfo {author} {\bibfnamefont {H.}~\bibnamefont {Minakata}},
  \bibinfo {author} {\bibfnamefont {B.}~\bibnamefont {Panes}}, \ and\ \bibinfo
  {author} {\bibfnamefont {R.}~\bibnamefont {Zukanovich~Funchal}},\ }\href
  {\doibase 10.1007/JHEP02(2015)189} {\bibfield  {journal} {\bibinfo  {journal}
  {JHEP}\ }\textbf {\bibinfo {volume} {02}},\ \bibinfo {pages} {189} (\bibinfo
  {year} {2015})},\ \Eprint {http://arxiv.org/abs/1411.5318} {arXiv:1411.5318
  [hep-ph]} \BibitemShut {NoStop}%
\bibitem [{\citenamefont {El~Aisati}\ \emph {et~al.}(2015)\citenamefont
  {El~Aisati}, \citenamefont {Gustafsson},\ and\ \citenamefont
  {Hambye}}]{Aisati:2015vma}%
  \BibitemOpen
  \bibfield  {author} {\bibinfo {author} {\bibfnamefont {C.}~\bibnamefont
  {El~Aisati}}, \bibinfo {author} {\bibfnamefont {M.}~\bibnamefont
  {Gustafsson}}, \ and\ \bibinfo {author} {\bibfnamefont {T.}~\bibnamefont
  {Hambye}},\ }\href {\doibase 10.1103/PhysRevD.92.123515} {\bibfield
  {journal} {\bibinfo  {journal} {Phys. Rev.}\ }\textbf {\bibinfo {volume}
  {D92}},\ \bibinfo {pages} {123515} (\bibinfo {year} {2015})},\ \Eprint
  {http://arxiv.org/abs/1506.02657} {arXiv:1506.02657 [hep-ph]} \BibitemShut
  {NoStop}%
\bibitem [{\citenamefont {Ko}\ and\ \citenamefont {Tang}(2015)}]{Ko:2015nma}%
  \BibitemOpen
  \bibfield  {author} {\bibinfo {author} {\bibfnamefont {P.}~\bibnamefont
  {Ko}}\ and\ \bibinfo {author} {\bibfnamefont {Y.}~\bibnamefont {Tang}},\
  }\href {\doibase 10.1016/j.physletb.2015.10.021} {\bibfield  {journal}
  {\bibinfo  {journal} {Phys. Lett.}\ }\textbf {\bibinfo {volume} {B751}},\
  \bibinfo {pages} {81} (\bibinfo {year} {2015})},\ \Eprint
  {http://arxiv.org/abs/1508.02500} {arXiv:1508.02500 [hep-ph]} \BibitemShut
  {NoStop}%
\bibitem [{\citenamefont {Dev}\ \emph {et~al.}(2016{\natexlab{a}})\citenamefont
  {Dev}, \citenamefont {Kazanas}, \citenamefont {Mohapatra}, \citenamefont
  {Teplitz},\ and\ \citenamefont {Zhang}}]{Dev:2016qbd}%
  \BibitemOpen
  \bibfield  {author} {\bibinfo {author} {\bibfnamefont {P.~S.~B.}\
  \bibnamefont {Dev}}, \bibinfo {author} {\bibfnamefont {D.}~\bibnamefont
  {Kazanas}}, \bibinfo {author} {\bibfnamefont {R.~N.}\ \bibnamefont
  {Mohapatra}}, \bibinfo {author} {\bibfnamefont {V.~L.}\ \bibnamefont
  {Teplitz}}, \ and\ \bibinfo {author} {\bibfnamefont {Y.}~\bibnamefont
  {Zhang}},\ }\href {\doibase 10.1088/1475-7516/2016/08/034} {\bibfield
  {journal} {\bibinfo  {journal} {JCAP}\ }\textbf {\bibinfo {volume} {1608}},\
  \bibinfo {pages} {034} (\bibinfo {year} {2016}{\natexlab{a}})},\ \Eprint
  {http://arxiv.org/abs/1606.04517} {arXiv:1606.04517 [hep-ph]} \BibitemShut
  {NoStop}%
\bibitem [{\citenamefont {Re~Fiorentin}\ \emph {et~al.}(2016)\citenamefont
  {Re~Fiorentin}, \citenamefont {Niro},\ and\ \citenamefont
  {Fornengo}}]{Fiorentin:2016avj}%
  \BibitemOpen
  \bibfield  {author} {\bibinfo {author} {\bibfnamefont {M.}~\bibnamefont
  {Re~Fiorentin}}, \bibinfo {author} {\bibfnamefont {V.}~\bibnamefont {Niro}},
  \ and\ \bibinfo {author} {\bibfnamefont {N.}~\bibnamefont {Fornengo}},\
  }\href {\doibase 10.1007/JHEP11(2016)022} {\bibfield  {journal} {\bibinfo
  {journal} {JHEP}\ }\textbf {\bibinfo {volume} {11}},\ \bibinfo {pages} {022}
  (\bibinfo {year} {2016})},\ \Eprint {http://arxiv.org/abs/1606.04445}
  {arXiv:1606.04445 [hep-ph]} \BibitemShut {NoStop}%
\bibitem [{\citenamefont {Di~Bari}\ \emph {et~al.}(2016)\citenamefont
  {Di~Bari}, \citenamefont {Ludl},\ and\ \citenamefont
  {Palomares-Ruiz}}]{DiBari:2016guw}%
  \BibitemOpen
  \bibfield  {author} {\bibinfo {author} {\bibfnamefont {P.}~\bibnamefont
  {Di~Bari}}, \bibinfo {author} {\bibfnamefont {P.~O.}\ \bibnamefont {Ludl}}, \
  and\ \bibinfo {author} {\bibfnamefont {S.}~\bibnamefont {Palomares-Ruiz}},\
  }\href {\doibase 10.1088/1475-7516/2016/11/044} {\bibfield  {journal}
  {\bibinfo  {journal} {JCAP}\ }\textbf {\bibinfo {volume} {1611}},\ \bibinfo
  {pages} {044} (\bibinfo {year} {2016})},\ \Eprint
  {http://arxiv.org/abs/1606.06238} {arXiv:1606.06238 [hep-ph]} \BibitemShut
  {NoStop}%
\bibitem [{\citenamefont {Anchordoqui}\ \emph {et~al.}(2015)\citenamefont
  {Anchordoqui}, \citenamefont {Barger}, \citenamefont {Goldberg},
  \citenamefont {Huang}, \citenamefont {Marfatia}, \citenamefont {da~Silva},\
  and\ \citenamefont {Weiler}}]{Anchordoqui:2015lqa}%
  \BibitemOpen
  \bibfield  {author} {\bibinfo {author} {\bibfnamefont {L.~A.}\ \bibnamefont
  {Anchordoqui}}, \bibinfo {author} {\bibfnamefont {V.}~\bibnamefont {Barger}},
  \bibinfo {author} {\bibfnamefont {H.}~\bibnamefont {Goldberg}}, \bibinfo
  {author} {\bibfnamefont {X.}~\bibnamefont {Huang}}, \bibinfo {author}
  {\bibfnamefont {D.}~\bibnamefont {Marfatia}}, \bibinfo {author}
  {\bibfnamefont {L.~H.~M.}\ \bibnamefont {da~Silva}}, \ and\ \bibinfo {author}
  {\bibfnamefont {T.~J.}\ \bibnamefont {Weiler}},\ }\href {\doibase
  10.1103/PhysRevD.92.061301, 10.1103/PhysRevD.94.069901} {\bibfield  {journal}
  {\bibinfo  {journal} {Phys. Rev.}\ }\textbf {\bibinfo {volume} {D92}},\
  \bibinfo {pages} {061301} (\bibinfo {year} {2015})},\ \bibinfo {note}
  {[Erratum: Phys. Rev.D94,no.6,069901(2016)]},\ \Eprint
  {http://arxiv.org/abs/1506.08788} {arXiv:1506.08788 [hep-ph]} \BibitemShut
  {NoStop}%
\bibitem [{\citenamefont {Dev}\ \emph {et~al.}(2016{\natexlab{b}})\citenamefont
  {Dev}, \citenamefont {Ghosh},\ and\ \citenamefont
  {Rodejohann}}]{Dev:2016uxj}%
  \BibitemOpen
  \bibfield  {author} {\bibinfo {author} {\bibfnamefont {P.~S.~B.}\
  \bibnamefont {Dev}}, \bibinfo {author} {\bibfnamefont {D.~K.}\ \bibnamefont
  {Ghosh}}, \ and\ \bibinfo {author} {\bibfnamefont {W.}~\bibnamefont
  {Rodejohann}},\ }\href {\doibase 10.1016/j.physletb.2016.08.066} {\bibfield
  {journal} {\bibinfo  {journal} {Phys. Lett.}\ }\textbf {\bibinfo {volume}
  {B762}},\ \bibinfo {pages} {116} (\bibinfo {year} {2016}{\natexlab{b}})},\
  \Eprint {http://arxiv.org/abs/1605.09743} {arXiv:1605.09743 [hep-ph]}
  \BibitemShut {NoStop}%
\bibitem [{\citenamefont {Chianese}\ and\ \citenamefont
  {Merle}(2017)}]{Chianese:2016smc}%
  \BibitemOpen
  \bibfield  {author} {\bibinfo {author} {\bibfnamefont {M.}~\bibnamefont
  {Chianese}}\ and\ \bibinfo {author} {\bibfnamefont {A.}~\bibnamefont
  {Merle}},\ }\href {\doibase 10.1088/1475-7516/2017/04/017} {\bibfield
  {journal} {\bibinfo  {journal} {JCAP}\ }\textbf {\bibinfo {volume} {1704}},\
  \bibinfo {pages} {017} (\bibinfo {year} {2017})},\ \Eprint
  {http://arxiv.org/abs/1607.05283} {arXiv:1607.05283 [hep-ph]} \BibitemShut
  {NoStop}%
\bibitem [{\citenamefont {Borah}\ \emph {et~al.}(2017)\citenamefont {Borah},
  \citenamefont {Dasgupta}, \citenamefont {Dey}, \citenamefont {Patra},\ and\
  \citenamefont {Tomar}}]{Borah:2017xgm}%
  \BibitemOpen
  \bibfield  {author} {\bibinfo {author} {\bibfnamefont {D.}~\bibnamefont
  {Borah}}, \bibinfo {author} {\bibfnamefont {A.}~\bibnamefont {Dasgupta}},
  \bibinfo {author} {\bibfnamefont {U.~K.}\ \bibnamefont {Dey}}, \bibinfo
  {author} {\bibfnamefont {S.}~\bibnamefont {Patra}}, \ and\ \bibinfo {author}
  {\bibfnamefont {G.}~\bibnamefont {Tomar}},\ }\href {\doibase
  10.1007/JHEP09(2017)005} {\bibfield  {journal} {\bibinfo  {journal} {JHEP}\
  }\textbf {\bibinfo {volume} {09}},\ \bibinfo {pages} {005} (\bibinfo {year}
  {2017})},\ \Eprint {http://arxiv.org/abs/1704.04138} {arXiv:1704.04138
  [hep-ph]} \BibitemShut {NoStop}%
\bibitem [{\citenamefont {Boucenna}\ \emph {et~al.}(2015)\citenamefont
  {Boucenna}, \citenamefont {Chianese}, \citenamefont {Mangano}, \citenamefont
  {Miele}, \citenamefont {Morisi}, \citenamefont {Pisanti},\ and\ \citenamefont
  {Vitagliano}}]{Boucenna:2015tra}%
  \BibitemOpen
  \bibfield  {author} {\bibinfo {author} {\bibfnamefont {S.~M.}\ \bibnamefont
  {Boucenna}}, \bibinfo {author} {\bibfnamefont {M.}~\bibnamefont {Chianese}},
  \bibinfo {author} {\bibfnamefont {G.}~\bibnamefont {Mangano}}, \bibinfo
  {author} {\bibfnamefont {G.}~\bibnamefont {Miele}}, \bibinfo {author}
  {\bibfnamefont {S.}~\bibnamefont {Morisi}}, \bibinfo {author} {\bibfnamefont
  {O.}~\bibnamefont {Pisanti}}, \ and\ \bibinfo {author} {\bibfnamefont
  {E.}~\bibnamefont {Vitagliano}},\ }\href {\doibase
  10.1088/1475-7516/2015/12/055} {\bibfield  {journal} {\bibinfo  {journal}
  {JCAP}\ }\textbf {\bibinfo {volume} {1512}},\ \bibinfo {pages} {055}
  (\bibinfo {year} {2015})},\ \Eprint {http://arxiv.org/abs/1507.01000}
  {arXiv:1507.01000 [hep-ph]} \BibitemShut {NoStop}%
\bibitem [{\citenamefont {Hiroshima}\ \emph {et~al.}(2018)\citenamefont
  {Hiroshima}, \citenamefont {Kitano}, \citenamefont {Kohri},\ and\
  \citenamefont {Murase}}]{Hiroshima:2017hmy}%
  \BibitemOpen
  \bibfield  {author} {\bibinfo {author} {\bibfnamefont {N.}~\bibnamefont
  {Hiroshima}}, \bibinfo {author} {\bibfnamefont {R.}~\bibnamefont {Kitano}},
  \bibinfo {author} {\bibfnamefont {K.}~\bibnamefont {Kohri}}, \ and\ \bibinfo
  {author} {\bibfnamefont {K.}~\bibnamefont {Murase}},\ }\href {\doibase
  10.1103/PhysRevD.97.023006} {\bibfield  {journal} {\bibinfo  {journal} {Phys.
  Rev.}\ }\textbf {\bibinfo {volume} {D97}},\ \bibinfo {pages} {023006}
  (\bibinfo {year} {2018})},\ \Eprint {http://arxiv.org/abs/1705.04419}
  {arXiv:1705.04419 [hep-ph]} \BibitemShut {NoStop}%
\bibitem [{\citenamefont {Bhattacharya}\ \emph {et~al.}(2017)\citenamefont
  {Bhattacharya}, \citenamefont {Esmaili}, \citenamefont {Palomares-Ruiz},\
  and\ \citenamefont {Sarcevic}}]{Bhattacharya:2017jaw}%
  \BibitemOpen
  \bibfield  {author} {\bibinfo {author} {\bibfnamefont {A.}~\bibnamefont
  {Bhattacharya}}, \bibinfo {author} {\bibfnamefont {A.}~\bibnamefont
  {Esmaili}}, \bibinfo {author} {\bibfnamefont {S.}~\bibnamefont
  {Palomares-Ruiz}}, \ and\ \bibinfo {author} {\bibfnamefont {I.}~\bibnamefont
  {Sarcevic}},\ }\href {\doibase 10.1088/1475-7516/2017/07/027} {\bibfield
  {journal} {\bibinfo  {journal} {JCAP}\ }\textbf {\bibinfo {volume} {1707}},\
  \bibinfo {pages} {027} (\bibinfo {year} {2017})},\ \Eprint
  {http://arxiv.org/abs/1706.05746} {arXiv:1706.05746 [hep-ph]} \BibitemShut
  {NoStop}%
\bibitem [{\citenamefont {Bhattacharya}\ \emph {et~al.}(2019)\citenamefont
  {Bhattacharya}, \citenamefont {Esmaili}, \citenamefont {Palomares-Ruiz},\
  and\ \citenamefont {Sarcevic}}]{Bhattacharya:2019ucd}%
  \BibitemOpen
  \bibfield  {author} {\bibinfo {author} {\bibfnamefont {A.}~\bibnamefont
  {Bhattacharya}}, \bibinfo {author} {\bibfnamefont {A.}~\bibnamefont
  {Esmaili}}, \bibinfo {author} {\bibfnamefont {S.}~\bibnamefont
  {Palomares-Ruiz}}, \ and\ \bibinfo {author} {\bibfnamefont {I.}~\bibnamefont
  {Sarcevic}},\ }\href {\doibase 10.1088/1475-7516/2019/05/051} {\bibfield
  {journal} {\bibinfo  {journal} {JCAP}\ }\textbf {\bibinfo {volume} {1905}},\
  \bibinfo {pages} {051} (\bibinfo {year} {2019})},\ \Eprint
  {http://arxiv.org/abs/1903.12623} {arXiv:1903.12623 [hep-ph]} \BibitemShut
  {NoStop}%
\bibitem [{\citenamefont {Kopp}\ \emph {et~al.}(2015)\citenamefont {Kopp},
  \citenamefont {Liu},\ and\ \citenamefont {Wang}}]{Kopp:2015bfa}%
  \BibitemOpen
  \bibfield  {author} {\bibinfo {author} {\bibfnamefont {J.}~\bibnamefont
  {Kopp}}, \bibinfo {author} {\bibfnamefont {J.}~\bibnamefont {Liu}}, \ and\
  \bibinfo {author} {\bibfnamefont {X.-P.}\ \bibnamefont {Wang}},\ }\href
  {\doibase 10.1007/JHEP04(2015)105} {\bibfield  {journal} {\bibinfo  {journal}
  {JHEP}\ }\textbf {\bibinfo {volume} {04}},\ \bibinfo {pages} {105} (\bibinfo
  {year} {2015})},\ \Eprint {http://arxiv.org/abs/1503.02669} {arXiv:1503.02669
  [hep-ph]} \BibitemShut {NoStop}%
\bibitem [{\citenamefont {Zavala}(2014)}]{Zavala:2014dla}%
  \BibitemOpen
  \bibfield  {author} {\bibinfo {author} {\bibfnamefont {J.}~\bibnamefont
  {Zavala}},\ }\href {\doibase 10.1103/PhysRevD.89.123516} {\bibfield
  {journal} {\bibinfo  {journal} {Phys. Rev.}\ }\textbf {\bibinfo {volume}
  {D89}},\ \bibinfo {pages} {123516} (\bibinfo {year} {2014})},\ \Eprint
  {http://arxiv.org/abs/1404.2932} {arXiv:1404.2932 [astro-ph.HE]} \BibitemShut
  {NoStop}%
\bibitem [{\citenamefont {El~Aisati}\ \emph {et~al.}(2017)\citenamefont
  {El~Aisati}, \citenamefont {Garcia-Cely}, \citenamefont {Hambye},\ and\
  \citenamefont {Vanderheyden}}]{ElAisati:2017ppn}%
  \BibitemOpen
  \bibfield  {author} {\bibinfo {author} {\bibfnamefont {C.}~\bibnamefont
  {El~Aisati}}, \bibinfo {author} {\bibfnamefont {C.}~\bibnamefont
  {Garcia-Cely}}, \bibinfo {author} {\bibfnamefont {T.}~\bibnamefont {Hambye}},
  \ and\ \bibinfo {author} {\bibfnamefont {L.}~\bibnamefont {Vanderheyden}},\
  }\href {\doibase 10.1088/1475-7516/2017/10/021} {\bibfield  {journal}
  {\bibinfo  {journal} {JCAP}\ }\textbf {\bibinfo {volume} {1710}},\ \bibinfo
  {pages} {021} (\bibinfo {year} {2017})},\ \Eprint
  {http://arxiv.org/abs/1706.06600} {arXiv:1706.06600 [hep-ph]} \BibitemShut
  {NoStop}%
\bibitem [{\citenamefont {Kachelriess}\ \emph {et~al.}(2018)\citenamefont
  {Kachelriess}, \citenamefont {Kalashev},\ and\ \citenamefont
  {Kuznetsov}}]{Kachelriess:2018rty}%
  \BibitemOpen
  \bibfield  {author} {\bibinfo {author} {\bibfnamefont {M.}~\bibnamefont
  {Kachelriess}}, \bibinfo {author} {\bibfnamefont {O.~E.}\ \bibnamefont
  {Kalashev}}, \ and\ \bibinfo {author} {\bibfnamefont {M.~{\relax Yu}.}\
  \bibnamefont {Kuznetsov}},\ }\href {\doibase 10.1103/PhysRevD.98.083016}
  {\bibfield  {journal} {\bibinfo  {journal} {Phys. Rev.}\ }\textbf {\bibinfo
  {volume} {D98}},\ \bibinfo {pages} {083016} (\bibinfo {year} {2018})},\
  \Eprint {http://arxiv.org/abs/1805.04500} {arXiv:1805.04500 [astro-ph.HE]}
  \BibitemShut {NoStop}%
\bibitem [{\citenamefont {Chianese}\ \emph {et~al.}(2018)\citenamefont
  {Chianese}, \citenamefont {Miele}, \citenamefont {Morisi},\ and\
  \citenamefont {Peinado}}]{Chianese:2018ijk}%
  \BibitemOpen
  \bibfield  {author} {\bibinfo {author} {\bibfnamefont {M.}~\bibnamefont
  {Chianese}}, \bibinfo {author} {\bibfnamefont {G.}~\bibnamefont {Miele}},
  \bibinfo {author} {\bibfnamefont {S.}~\bibnamefont {Morisi}}, \ and\ \bibinfo
  {author} {\bibfnamefont {E.}~\bibnamefont {Peinado}},\ }\href {\doibase
  10.1088/1475-7516/2018/12/016} {\bibfield  {journal} {\bibinfo  {journal}
  {JCAP}\ }\textbf {\bibinfo {volume} {1812}},\ \bibinfo {pages} {016}
  (\bibinfo {year} {2018})},\ \Eprint {http://arxiv.org/abs/1808.02486}
  {arXiv:1808.02486 [hep-ph]} \BibitemShut {NoStop}%
\bibitem [{\citenamefont {Aartsen}\ \emph
  {et~al.}(2018{\natexlab{c}})\citenamefont {Aartsen} \emph
  {et~al.}}]{Aartsen:2018mxl}%
  \BibitemOpen
  \bibfield  {author} {\bibinfo {author} {\bibfnamefont {M.~G.}\ \bibnamefont
  {Aartsen}} \emph {et~al.} (\bibinfo {collaboration} {IceCube}),\ }\href
  {\doibase 10.1140/epjc/s10052-018-6273-3} {\bibfield  {journal} {\bibinfo
  {journal} {Eur. Phys. J.}\ }\textbf {\bibinfo {volume} {C78}},\ \bibinfo
  {pages} {831} (\bibinfo {year} {2018}{\natexlab{c}})},\ \Eprint
  {http://arxiv.org/abs/1804.03848} {arXiv:1804.03848 [astro-ph.HE]}
  \BibitemShut {NoStop}%
\bibitem [{\citenamefont {Abeysekara}\ \emph {et~al.}(2018)\citenamefont
  {Abeysekara} \emph {et~al.}}]{Abeysekara:2017jxs}%
  \BibitemOpen
  \bibfield  {author} {\bibinfo {author} {\bibfnamefont {A.~U.}\ \bibnamefont
  {Abeysekara}} \emph {et~al.} (\bibinfo {collaboration} {HAWC}),\ }\href
  {\doibase 10.1088/1475-7516/2018/02/049} {\bibfield  {journal} {\bibinfo
  {journal} {JCAP}\ }\textbf {\bibinfo {volume} {1802}},\ \bibinfo {pages}
  {049} (\bibinfo {year} {2018})},\ \Eprint {http://arxiv.org/abs/1710.10288}
  {arXiv:1710.10288 [astro-ph.HE]} \BibitemShut {NoStop}%
\bibitem [{\citenamefont {Albert}\ \emph {et~al.}(2017)\citenamefont {Albert}
  \emph {et~al.}}]{Albert:2016emp}%
  \BibitemOpen
  \bibfield  {author} {\bibinfo {author} {\bibfnamefont {A.}~\bibnamefont
  {Albert}} \emph {et~al.},\ }\href {\doibase 10.1016/j.physletb.2019.05.022,
  10.1016/j.physletb.2017.03.063} {\bibfield  {journal} {\bibinfo  {journal}
  {Phys. Lett.}\ }\textbf {\bibinfo {volume} {B769}},\ \bibinfo {pages} {249}
  (\bibinfo {year} {2017})},\ \bibinfo {note} {[Erratum: Phys. Lett.B(2019)]},\
  \Eprint {http://arxiv.org/abs/1612.04595} {arXiv:1612.04595 [astro-ph.HE]}
  \BibitemShut {NoStop}%
\bibitem [{\citenamefont {Gozzini}\ \emph {et~al.}(2019)\citenamefont
  {Gozzini}, \citenamefont {Iovine}, \citenamefont {Sánchez}, \citenamefont
  {Baur},\ and\ \citenamefont {de~Dios Zornoza~Gómez}}]{Gozzini:2019esk}%
  \BibitemOpen
  \bibfield  {author} {\bibinfo {author} {\bibfnamefont {S.~R.}\ \bibnamefont
  {Gozzini}}, \bibinfo {author} {\bibfnamefont {N.}~\bibnamefont {Iovine}},
  \bibinfo {author} {\bibfnamefont {J.~A.~A.}\ \bibnamefont {Sánchez}},
  \bibinfo {author} {\bibfnamefont {S.}~\bibnamefont {Baur}}, \ and\ \bibinfo
  {author} {\bibfnamefont {J.}~\bibnamefont {de~Dios Zornoza~Gómez}} (\bibinfo
  {collaboration} {ANTARES, IceCube}),\ }\bibfield  {booktitle} {\emph
  {\bibinfo {booktitle} {{Proceedings, 8th Very Large Volume Neutrino Telescope
  Workshop (VLVnT-2018): Dubna, Russia, October 2-4, 2018}}},\ }\href {\doibase
  10.1051/epjconf/201920704007} {\bibfield  {journal} {\bibinfo  {journal} {EPJ
  Web Conf.}\ }\textbf {\bibinfo {volume} {207}},\ \bibinfo {pages} {04007}
  (\bibinfo {year} {2019})}\BibitemShut {NoStop}%
\bibitem [{\citenamefont {Cohen}\ \emph {et~al.}(2017)\citenamefont {Cohen},
  \citenamefont {Murase}, \citenamefont {Rodd}, \citenamefont {Safdi},\ and\
  \citenamefont {Soreq}}]{Cohen:2016uyg}%
  \BibitemOpen
  \bibfield  {author} {\bibinfo {author} {\bibfnamefont {T.}~\bibnamefont
  {Cohen}}, \bibinfo {author} {\bibfnamefont {K.}~\bibnamefont {Murase}},
  \bibinfo {author} {\bibfnamefont {N.~L.}\ \bibnamefont {Rodd}}, \bibinfo
  {author} {\bibfnamefont {B.~R.}\ \bibnamefont {Safdi}}, \ and\ \bibinfo
  {author} {\bibfnamefont {Y.}~\bibnamefont {Soreq}},\ }\href {\doibase
  10.1103/PhysRevLett.119.021102} {\bibfield  {journal} {\bibinfo  {journal}
  {Phys. Rev. Lett.}\ }\textbf {\bibinfo {volume} {119}},\ \bibinfo {pages}
  {021102} (\bibinfo {year} {2017})},\ \Eprint
  {http://arxiv.org/abs/1612.05638} {arXiv:1612.05638 [hep-ph]} \BibitemShut
  {NoStop}%
\bibitem [{\citenamefont {Blanco}\ and\ \citenamefont
  {Hooper}(2019)}]{Blanco:2018esa}%
  \BibitemOpen
  \bibfield  {author} {\bibinfo {author} {\bibfnamefont {C.}~\bibnamefont
  {Blanco}}\ and\ \bibinfo {author} {\bibfnamefont {D.}~\bibnamefont
  {Hooper}},\ }\href {\doibase 10.1088/1475-7516/2019/03/019} {\bibfield
  {journal} {\bibinfo  {journal} {JCAP}\ }\textbf {\bibinfo {volume} {1903}},\
  \bibinfo {pages} {019} (\bibinfo {year} {2019})},\ \Eprint
  {http://arxiv.org/abs/1811.05988} {arXiv:1811.05988 [astro-ph.HE]}
  \BibitemShut {NoStop}%
\bibitem [{\citenamefont {Atwood}\ \emph {et~al.}(2009)\citenamefont {Atwood}
  \emph {et~al.}}]{Atwood:2009ez}%
  \BibitemOpen
  \bibfield  {author} {\bibinfo {author} {\bibfnamefont {W.~B.}\ \bibnamefont
  {Atwood}} \emph {et~al.} (\bibinfo {collaboration} {Fermi-LAT}),\ }\href
  {\doibase 10.1088/0004-637X/697/2/1071} {\bibfield  {journal} {\bibinfo
  {journal} {Astrophys. J.}\ }\textbf {\bibinfo {volume} {697}},\ \bibinfo
  {pages} {1071} (\bibinfo {year} {2009})},\ \Eprint
  {http://arxiv.org/abs/0902.1089} {arXiv:0902.1089 [astro-ph.IM]} \BibitemShut
  {NoStop}%
\bibitem [{\citenamefont {Kalashev}\ and\ \citenamefont
  {Kuznetsov}(2016)}]{Kalashev:2016cre}%
  \BibitemOpen
  \bibfield  {author} {\bibinfo {author} {\bibfnamefont {O.~K.}\ \bibnamefont
  {Kalashev}}\ and\ \bibinfo {author} {\bibfnamefont {M.~{\relax Yu}.}\
  \bibnamefont {Kuznetsov}},\ }\href {\doibase 10.1103/PhysRevD.94.063535}
  {\bibfield  {journal} {\bibinfo  {journal} {Phys. Rev.}\ }\textbf {\bibinfo
  {volume} {D94}},\ \bibinfo {pages} {063535} (\bibinfo {year} {2016})},\
  \Eprint {http://arxiv.org/abs/1606.07354} {arXiv:1606.07354 [astro-ph.HE]}
  \BibitemShut {NoStop}%
\bibitem [{Aab(2015)}]{Aab:2015bza}%
  \BibitemOpen
  \href
  {http://lss.fnal.gov/archive/2015/conf/fermilab-conf-15-396-ad-ae-cd-td.pdf}
  {\emph {\bibinfo {title} {{The Pierre Auger Observatory: Contributions to the
  34th International Cosmic Ray Conference (ICRC 2015)}}}}\ (\bibinfo {year}
  {2015})\ \Eprint {http://arxiv.org/abs/1509.03732} {arXiv:1509.03732
  [astro-ph.HE]} \BibitemShut {NoStop}%
\bibitem [{\citenamefont {Chantell}\ \emph {et~al.}(1997)\citenamefont
  {Chantell} \emph {et~al.}}]{Chantell:1997gs}%
  \BibitemOpen
  \bibfield  {author} {\bibinfo {author} {\bibfnamefont {M.~C.}\ \bibnamefont
  {Chantell}} \emph {et~al.} (\bibinfo {collaboration} {CASA-MIA}),\ }\href
  {\doibase 10.1103/PhysRevLett.79.1805} {\bibfield  {journal} {\bibinfo
  {journal} {Phys. Rev. Lett.}\ }\textbf {\bibinfo {volume} {79}},\ \bibinfo
  {pages} {1805} (\bibinfo {year} {1997})},\ \Eprint
  {http://arxiv.org/abs/astro-ph/9705246} {arXiv:astro-ph/9705246 [astro-ph]}
  \BibitemShut {NoStop}%
\bibitem [{\citenamefont {Kang}\ \emph {et~al.}(2015)\citenamefont {Kang} \emph
  {et~al.}}]{Kang:2015gpa}%
  \BibitemOpen
  \bibfield  {author} {\bibinfo {author} {\bibfnamefont {D.}~\bibnamefont
  {Kang}} \emph {et~al.},\ }\bibfield  {booktitle} {\emph {\bibinfo {booktitle}
  {{Proceedings, 24th European Cosmic Ray Symposium (ECRS 2014): Kiel, Germany,
  September 1-5, 2014}}},\ }\href {\doibase 10.1088/1742-6596/632/1/012013}
  {\bibfield  {journal} {\bibinfo  {journal} {J. Phys. Conf. Ser.}\ }\textbf
  {\bibinfo {volume} {632}},\ \bibinfo {pages} {012013} (\bibinfo {year}
  {2015})}\BibitemShut {NoStop}%
\bibitem [{\citenamefont {Cao}(2010)}]{Cao:2010zz}%
  \BibitemOpen
  \bibfield  {author} {\bibinfo {author} {\bibfnamefont {Z.}~\bibnamefont
  {Cao}} (\bibinfo {collaboration} {LHAASO}),\ }\href {\doibase
  10.1088/1674-1137/34/2/018} {\bibfield  {journal} {\bibinfo  {journal} {Chin.
  Phys.}\ }\textbf {\bibinfo {volume} {C34}},\ \bibinfo {pages} {249} (\bibinfo
  {year} {2010})}\BibitemShut {NoStop}%
\bibitem [{\citenamefont {Acciari}\ \emph {et~al.}(2018)\citenamefont {Acciari}
  \emph {et~al.}}]{Acciari:2018sjn}%
  \BibitemOpen
  \bibfield  {author} {\bibinfo {author} {\bibfnamefont {V.~A.}\ \bibnamefont
  {Acciari}} \emph {et~al.} (\bibinfo {collaboration} {MAGIC}),\ }\href
  {\doibase 10.1016/j.dark.2018.08.002} {\bibfield  {journal} {\bibinfo
  {journal} {Phys. Dark Univ.}\ }\textbf {\bibinfo {volume} {22}},\ \bibinfo
  {pages} {38} (\bibinfo {year} {2018})},\ \Eprint
  {http://arxiv.org/abs/1806.11063} {arXiv:1806.11063 [astro-ph.HE]}
  \BibitemShut {NoStop}%
\bibitem [{\citenamefont {Rinchiuso}(2019)}]{Rinchiuso:2019rrh}%
  \BibitemOpen
  \bibfield  {author} {\bibinfo {author} {\bibfnamefont {L.}~\bibnamefont
  {Rinchiuso}} (\bibinfo {collaboration} {H.E.S.S.}),\ }\bibfield  {booktitle}
  {\emph {\bibinfo {booktitle} {{Proceedings, 7th Roma International Conference
  on Astroparticle Physic (RICAP18): Rome, Italy, September 4-7, 2018}}},\
  }\href {\doibase 10.1051/epjconf/201920901023} {\bibfield  {journal}
  {\bibinfo  {journal} {EPJ Web Conf.}\ }\textbf {\bibinfo {volume} {209}},\
  \bibinfo {pages} {01023} (\bibinfo {year} {2019})},\ \Eprint
  {http://arxiv.org/abs/1901.05299} {arXiv:1901.05299 [astro-ph.HE]}
  \BibitemShut {NoStop}%
\bibitem [{\citenamefont {Zitzer}(2018)}]{Zitzer:2017xlo}%
  \BibitemOpen
  \bibfield  {author} {\bibinfo {author} {\bibfnamefont {B.}~\bibnamefont
  {Zitzer}} (\bibinfo {collaboration} {VERITAS}),\ }\bibfield  {booktitle}
  {\emph {\bibinfo {booktitle} {{The Fluorescence detector Array of
  Single-pixel Telescopes: Contributions to the 35th International Cosmic Ray
  Conference (ICRC 2017)}}},\ }\href {\doibase 10.22323/1.301.0904} {\bibfield
  {journal} {\bibinfo  {journal} {PoS}\ }\textbf {\bibinfo {volume}
  {ICRC2017}},\ \bibinfo {pages} {904} (\bibinfo {year} {2018})},\ \bibinfo
  {note} {[35,904(2017)]},\ \Eprint {http://arxiv.org/abs/1708.07447}
  {arXiv:1708.07447 [astro-ph.HE]} \BibitemShut {NoStop}%
\bibitem{Dzhappuev:2018bnl} 
  D.~D.~Dzhappuev {\it et al.},
  EPJ Web Conf.\  {\bf 207}, 03004 (2019)
  doi:10.1051/epjconf/201920703004
  [arXiv:1812.02663 [astro-ph.HE]].
  
\bibitem [{\citenamefont {Ong}(2019)}]{Ong:2019zyq}%
  \BibitemOpen
  \bibfield  {author} {\bibinfo {author} {\bibfnamefont {R.~A.}\ \bibnamefont
  {Ong}} (\bibinfo {collaboration} {CTA Consortium}),\ }\bibfield  {booktitle}
  {\emph {\bibinfo {booktitle} {{Proceedings, 7th Roma International Conference
  on Astroparticle Physic (RICAP18): Rome, Italy, September 4-7, 2018}}},\
  }\href {\doibase 10.1051/epjconf/201920901038} {\bibfield  {journal}
  {\bibinfo  {journal} {EPJ Web Conf.}\ }\textbf {\bibinfo {volume} {209}},\
  \bibinfo {pages} {01038} (\bibinfo {year} {2019})},\ \Eprint
  {http://arxiv.org/abs/1904.12196} {arXiv:1904.12196 [astro-ph.HE]}
  \BibitemShut {NoStop}%
\bibitem [{\citenamefont {Ibarra}\ \emph {et~al.}(2015)\citenamefont {Ibarra},
  \citenamefont {Lamperstorfer}, \citenamefont {López-Gehler}, \citenamefont
  {Pato},\ and\ \citenamefont {Bertone}}]{Ibarra:2015tya}%
  \BibitemOpen
  \bibfield  {author} {\bibinfo {author} {\bibfnamefont {A.}~\bibnamefont
  {Ibarra}}, \bibinfo {author} {\bibfnamefont {A.~S.}\ \bibnamefont
  {Lamperstorfer}}, \bibinfo {author} {\bibfnamefont {S.}~\bibnamefont
  {López-Gehler}}, \bibinfo {author} {\bibfnamefont {M.}~\bibnamefont {Pato}},
  \ and\ \bibinfo {author} {\bibfnamefont {G.}~\bibnamefont {Bertone}},\ }\href
  {\doibase 10.1088/1475-7516/2016/06/E02, 10.1088/1475-7516/2015/09/048}
  {\bibfield  {journal} {\bibinfo  {journal} {JCAP}\ }\textbf {\bibinfo
  {volume} {1509}},\ \bibinfo {pages} {048} (\bibinfo {year} {2015})},\
  \bibinfo {note} {[Erratum: JCAP1606,no.06,E02(2016)]},\ \Eprint
  {http://arxiv.org/abs/1503.06797} {arXiv:1503.06797 [astro-ph.HE]}
  \BibitemShut {NoStop}%
\bibitem [{\citenamefont {Silverwood}\ \emph {et~al.}(2015)\citenamefont
  {Silverwood}, \citenamefont {Weniger}, \citenamefont {Scott},\ and\
  \citenamefont {Bertone}}]{Silverwood:2014yza}%
  \BibitemOpen
  \bibfield  {author} {\bibinfo {author} {\bibfnamefont {H.}~\bibnamefont
  {Silverwood}}, \bibinfo {author} {\bibfnamefont {C.}~\bibnamefont {Weniger}},
  \bibinfo {author} {\bibfnamefont {P.}~\bibnamefont {Scott}}, \ and\ \bibinfo
  {author} {\bibfnamefont {G.}~\bibnamefont {Bertone}},\ }\href {\doibase
  10.1088/1475-7516/2015/03/055} {\bibfield  {journal} {\bibinfo  {journal}
  {JCAP}\ }\textbf {\bibinfo {volume} {1503}},\ \bibinfo {pages} {055}
  (\bibinfo {year} {2015})},\ \Eprint {http://arxiv.org/abs/1408.4131}
  {arXiv:1408.4131 [astro-ph.HE]} \BibitemShut {NoStop}%
\bibitem [{\citenamefont {Hryczuk}\ \emph {et~al.}(2019)\citenamefont
  {Hryczuk}, \citenamefont {Jodlowski}, \citenamefont {Moulin}, \citenamefont
  {Rinchiuso}, \citenamefont {Roszkowski}, \citenamefont {Sessolo},\ and\
  \citenamefont {Trojanowski}}]{Hryczuk:2019nql}%
  \BibitemOpen
  \bibfield  {author} {\bibinfo {author} {\bibfnamefont {A.}~\bibnamefont
  {Hryczuk}}, \bibinfo {author} {\bibfnamefont {K.}~\bibnamefont {Jodlowski}},
  \bibinfo {author} {\bibfnamefont {E.}~\bibnamefont {Moulin}}, \bibinfo
  {author} {\bibfnamefont {L.}~\bibnamefont {Rinchiuso}}, \bibinfo {author}
  {\bibfnamefont {L.}~\bibnamefont {Roszkowski}}, \bibinfo {author}
  {\bibfnamefont {E.~M.}\ \bibnamefont {Sessolo}}, \ and\ \bibinfo {author}
  {\bibfnamefont {S.}~\bibnamefont {Trojanowski}},\ }\href@noop {} {\
  (\bibinfo {year} {2019})},\ \Eprint {http://arxiv.org/abs/1905.00315}
  {arXiv:1905.00315 [hep-ph]} \BibitemShut {NoStop}%
\bibitem [{\citenamefont {Pierre}\ \emph {et~al.}(2014)\citenamefont {Pierre},
  \citenamefont {Siegal-Gaskins},\ and\ \citenamefont
  {Scott}}]{Pierre:2014tra}%
  \BibitemOpen
  \bibfield  {author} {\bibinfo {author} {\bibfnamefont {M.}~\bibnamefont
  {Pierre}}, \bibinfo {author} {\bibfnamefont {J.~M.}\ \bibnamefont
  {Siegal-Gaskins}}, \ and\ \bibinfo {author} {\bibfnamefont {P.}~\bibnamefont
  {Scott}},\ }\href {\doibase 10.1088/1475-7516/2014/10/E01,
  10.1088/1475-7516/2014/06/024} {\bibfield  {journal} {\bibinfo  {journal}
  {JCAP}\ }\textbf {\bibinfo {volume} {1406}},\ \bibinfo {pages} {024}
  (\bibinfo {year} {2014})},\ \bibinfo {note} {[Erratum: JCAP1410,E01(2014)]},\
  \Eprint {http://arxiv.org/abs/1401.7330} {arXiv:1401.7330 [astro-ph.HE]}
  \BibitemShut {NoStop}%
 
\bibitem{Neronov:2018ibl} 
  A.~Neronov, M.~Kachelrieß and D.~V.~Semikoz,
  Phys.\ Rev.\ D {\bf 98}, no. 2, 023004 (2018)
  doi:10.1103/PhysRevD.98.023004
  [arXiv:1802.09983 [astro-ph.HE]].
\bibitem{Neronov:2019ncc} 
  A.~Neronov and D.~Semikoz,
  arXiv:1907.06061 [astro-ph.HE].
\bibitem{Ackermann:2014usa} 
  M.~Ackermann {\it et al.} [Fermi-LAT Collaboration],
  Astrophys.\ J.\  {\bf 799}, 86 (2015)
  doi:10.1088/0004-637X/799/1/86
  [arXiv:1410.3696 [astro-ph.HE]].
 
\bibitem [{\citenamefont {Bertone}\ \emph {et~al.}(2005)\citenamefont
  {Bertone}, \citenamefont {Hooper},\ and\ \citenamefont
  {Silk}}]{Bertone:2004pz}%
  \BibitemOpen
  \bibfield  {author} {\bibinfo {author} {\bibfnamefont {G.}~\bibnamefont
  {Bertone}}, \bibinfo {author} {\bibfnamefont {D.}~\bibnamefont {Hooper}}, \
  and\ \bibinfo {author} {\bibfnamefont {J.}~\bibnamefont {Silk}},\ }\href
  {\doibase 10.1016/j.physrep.2004.08.031} {\bibfield  {journal} {\bibinfo
  {journal} {Phys. Rept.}\ }\textbf {\bibinfo {volume} {405}},\ \bibinfo
  {pages} {279} (\bibinfo {year} {2005})},\ \Eprint
  {http://arxiv.org/abs/hep-ph/0404175} {arXiv:hep-ph/0404175 [hep-ph]}
  \BibitemShut {NoStop}%
\bibitem [{\citenamefont {Cirelli}\ \emph {et~al.}(2011)\citenamefont
  {Cirelli}, \citenamefont {Corcella}, \citenamefont {Hektor}, \citenamefont
  {Hutsi}, \citenamefont {Kadastik}, \citenamefont {Panci}, \citenamefont
  {Raidal}, \citenamefont {Sala},\ and\ \citenamefont
  {Strumia}}]{Cirelli:2010xx}%
  \BibitemOpen
  \bibfield  {author} {\bibinfo {author} {\bibfnamefont {M.}~\bibnamefont
  {Cirelli}}, \bibinfo {author} {\bibfnamefont {G.}~\bibnamefont {Corcella}},
  \bibinfo {author} {\bibfnamefont {A.}~\bibnamefont {Hektor}}, \bibinfo
  {author} {\bibfnamefont {G.}~\bibnamefont {Hutsi}}, \bibinfo {author}
  {\bibfnamefont {M.}~\bibnamefont {Kadastik}}, \bibinfo {author}
  {\bibfnamefont {P.}~\bibnamefont {Panci}}, \bibinfo {author} {\bibfnamefont
  {M.}~\bibnamefont {Raidal}}, \bibinfo {author} {\bibfnamefont
  {F.}~\bibnamefont {Sala}}, \ and\ \bibinfo {author} {\bibfnamefont
  {A.}~\bibnamefont {Strumia}},\ }\href {\doibase
  10.1088/1475-7516/2012/10/E01, 10.1088/1475-7516/2011/03/051} {\bibfield
  {journal} {\bibinfo  {journal} {JCAP}\ }\textbf {\bibinfo {volume} {1103}},\
  \bibinfo {pages} {051} (\bibinfo {year} {2011})},\ \bibinfo {note} {[Erratum:
  JCAP1210,E01(2012)]},\ \Eprint {http://arxiv.org/abs/1012.4515}
  {arXiv:1012.4515 [hep-ph]} \BibitemShut {NoStop}%
\bibitem [{\citenamefont {Bahr}\ \emph {et~al.}(2008)\citenamefont {Bahr} \emph
  {et~al.}}]{Bahr:2008pv}%
  \BibitemOpen
  \bibfield  {author} {\bibinfo {author} {\bibfnamefont {M.}~\bibnamefont
  {Bahr}} \emph {et~al.},\ }\href {\doibase 10.1140/epjc/s10052-008-0798-9}
  {\bibfield  {journal} {\bibinfo  {journal} {Eur. Phys. J.}\ }\textbf
  {\bibinfo {volume} {C58}},\ \bibinfo {pages} {639} (\bibinfo {year}
  {2008})},\ \Eprint {http://arxiv.org/abs/0803.0883} {arXiv:0803.0883
  [hep-ph]} \BibitemShut {NoStop}%
\bibitem [{\citenamefont {Sjöstrand}\ \emph {et~al.}(2015)\citenamefont
  {Sjöstrand}, \citenamefont {Ask}, \citenamefont {Christiansen},
  \citenamefont {Corke}, \citenamefont {Desai}, \citenamefont {Ilten},
  \citenamefont {Mrenna}, \citenamefont {Prestel}, \citenamefont {Rasmussen},\
  and\ \citenamefont {Skands}}]{Sjostrand:2014zea}%
  \BibitemOpen
  \bibfield  {author} {\bibinfo {author} {\bibfnamefont {T.}~\bibnamefont
  {Sjöstrand}}, \bibinfo {author} {\bibfnamefont {S.}~\bibnamefont {Ask}},
  \bibinfo {author} {\bibfnamefont {J.~R.}\ \bibnamefont {Christiansen}},
  \bibinfo {author} {\bibfnamefont {R.}~\bibnamefont {Corke}}, \bibinfo
  {author} {\bibfnamefont {N.}~\bibnamefont {Desai}}, \bibinfo {author}
  {\bibfnamefont {P.}~\bibnamefont {Ilten}}, \bibinfo {author} {\bibfnamefont
  {S.}~\bibnamefont {Mrenna}}, \bibinfo {author} {\bibfnamefont
  {S.}~\bibnamefont {Prestel}}, \bibinfo {author} {\bibfnamefont {C.~O.}\
  \bibnamefont {Rasmussen}}, \ and\ \bibinfo {author} {\bibfnamefont {P.~Z.}\
  \bibnamefont {Skands}},\ }\href {\doibase 10.1016/j.cpc.2015.01.024}
  {\bibfield  {journal} {\bibinfo  {journal} {Comput. Phys. Commun.}\ }\textbf
  {\bibinfo {volume} {191}},\ \bibinfo {pages} {159} (\bibinfo {year}
  {2015})},\ \Eprint {http://arxiv.org/abs/1410.3012} {arXiv:1410.3012
  [hep-ph]} \BibitemShut {NoStop}%
\bibitem [{\citenamefont {Christiansen}\ and\ \citenamefont
  {Sjöstrand}(2014)}]{Christiansen:2014kba}%
  \BibitemOpen
  \bibfield  {author} {\bibinfo {author} {\bibfnamefont {J.~R.}\ \bibnamefont
  {Christiansen}}\ and\ \bibinfo {author} {\bibfnamefont {T.}~\bibnamefont
  {Sjöstrand}},\ }\href {\doibase 10.1007/JHEP04(2014)115} {\bibfield
  {journal} {\bibinfo  {journal} {JHEP}\ }\textbf {\bibinfo {volume} {04}},\
  \bibinfo {pages} {115} (\bibinfo {year} {2014})},\ \Eprint
  {http://arxiv.org/abs/1401.5238} {arXiv:1401.5238 [hep-ph]} \BibitemShut
  {NoStop}%
\bibitem [{\citenamefont {Berezinsky}\ \emph {et~al.}(2002)\citenamefont
  {Berezinsky}, \citenamefont {Kachelriess},\ and\ \citenamefont
  {Ostapchenko}}]{Berezinsky:2002hq}%
  \BibitemOpen
  \bibfield  {author} {\bibinfo {author} {\bibfnamefont {V.}~\bibnamefont
  {Berezinsky}}, \bibinfo {author} {\bibfnamefont {M.}~\bibnamefont
  {Kachelriess}}, \ and\ \bibinfo {author} {\bibfnamefont {S.}~\bibnamefont
  {Ostapchenko}},\ }\href {\doibase 10.1103/PhysRevLett.89.171802} {\bibfield
  {journal} {\bibinfo  {journal} {Phys. Rev. Lett.}\ }\textbf {\bibinfo
  {volume} {89}},\ \bibinfo {pages} {171802} (\bibinfo {year} {2002})},\
  \Eprint {http://arxiv.org/abs/hep-ph/0205218} {arXiv:hep-ph/0205218 [hep-ph]}
  \BibitemShut {NoStop}%
\bibitem [{\citenamefont {Esmaili}\ and\ \citenamefont
  {Serpico}(2015)}]{Esmaili:2015xpa}%
  \BibitemOpen
  \bibfield  {author} {\bibinfo {author} {\bibfnamefont {A.}~\bibnamefont
  {Esmaili}}\ and\ \bibinfo {author} {\bibfnamefont {P.~D.}\ \bibnamefont
  {Serpico}},\ }\href {\doibase 10.1088/1475-7516/2015/10/014} {\bibfield
  {journal} {\bibinfo  {journal} {JCAP}\ }\textbf {\bibinfo {volume} {1510}},\
  \bibinfo {pages} {014} (\bibinfo {year} {2015})},\ \Eprint
  {http://arxiv.org/abs/1505.06486} {arXiv:1505.06486 [hep-ph]} \BibitemShut
  {NoStop}%
\bibitem [{\citenamefont {Buch}\ \emph {et~al.}(2015)\citenamefont {Buch},
  \citenamefont {Cirelli}, \citenamefont {Giesen},\ and\ \citenamefont
  {Taoso}}]{Buch:2015iya}%
  \BibitemOpen
  \bibfield  {author} {\bibinfo {author} {\bibfnamefont {J.}~\bibnamefont
  {Buch}}, \bibinfo {author} {\bibfnamefont {M.}~\bibnamefont {Cirelli}},
  \bibinfo {author} {\bibfnamefont {G.}~\bibnamefont {Giesen}}, \ and\ \bibinfo
  {author} {\bibfnamefont {M.}~\bibnamefont {Taoso}},\ }\href {\doibase
  10.1088/1475-7516/2015/9/037, 10.1088/1475-7516/2015/09/037} {\bibfield
  {journal} {\bibinfo  {journal} {JCAP}\ }\textbf {\bibinfo {volume} {1509}},\
  \bibinfo {pages} {037} (\bibinfo {year} {2015})},\ \Eprint
  {http://arxiv.org/abs/1505.01049} {arXiv:1505.01049 [hep-ph]} \BibitemShut
  {NoStop}%
\bibitem [{\citenamefont {Palomares-Ruiz}\ \emph {et~al.}(2015)\citenamefont
  {Palomares-Ruiz}, \citenamefont {Vincent},\ and\ \citenamefont
  {Mena}}]{Palomares-Ruiz:2015mka}%
  \BibitemOpen
  \bibfield  {author} {\bibinfo {author} {\bibfnamefont {S.}~\bibnamefont
  {Palomares-Ruiz}}, \bibinfo {author} {\bibfnamefont {A.~C.}\ \bibnamefont
  {Vincent}}, \ and\ \bibinfo {author} {\bibfnamefont {O.}~\bibnamefont
  {Mena}},\ }\href {\doibase 10.1103/PhysRevD.91.103008} {\bibfield  {journal}
  {\bibinfo  {journal} {Phys. Rev.}\ }\textbf {\bibinfo {volume} {D91}},\
  \bibinfo {pages} {103008} (\bibinfo {year} {2015})},\ \Eprint
  {http://arxiv.org/abs/1502.02649} {arXiv:1502.02649 [astro-ph.HE]}
  \BibitemShut {NoStop}%
\bibitem [{\citenamefont {D'Amico}(2018)}]{DAmico:2017dwq}%
  \BibitemOpen
  \bibfield  {author} {\bibinfo {author} {\bibfnamefont {G.}~\bibnamefont
  {D'Amico}},\ }\href {\doibase 10.1016/j.astropartphys.2018.04.002} {\bibfield
   {journal} {\bibinfo  {journal} {Astropart. Phys.}\ }\textbf {\bibinfo
  {volume} {101}},\ \bibinfo {pages} {8} (\bibinfo {year} {2018})},\ \Eprint
  {http://arxiv.org/abs/1712.04979} {arXiv:1712.04979 [astro-ph.HE]}
  \BibitemShut {NoStop}%
\bibitem [{\citenamefont {Palladino}\ \emph {et~al.}(2015)\citenamefont
  {Palladino}, \citenamefont {Pagliaroli}, \citenamefont {Villante},\ and\
  \citenamefont {Vissani}}]{Palladino:2015zua}%
  \BibitemOpen
  \bibfield  {author} {\bibinfo {author} {\bibfnamefont {A.}~\bibnamefont
  {Palladino}}, \bibinfo {author} {\bibfnamefont {G.}~\bibnamefont
  {Pagliaroli}}, \bibinfo {author} {\bibfnamefont {F.~L.}\ \bibnamefont
  {Villante}}, \ and\ \bibinfo {author} {\bibfnamefont {F.}~\bibnamefont
  {Vissani}},\ }\href {\doibase 10.1103/PhysRevLett.114.171101} {\bibfield
  {journal} {\bibinfo  {journal} {Phys. Rev. Lett.}\ }\textbf {\bibinfo
  {volume} {114}},\ \bibinfo {pages} {171101} (\bibinfo {year} {2015})},\
  \Eprint {http://arxiv.org/abs/1502.02923} {arXiv:1502.02923 [astro-ph.HE]}
  \BibitemShut {NoStop}%
\bibitem [{\citenamefont {Gandhi}\ \emph {et~al.}(1998)\citenamefont {Gandhi},
  \citenamefont {Quigg}, \citenamefont {Reno},\ and\ \citenamefont
  {Sarcevic}}]{Gandhi:1998ri}%
  \BibitemOpen
  \bibfield  {author} {\bibinfo {author} {\bibfnamefont {R.}~\bibnamefont
  {Gandhi}}, \bibinfo {author} {\bibfnamefont {C.}~\bibnamefont {Quigg}},
  \bibinfo {author} {\bibfnamefont {M.~H.}\ \bibnamefont {Reno}}, \ and\
  \bibinfo {author} {\bibfnamefont {I.}~\bibnamefont {Sarcevic}},\ }\href
  {\doibase 10.1103/PhysRevD.58.093009} {\bibfield  {journal} {\bibinfo
  {journal} {Phys. Rev.}\ }\textbf {\bibinfo {volume} {D58}},\ \bibinfo {pages}
  {093009} (\bibinfo {year} {1998})},\ \Eprint
  {http://arxiv.org/abs/hep-ph/9807264} {arXiv:hep-ph/9807264 [hep-ph]}
  \BibitemShut {NoStop}%
\bibitem [{\citenamefont {Ishihara}(2016)}]{Ishihara:2015vqt}%
  \BibitemOpen
  \bibfield  {author} {\bibinfo {author} {\bibfnamefont {A.}~\bibnamefont
  {Ishihara}} (\bibinfo {collaboration} {IceCube}),\ }\bibfield  {booktitle}
  {\emph {\bibinfo {booktitle} {{Proceedings, 34th International Cosmic Ray
  Conference (ICRC 2015): The Hague, The Netherlands, July 30-August 6,
  2015}}},\ }\href {\doibase 10.22323/1.236.1064} {\bibfield  {journal}
  {\bibinfo  {journal} {PoS}\ }\textbf {\bibinfo {volume} {ICRC2015}},\
  \bibinfo {pages} {1064} (\bibinfo {year} {2016})}\BibitemShut {NoStop}%
\bibitem [{\citenamefont {Wilks}(1938)}]{Wilks:1938dza}%
  \BibitemOpen
  \bibfield  {author} {\bibinfo {author} {\bibfnamefont {S.~S.}\ \bibnamefont
  {Wilks}},\ }\href {\doibase 10.1214/aoms/1177732360} {\bibfield  {journal}
  {\bibinfo  {journal} {Annals Math. Statist.}\ }\textbf {\bibinfo {volume}
  {9}},\ \bibinfo {pages} {60} (\bibinfo {year} {1938})}\BibitemShut {NoStop}%
\bibitem [{\citenamefont {Schatz}\ \emph {et~al.}(2003)\citenamefont {Schatz}
  \emph {et~al.}}]{Schatz:2003aw}%
  \BibitemOpen
  \bibfield  {author} {\bibinfo {author} {\bibfnamefont {G.}~\bibnamefont
  {Schatz}} \emph {et~al.},\ }in\ \href
  {http://www-rccn.icrr.u-tokyo.ac.jp/icrc2003/PROCEEDINGS/PDF/566.pdf} {\emph
  {\bibinfo {booktitle} {{Proceedings, 28th International Cosmic Ray Conference
  (ICRC 2003): Tsukuba, Japan, July 31-August 7, 2003}}}}\ (\bibinfo {year}
  {2003})\ pp.\ \bibinfo {pages} {2293--2296}\BibitemShut {NoStop}%
\bibitem [{\citenamefont {Hoerandel}(2003)}]{Hoerandel:2002yg}%
  \BibitemOpen
  \bibfield  {author} {\bibinfo {author} {\bibfnamefont {J.~R.}\ \bibnamefont
  {Hoerandel}},\ }\href {\doibase 10.1016/S0927-6505(02)00198-6} {\bibfield
  {journal} {\bibinfo  {journal} {Astropart. Phys.}\ }\textbf {\bibinfo
  {volume} {19}},\ \bibinfo {pages} {193} (\bibinfo {year} {2003})},\ \Eprint
  {http://arxiv.org/abs/astro-ph/0210453} {arXiv:astro-ph/0210453 [astro-ph]}
  \BibitemShut {NoStop}%
\bibitem [{\citenamefont {Aguilar}\ \emph {et~al.}(2014)\citenamefont {Aguilar}
  \emph {et~al.}}]{Aguilar:2014fea}%
  \BibitemOpen
  \bibfield  {author} {\bibinfo {author} {\bibfnamefont {M.}~\bibnamefont
  {Aguilar}} \emph {et~al.} (\bibinfo {collaboration} {AMS}),\ }\href {\doibase
  10.1103/PhysRevLett.113.221102} {\bibfield  {journal} {\bibinfo  {journal}
  {Phys. Rev. Lett.}\ }\textbf {\bibinfo {volume} {113}},\ \bibinfo {pages}
  {221102} (\bibinfo {year} {2014})}\BibitemShut {NoStop}%
\bibitem [{\citenamefont {Aharonian}\ \emph
  {et~al.}(2006{\natexlab{a}})\citenamefont {Aharonian} \emph
  {et~al.}}]{Aharonian:2006au}%
  \BibitemOpen
  \bibfield  {author} {\bibinfo {author} {\bibfnamefont {F.}~\bibnamefont
  {Aharonian}} \emph {et~al.} (\bibinfo {collaboration} {H.E.S.S.}),\ }\href
  {\doibase 10.1038/nature04467} {\bibfield  {journal} {\bibinfo  {journal}
  {Nature}\ }\textbf {\bibinfo {volume} {439}},\ \bibinfo {pages} {695}
  (\bibinfo {year} {2006}{\natexlab{a}})},\ \Eprint
  {http://arxiv.org/abs/astro-ph/0603021} {arXiv:astro-ph/0603021 [astro-ph]}
  \BibitemShut {NoStop}%
\bibitem [{\citenamefont {Chernyakova}\ \emph {et~al.}(2011)\citenamefont
  {Chernyakova}, \citenamefont {Malyshev}, \citenamefont {Aharonian},
  \citenamefont {Crocker},\ and\ \citenamefont {Jones}}]{Chernyakova:2011zz}%
  \BibitemOpen
  \bibfield  {author} {\bibinfo {author} {\bibfnamefont {M.}~\bibnamefont
  {Chernyakova}}, \bibinfo {author} {\bibfnamefont {D.}~\bibnamefont
  {Malyshev}}, \bibinfo {author} {\bibfnamefont {F.~A.}\ \bibnamefont
  {Aharonian}}, \bibinfo {author} {\bibfnamefont {R.~M.}\ \bibnamefont
  {Crocker}}, \ and\ \bibinfo {author} {\bibfnamefont {D.~I.}\ \bibnamefont
  {Jones}},\ }\href {\doibase 10.1088/0004-637X/726/2/60} {\bibfield  {journal}
  {\bibinfo  {journal} {Astrophys. J.}\ }\textbf {\bibinfo {volume} {726}},\
  \bibinfo {pages} {60} (\bibinfo {year} {2011})},\ \Eprint
  {http://arxiv.org/abs/1009.2630} {arXiv:1009.2630 [astro-ph.HE]} \BibitemShut
  {NoStop}%
\bibitem [{\citenamefont {Aharonian}\ \emph
  {et~al.}(2006{\natexlab{b}})\citenamefont {Aharonian} \emph
  {et~al.}}]{Aharonian:2006wh}%
  \BibitemOpen
  \bibfield  {author} {\bibinfo {author} {\bibfnamefont {F.}~\bibnamefont
  {Aharonian}} \emph {et~al.} (\bibinfo {collaboration} {H.E.S.S.}),\ }\href
  {\doibase 10.1103/PhysRevLett.97.221102, 10.1103/PhysRevLett.97.249901}
  {\bibfield  {journal} {\bibinfo  {journal} {Phys. Rev. Lett.}\ }\textbf
  {\bibinfo {volume} {97}},\ \bibinfo {pages} {221102} (\bibinfo {year}
  {2006}{\natexlab{b}})},\ \bibinfo {note} {[Erratum: Phys. Rev.
  Lett.97,249901(2006)]},\ \Eprint {http://arxiv.org/abs/astro-ph/0610509}
  {arXiv:astro-ph/0610509 [astro-ph]} \BibitemShut {NoStop}%
\end{thebibliography}
\end{document}